\documentclass[11pt]{article}
\usepackage{jheppub}
\usepackage{amsmath,amssymb,amsfonts,graphicx,slashed,amsthm,mathtools,upgreek, enumerate, tensor}
\usepackage[dvipsnames]{xcolor}
\usepackage{arydshln}

\usepackage{booktabs}

\usepackage{float}

\usepackage{caption}

\usepackage{comment}
\usepackage{hyperref}
\usepackage[utf8]{inputenc}
\usepackage[titletoc]{appendix}

\usepackage{cleveref}

\usepackage{braket}
\usepackage{bm}
\usepackage{cancel}
\usepackage{multirow}

\usepackage{subcaption}

\numberwithin{equation}{section}

\allowdisplaybreaks

\newcommand{\q}[0]{\mathtt{q}}
\newcommand{\km}[0]{\mathtt{k}}
\newcommand{\tr}[0]{\mathrm{Tr}}

\title{Multipartite Markov Gaps and Entanglement Wedge Multiway Cuts}

\author[a, b]{Norihiro Iizuka}
 
\author[b]{Akihiro Miyata} 
 
\author[c]{and Mitsuhiro Nishida}
 
\affiliation[a]{\it Department of Physics, National Tsing Hua University, Hsinchu 300044, Taiwan}

\affiliation[b]{\it Yukawa Institute for Theoretical Physics, Kyoto University, Kyoto 606-8502, Japan}

\affiliation[c]{\it National Institute of Technology, Yuge College, Ehime 794-2593, Japan}
 
\emailAdd{iizuka@phys.nthu.edu.tw}
\emailAdd{akihiro.miyata@yukawa.kyoto-u.ac.jp}
\emailAdd{mnishida124@gmail.com}

\abstract{
The Markov gap, defined as the difference between reflected entropy and mutual information, serves as a diagnostic for quantum recoverability and multipartite entanglement. In holographic settings, it admits a geometric interpretation as the deviation between entanglement wedge cross-sections and RT surfaces. 
Motivated by this holographic perspective, we propose a generalization of the Markov gap to multipartite systems by using a reflected multi-entropy. The resulting Multipartite Markov gap can capture geometric obstructions to bulk reconstruction.
We investigate the properties of this quantity from both information-theoretic and holographic viewpoints, and examine its potential operational significance through candidate recovery maps. We further introduce the  genuine reflected multi-entropy, which is designed to vanish for states containing only lower-partite entanglement. Together, these quantities offer complementary probes of recoverability and multipartite structure in holographic quantum systems.}

\preprint{}

\begin{document}

\maketitle

\parskip=10pt

\section{Introduction}

In recent developments of quantum information theory and holography, the notion of the Markov gap has emerged as a powerful diagnostic for evaluating the optimal fidelity of a specific quantum recovery process~\cite{Akers:2019gcv, Hayden:2021gno}.  
The Markov gap is defined in terms of the reflected entropy~\cite{Dutta:2019gen}, which is constructed from the canonical purification of a bipartite mixed state derived from a tripartite pure state \( \rho_{ABC} \).  
By tracing out subsystem \( C \), one obtains a reduced density matrix \( \rho_{AB} \), whose canonical purification resides in a doubled Hilbert space \( \mathcal{H}_A \otimes \mathcal{H}_{A^*} \otimes \mathcal{H}_B \otimes \mathcal{H}_{B^*} \).  
The reflected entropy \( S_R(A:B) \) is then defined as the entanglement entropy of the bipartition between \( AA^* \) and \( BB^* \) in this purified state.  
The Markov gap is given by the difference between the reflected entropy and the mutual information:
\begin{equation}
\begin{aligned}
\text{Markov gap} &= S_R(A:B) - I(A:B) \,, \\
I(A:B)&= S(A) + S(B) - S(AB) \,.
\end{aligned}
\end{equation}

The Markov gap quantifies the deviation of a tripartite state \( \rho_{ABC} \) from satisfying the quantum Markov condition~\cite{Hayden:2021gno}, and is operationally significant in the context of quantum state reconstruction.  
When the Markov gap vanishes, \textit{i.e.}, \( S_R(A:B) = I(A:B) \), the conditional mutual information \( I(A:B^*|B) \) is zero, and the canonical purification of \( \rho_{AB} \) exhibits an exact quantum Markov structure.  
In this case, a perfect recovery map (such as the Petz map~\cite{Petz:1986tvy,Petz:1988usv}) exists that enables full reconstruction of the tripartite purification \( \rho_{ABB^*} \) from the reduced state \( \rho_{AB} \), by acting only on subsystem \( B \).  
This implies that there is no obstruction to bulk reconstruction in the corresponding holographic setting.

In holographic settings, the Markov gap admits a geometric interpretation: the reflected entropy corresponds to the area of the entanglement wedge cross-section, while the mutual information corresponds to the area of boundary-anchored minimal surfaces as prescribed by the RT (Ryu--Takayanagi) formula \cite{Ryu:2006bv}.  
Thus, the Markov gap geometrically represents the difference between a non-minimal surface passing through the entanglement wedge and the RT surfaces, serving as a geometric obstruction to perfect recovery.  
Moreover, since the Markov gap captures the amount of tripartite entanglement, its vanishing indicates that all entanglement can be captured by bipartite correlations, while a nonzero Markov gap implies the presence of genuine multipartite entanglement~\cite{Akers:2019gcv}.

From an operational standpoint, a nonzero Markov gap (\( S_R > I \)) reflects a fundamental limitation in the fidelity of any recovery channel.  
In holography, this obstruction manifests geometrically when the entanglement wedge cross-section develops nontrivial boundaries—such as ``corners'' where bulk extremal surfaces intersect.  
These geometric features correspond to multipartite entanglement patterns in the boundary state that cannot be captured by any local recovery map, thus signaling a breakdown of reconstructability in certain regions of the bulk.

Motivated by the intimate connection between the Markov gap and reconstructability of the holographic bulk, the central goal of this paper is to extend the notion of the Markov gap beyond tripartite systems.
Since the original Markov gap is defined as the difference between reflected entropy and a minimal combination of mutual informations, this construction allows a natural extension: the generalized Markov gap can be formulated as the deviation of the multi-partite generalized reflected entropy from its minimal consistent value under bipartite decompositions. In this paper, we initiate a systematic study of such generalized Markov gaps.

A key ingredient in this generalization is the extension of reflected entropy to multipartite systems. As previously mentioned, the reflected entropy \( S_R(A:B) \) can be viewed as the entanglement entropy between the subsystems \( AA^* \) and \( BB^* \) in the canonical purification of a bipartite state \( \rho_{AB} \). It is therefore natural to expect that in the multipartite setting, a generalization of reflected entropy should involve some form of multipartite entanglement entropy. Indeed, as proposed in~\cite{Yuan:2024yfg}, for a general \( \q \)-partite state \( \rho_{A_1 A_2 \cdots A_{\q}} \), one may trace out the final subsystem \( A_{\q} \) and consider a canonical purification of the remaining reduced state \( \rho_{A_1 A_2 \cdots A_{\q-1}} \). This yields a purified state on the Hilbert space \( \mathcal{H}_{A_1} \otimes \mathcal{H}_{A_1^*}  \otimes \cdots \otimes \mathcal{H}_{A_{\q-1}} \otimes \mathcal{H}_{ A_{\q-1}^*} \), where each \( A_i^* \) is an auxiliary system for canonical purification. One can then define a multipartite version of reflected entropy by the multi-entropy \cite{Gadde:2022cqi,Penington:2022dhr,Gadde:2023zzj,Gadde:2023zni}, a generalization of the entanglement entropy to the multi-partite system, among the \( \q-1 \) subsystems $ A_1 A_1^*$, $ \ldots$ , $A_{\q-1} A_{\q-1}^* $, such that it captures their total or genuine multipartite correlations. Given this, one can define a generalized Markov gap as the difference between the $\mathtt{q}$-partite reflected entropy and its minimal in holographic settings\footnote{Note that this minimal is different from the proposed minimal in \cite{Yuan:2024yfg}.}.

By developing an information-theoretic framework for these generalized Markov gaps and exploring their behavior in holographic states, we aim to shed light on the structure of multipartite entanglement and its reconstructability in the holographic settings. In particular, we investigate whether such generalized gaps admit lower bounds analogous to the tripartite case, and how these bounds are encoded in the bulk geometry, especially in the presence of non-trivial entanglement wedges.

An alternative perspective on the Markov gap is that it selectively captures only the contribution from \emph{genuine} tripartite entanglement in a three-party pure state~\cite{Akers:2019gcv}. If the entanglement in \( \rho_{ABC} \) is purely bipartite {\it i.e.}, it can be decomposed as a sum of two-party correlations, then the reflected entropy saturates its lower bound given by the mutual information, and the Markov gap vanishes. Thus, a nonzero Markov gap serves as a precise indicator of irreducible tripartite entanglement in the state. This feature suggests a second route to generalization: one may attempt to define $\mathtt{q}$-partite Markov gaps as quantities that vanish for states whose entanglement structure is reducible to correlations among fewer than $\mathtt{q}$-parties, but become nonzero only in the presence of irreducible $\mathtt{q}$-partite entanglement.

While such entanglement-theoretic extensions using multi-entropy offer valuable conceptual insight, it also presents several challenges. In particular, beyond the tripartite case, it remains unclear—at least to the best of our knowledge—whether the multi-partite generalized Markov gap always admits a well-defined operational meaning. In bipartite and tripartite systems, the recovery maps and the associated fidelity bounds are well understood. However, in the multipartite setting, as far as we are aware, there exists no unique notion of multipartite Markov chain or its corresponding recovery process. To address this issue, in this paper we also examine the possibility of reconstruction using several variants of the Petz map (although it remains unclear whether they yield the optimal recovery), and evaluate the corresponding Markov gaps. Through this approach, we further explore the potential operational significance of the generalized multi-partite Markov gap from this reconstruction-based perspective.

The structure of this paper is as follows. In Section~\ref{sec:preliminaries}, we begin with a brief review of the original Markov gap, including its definition via reflected entropy and its geometric interpretation in holographic settings. We then summarize relevant aspects of multipartite entanglement entropy, especially multi-entropy measures, as these provide the natural framework in which to generalize reflected entropy to multi-party systems.
Section~\ref{sec:MMG} contains the core contribution of this paper. There, using a generalization of reflected entropy to multipartite systems through the canonical purification and multi-entropy, we define the corresponding generalized Markov gap as the deviation from a natural minimal configuration. This generalization is motivated by holography, and we call this generalized Markov gap as {\it Multipartite Markov gap}. We analyze its properties both from information-theoretic and holographic perspectives.
In Section~\ref{genuineMG}, we explore an alternative approach to generalizing the Markov gap, based on its role as a detector of genuine tripartite entanglement. By extending this characterization to higher-party systems, we obtain quantities sensitive only to irreducible multipartite entanglement just like genuine multi-entropy \cite{Iizuka:2025ioc, Iizuka:2025caq}. As we will discuss, this perspective leads to measures that capture complementary aspects of multipartite entanglement. We call this quantity {\it genuine reflected multi-entropy}. 
Finally, Section~\ref{sec:discussion} concludes with a summary of our findings and a discussion of open problems and potential directions for future research.

\section{Review of tripartite Markov Gap and reflected multi-entropy}\label{sec:preliminaries}
In this section, we review several multipartite entanglement measures that will serve as the basis for our construction of a generalized Markov gap for multipartite systems. In particular, we focus on reflected entropy, the Markov gap, multi-entropy, and their reflected counterparts. We briefly recall their definitions, properties, and interrelations, with an emphasis on the tripartite case, which will be directly relevant to our proposal in the next section.

\subsection{Reflected entropy}\label{subsec:reflected-entropy}
We begin by briefly reviewing the concept of reflected entropy~\cite{Dutta:2019gen}.
Let $\rho_{AB}$ be a mixed state on a bipartite system, obtained by tracing out subsystem $C$ from a tripartite pure state $\ket{\psi}_{ABC}$:
\begin{equation}\label{21rhoAB}
	\rho_{AB}=\tr_{C} \left[ \ket{\psi}_{ABC} \bra{\psi}\right].
\end{equation}

\begin{figure}[h]
    \centering
    \begin{subfigure}[b]{0.42\linewidth}
        \includegraphics[width=\linewidth]{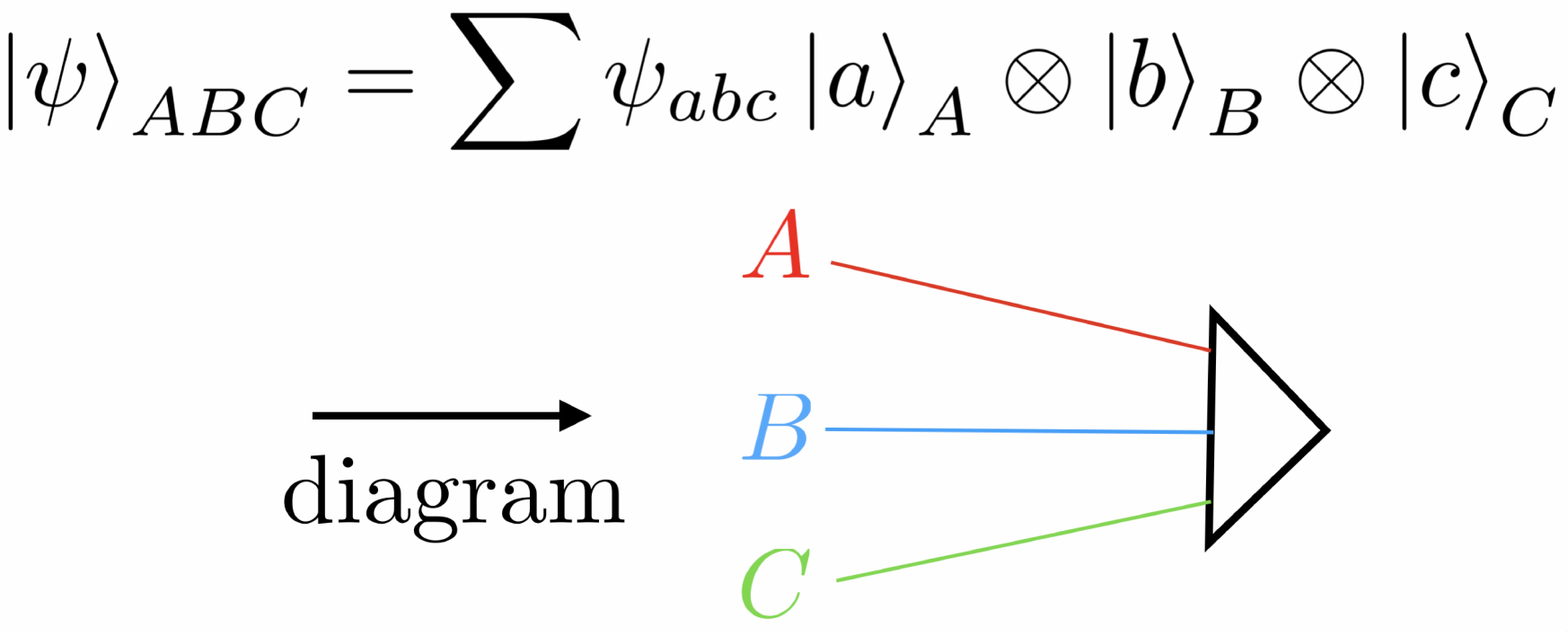}
        \caption{Diagram of the original pure state}
        \label{fig:diagram-purified-state-q=3-1}
    \end{subfigure}
    \hfill
    \begin{subfigure}[b]{0.48\linewidth}
        \includegraphics[width=\linewidth]{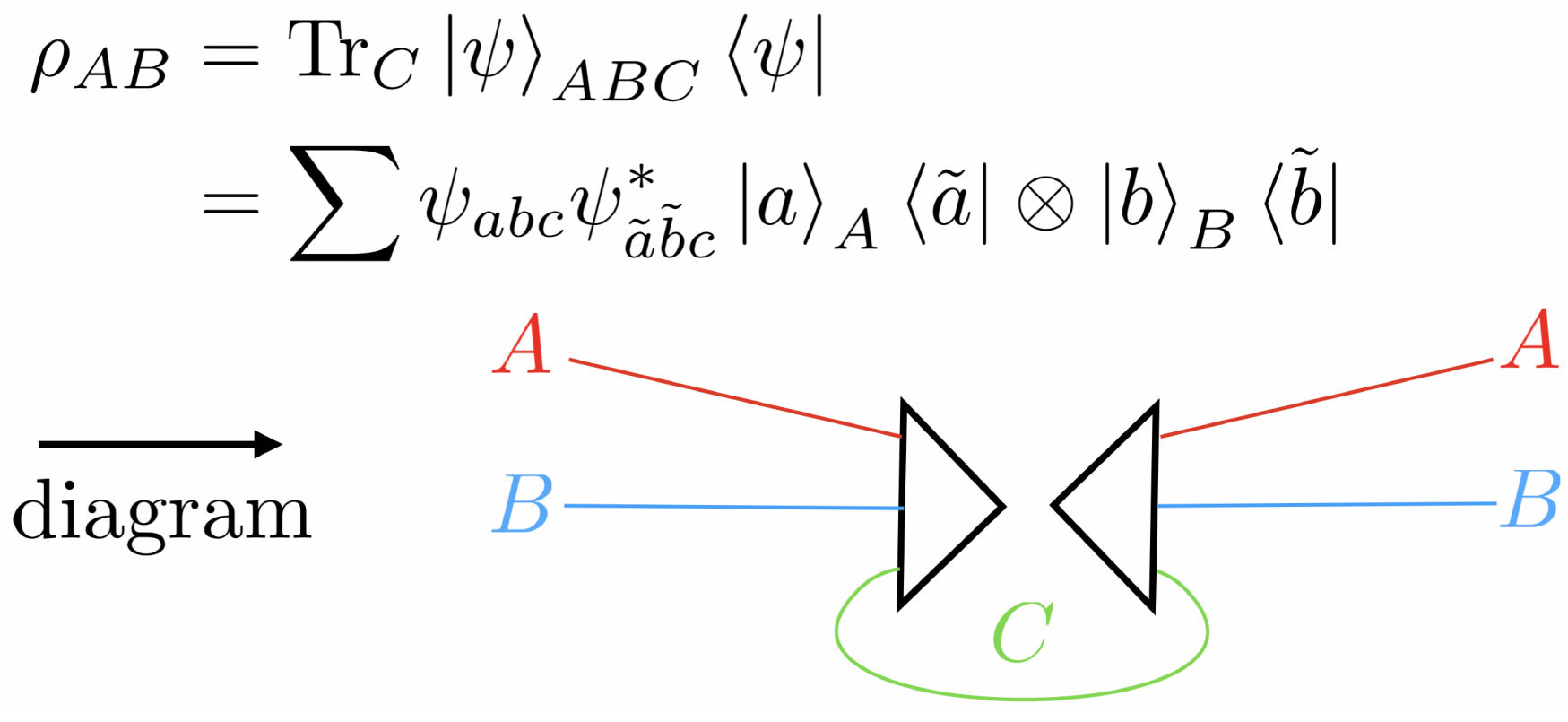}
        \caption{Diagram of the mixed state}
        \label{fig:diagram-purified-state-q=3-2}
    \end{subfigure}
    \vskip\baselineskip
    \begin{subfigure}[b]{1\linewidth}
        \includegraphics[width=0.9\linewidth]{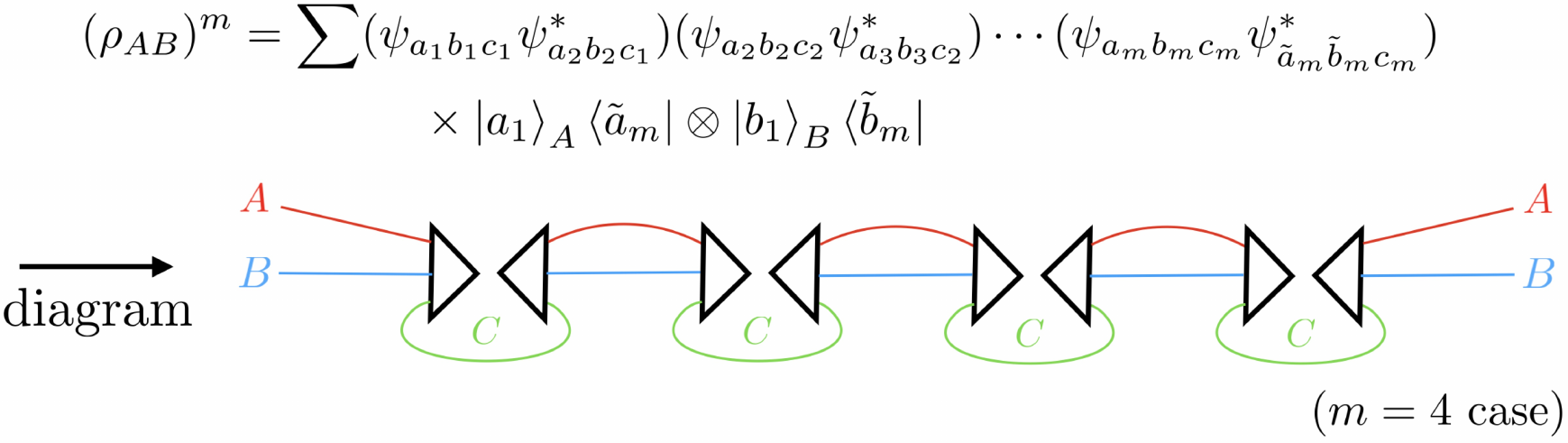}
        \caption{Diagram of the $m$-th power of the mixed state}
        \label{fig:diagram-purified-state-q=3-3}
        \vskip\baselineskip
        \includegraphics[width=0.9\linewidth]{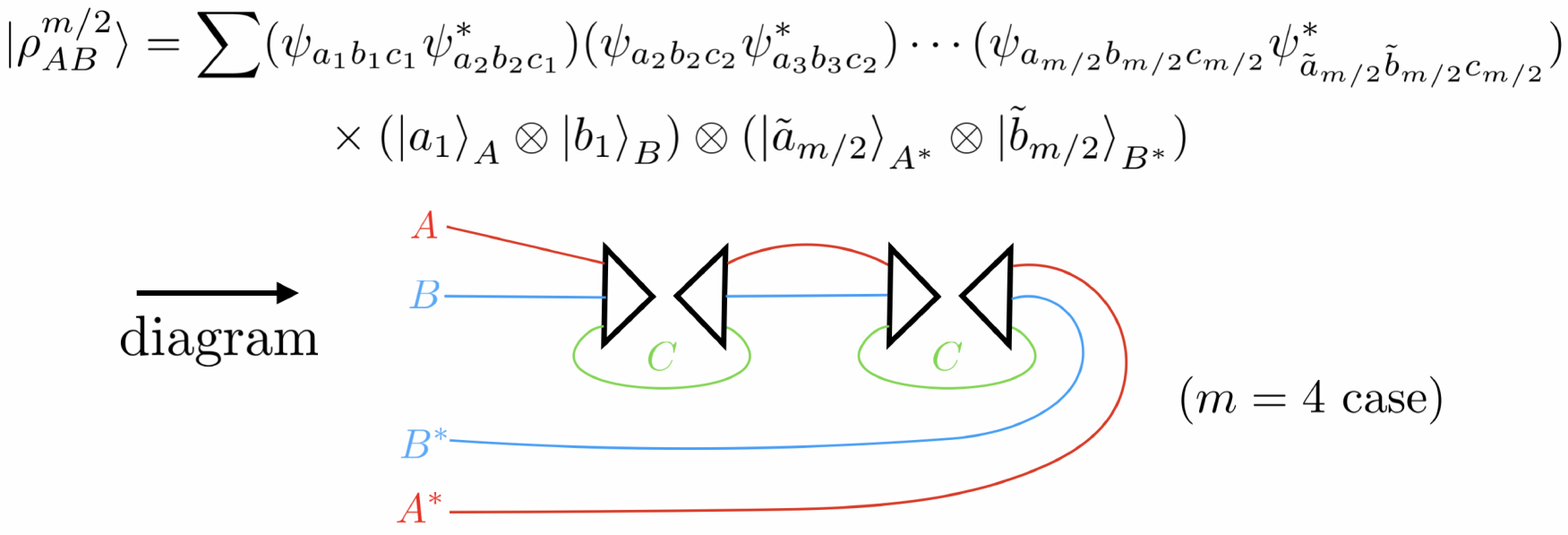}
        \caption{Diagram of the purification of the $m$-th power of the mixed state}
        \label{fig:diagram-purified-state-q=3-4}
    \end{subfigure}
    
    \caption{Diagrammatic representation of the purified state (\ref{eq:renyi-purified-state-q=3}) for $m=4$ case.}
    \label{fig:diagram-purified-state-q=3}
\end{figure}

Consider $(\rho_{AB})^m$, where $m$ is a positive even integer, $m \in 2\mathbb{N}$.  
To construct its canonical purification, we introduce auxiliary systems $A^*$ and $B^*$ that canonically purify subsystems $A$ and $B$, respectively. Following~\cite{Dutta:2019gen}, the canonical purified state is given by
\begin{equation}\label{eq:renyi-purified-state-q=3}
	\ket{ (\rho_{AB})^{m/2} }_{\substack{AB \\ A^{*}B^{*}}},
\end{equation}
where it satisfies the following condition, 
\begin{equation}
	\tr_{A^{*}B^{*}}\left[ \ket{ (\rho_{AB})^{m/2} }_{\substack{AB \\ A^{*}B^{*}}}  \bra{ (\rho_{AB})^{m/2} }  \right]=(\rho_{AB})^{m}.
\end{equation}
The procedure for the R\'enyi version of the canonical purification is illustrated schematically in Figure~\ref{fig:diagram-purified-state-q=3}.

For the purified state \eqref{eq:renyi-purified-state-q=3}, we define the $(m,n)$ R\'enyi reflected entropy~\cite{Dutta:2019gen} as the $n$-th R\'enyi entropy of the subsystem $AA^*$:
\begin{equation}\label{eq:def-renyi-reflected-entropy}
	 \begin{aligned}
	 	S_{R}^{(m,n)}(A:B) &\coloneqq \left.S_{n}(AA^{*})\right|_{\ket{ (\rho_{AB})^{m/2} }}\\
	 	&\coloneqq  \frac{1}{1-n} \log \frac{Z_{n \, (m)}}{(Z_{1 \, (m)})^{n}} \,,
	 \end{aligned}
\end{equation}
Here, $Z_{n\,(m)}$ denotes the partition function to evaluate the $n$-th R\'enyi entanglement entropy between the subsystems $AA^*$ and $BB^*$ for the purified state~\eqref{eq:renyi-purified-state-q=3}:  
\begin{equation}\label{eq:def-Zn(m)}
	Z_{n \, (m)}\coloneqq \tr_{AA^{*}}\left[ \left( \tr_{BB^{*}} \left[ \ket{ (\rho_{AB})^{m/2} }_{ \substack{AB\\ A^{*}B^{*} }}  \bra{ (\rho_{AB})^{m/2} }  \right]  \right)^{n} \right].
\end{equation}
The $m$-dependence arises from the fact that the purified state~\eqref{eq:renyi-purified-state-q=3} itself is defined in terms of $\ket{(\rho_{AB})^{m/2}}$. 
Note that this state \eqref{eq:renyi-purified-state-q=3} is not normalized to be unity unless $\rho_{AB}$ is a pure state,
\begin{equation}\label{eq:normalization-purified-state-q=3}
	\tr_{ABA^{*}B^{*}}\left[ \ket{ (\rho_{AB})^{m/2} }_{\substack{AB \\ A^{*}B^{*}}}  \bra{ (\rho_{AB})^{m/2} }  \right]=\tr_{AB}\left[(\rho_{AB})^{m}\right] = Z_{1 \, (m)}.
\end{equation}
For this reason, the $n$-th power of the quantity is included in the denominator of the logarithm in \eqref{eq:def-renyi-reflected-entropy} as a normalization.

\begin{figure}[t]
    \centering
    \includegraphics[width=0.9\linewidth]{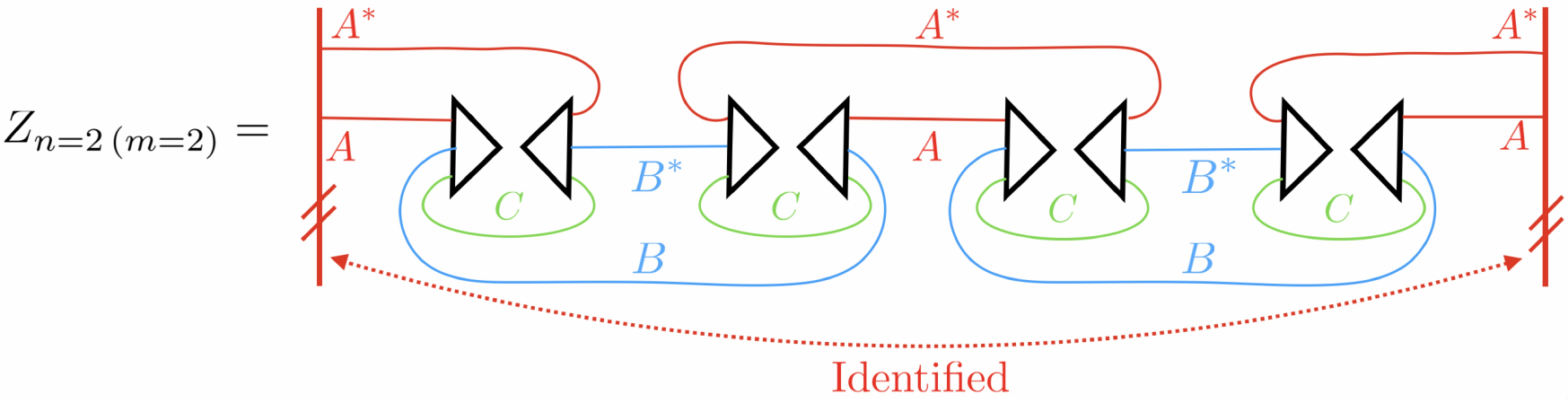}
    \caption{Diagrammatic representation of $Z_{n=2 \, (m=2)}$, (\ref{eq:def-Zn(m)}) for $m=n=2$ case.}
    \label{fig:Renyi-reflected-multi-q=3-n=2}
\end{figure}

We can visually represent the partition function $Z_{n \, (m)}$ by using the diagrams used in Figures \ref{fig:diagram-purified-state-q=3}. For instance, $Z_{n=2 \, (m=2)}$ can be represented by the diagram as illustrated in Figure \ref{fig:Renyi-reflected-multi-q=3-n=2}.

Taking the analytic continuation $m \to 1$, $n \to 1$, this reduces to the reflected entropy \cite{Dutta:2019gen} 
\begin{equation}
	S_{R}(A:B)\coloneqq \lim_{m\to 1,\\ n\to 1}S_{R}^{(m,n)}(A:B).
\end{equation}

\subsection{Tripartite Markov gap}
From the reflected entropy, we can define the Markov gap \cite{Akers:2019gcv,Hayden:2021gno}.
For an arbitrary state $\rho_{AB}$ on systems $AB$, the reflected entropy $S_{R}(A:B)$ is known to be lower bounded by the bi-partite mutual information $I^{(2)}(A:B)$ \cite{Dutta:2019gen},
\begin{equation}\label{eq:inequality-q=3-reflected-entropy-and-mutual-information}
	S_{R}(A:B)\geq I^{(2)}(A:B),
\end{equation}
where the bi-partite mutual information $I^{(2)}(A:B)$ is given by
\begin{equation}
	I^{(2)}(A:B)=S(A)+S(B)-S(AB).
\end{equation}
The inequality \eqref{eq:inequality-q=3-reflected-entropy-and-mutual-information} can be derived from the strong subadditivity in the following form \cite{Dutta:2019gen}\footnote{Or equivalently, one can start with the strong subadditivity using mutual information
\begin{equation}
    I(A^{*}:AB) \geq I(A^{*}:A).
\end{equation}
By rewriting the mutual information in terms of entanglement entropy, we get the inequality \eqref{eq:SSA-q=3-purified-situation}.},
\begin{equation}\label{eq:SSA-q=3-purified-situation}
	S(AA^{*})+S(AB)\geq S(A)+S(B^{*}).
\end{equation}
Rearranging the terms in the above inequality and using the fact $S(B^{*})=S(B)$ and $S_R(A:B) = S(AA^{*})$, we get the inequality \eqref{eq:inequality-q=3-reflected-entropy-and-mutual-information}. 

Motivated by this inequality, the (tripartite) Markov gap $MG(A:G)$ \cite{Akers:2019gcv,Hayden:2021gno} is defined by
\begin{equation}\label{eq:def-Markov-gap}
	MG(A:B)=S_{R}(A:B)-I^{(2)}(A:B),
\end{equation}
and it always satisfies
\begin{equation}\label{eq:positivity-Markov-Gap}
	MG(A:B)\geq 0.
\end{equation}
A remarkable property of the Markov gap is that it vanishes for original tripartite state $\ket{\psi}_{ABC}$\footnote{Here, original tripartite state means the state $\ket{\psi}_{ABC}$ before we trace over $C$ in \eqref{21rhoAB}.} containing only bipartite entanglement. In this sense, the Markov gap is a  good diagnostic for tripartite entanglement.

We can generalize the Markov gap to the R\'enyi case \cite{Liu:2024ulq,Sohal:2023hst,Berthiere:2023gkx}, that is, the R\'enyi Markov gap defined by
\begin{equation}\label{eq:def-Renyi-Markov-gap}
	MG_{m,n}(A:B)=S_{R}^{(m,n)}(A:B)-I_{n}^{(2)}(A:B),
\end{equation} 
where $I_{n}^{(2)}(A:B)$ is the bipartite R\'enyi mutual information given by
\begin{equation}\label{eq:def-renyi-bipartite-mutual-information}
	\begin{aligned}
		I_{n}^{(2)}(A:B)&\coloneqq S_{n}(A)+S_{n}(B)-S_{n}(AB).
	\end{aligned}
\end{equation}
Clearly by taking the limits $m \to 1$, $n\to 1$, the R\'enyi Markov gap is reduced to the Markov gap:
\begin{equation}
	\lim_{m\to 1}\lim_{n\to 1} MG_{m,n}(A:B)=MG(A:B).
\end{equation}
Note that, unlike in the case of \eqref{eq:positivity-Markov-Gap}, the R\'enyi Markov gap can be negative, since the R\'enyi version of the strong subadditivity \eqref{eq:SSA-q=3-purified-situation} does not generally hold \cite{Linden_2013}.

Also, note that the R\'enyi Markov gap has the property that it vanishes for the GHZ state. This is because the R\'enyi reflected entropy and the R\'enyi mutual information for the GHZ state are given by
\begin{equation}
	\left.S_{R}^{(m,n)}(A:B)\right._{\ket{\mathrm{GHZ}_{3}}}=\left.I_{n}^{(2)}(A:B)\right._{\ket{\mathrm{GHZ}_{3}}}=\log 2 \qquad \Longrightarrow \qquad \left.MG_{m,n}(A:B)\right._{\ket{\mathrm{GHZ}_{3}}}=0
\end{equation}
where $\ket{\mathrm{GHZ}_{3}}$ is given by
\begin{equation}\label{eq:def-tripartie-GHZ}
	\ket{\mathrm{GHZ}_{3}}_{ABC}=\frac{1}{\sqrt{2}}\big( \ket{000}_{ABC}+ \ket{111}_{ABC} \big).
\end{equation}
Although the (R\'enyi) Markov gap is a good diagnostic for tripartite entanglement, since it cannot quantify the GHZ state, it only detects tripartite entanglement such as the W state \cite{Hayden:2021gno}.

\subsection{R\'enyi Multi-entropy}
The R\'enyi multi-entropy is a generalization of the  R\'enyi entanglement entropy for $\q$-partite subsystems.
For a pure state $\ket{\psi}_{A_{1},A_{2},\cdots,A_{\q}}$ on $\q$-partite subsystems, the $n$-th R\'enyi $\q$-partite multi-entropy is defined by \cite{Gadde:2022cqi,Penington:2022dhr,Gadde:2023zzj,Gadde:2023zni}
\begin{equation}\label{eq:def-renyi-multi-entropy}
S_n^{(\q )}\left(A_1: A_2: \cdots: A_{\q}\right)\coloneqq\frac{1}{1-n}\cdot  \frac{1}{n^{\q-2}} \log \frac{Z_n^{(\q )}}{\left(Z_1^{(\q )}\right)^{n^{\q-1}}},
\end{equation}
\begin{equation}\label{eq:def-Zn(m)^{q}}
	Z_n^{(\q )}\coloneqq\left\langle\psi\right|^{\otimes n^{\q -1}} \Sigma_1\left(g_1\right) \Sigma_2\left(g_2\right) \ldots \Sigma_{\q }\left(g_{\q }\right) \left| \psi\right\rangle^{\otimes n^{\q -1}},
\end{equation}
where $\Sigma_i\left(g_{\km}\right)$ are the twist operators inducing the permutation action of $g_{\km}$ on indices of density matrices for subsystem $A_{\km}$\footnote{Rigorously speaking, such twist operators are most robustly defined in 2d CFTs~\cite{Yuan:2024yfg}. However, even in higher dimensions we assume a similar operator plays the role.}. Here, the permutation of $g_{\km}$ is defined by the discrete translation along ${\km}$-th direction on a $(\q-1)$-dimensional hypercube of length $n$;
\begin{equation}
	g_{\km}\cdot \left(x_{1},x_{2},\dots, x_{\km},\dots,x_{\q-1}\right)= \left(x_{1},x_{2},\dots, x_{\km}+1,\dots,x_{\q-1}\right), \qquad 1\leq \km \leq \q-1,
\end{equation}
\begin{equation}
	g_{\q}\cdot \left(x_{1},x_{2},\dots,x_{\q-1}\right) =\left(x_{1},x_{2},\dots,x_{\q-1}\right),
\end{equation}
where $\left(x_{1},x_{2},\dots,x_{\q-1}\right)$ denotes an integer lattice point on the $(\q-1)$-dimensional hypercube of length $n$ with the periodic boundary condition $x_{\km}=n+1\sim 1$, and $g_{\q}$ is the identity mapping.

The $n$-th R\'enyi $\q$-partite multi-entropy defined in~\eqref{eq:def-renyi-multi-entropy} is a generalization of the conventional bipartite $n$-th R\'enyi entanglement entropy.  
Indeed, for the case $\q = 2$, equation~\eqref{eq:def-renyi-multi-entropy} reduces to the standard R\'enyi entanglement entropy.
\begin{equation}\label{eq:def-renyi-EE}
		S_n^{(2)}\left(A_1: A_2\right)=S_n(A_{1})=S_n(A_{2})=\frac{1}{1-n}\log \frac{ \tr_{A_{1}} \left[ \left(\rho_{A_{1}}\right)^{n} \right] }{ \left( \tr_{A_{1}} \left[ \rho_{A_{1}}   \right]  \right)^{n} },
\end{equation}
\begin{equation}
	\rho_{A_{1}}=\tr_{A_{2}} \left[ \ket{\psi}_{A_{1},A_{2}} \bra{\psi}   \right].\label{ReducedMatrix}
\end{equation}
This reduces to the purity in the case $n=2$ and in the $n\to1 $ limit, it reduces to the conventional entanglement entropy
\begin{equation}
	S(A)=\lim_{n\to 1}S_{n}(A).
\end{equation}
For the bipartite case $\mathtt{q}=2$, after tracing out one subsystem, only $\mathtt{q}-1 = 1$ subsystem remains, as shown in equation~\eqref{ReducedMatrix}.  
Accordingly, there is only one direction in which the matrix indices can be contracted to form the purity.

However, for the multi-partite case with $\q \ge 3$, tracing out one subsystem leaves $\mathtt{q}-1 \ge 2$ subsystems.  
In this case, there are $\mathtt{q}-1 \ge 2$ orthogonal directions along which the matrix indices can be contracted.  
These contractions form a hypercubic structure with $n^{\mathtt{q}-1}$ replicas, which explains the appearance of $\otimes n^{\q -1}$ and the twist operators in equation~\eqref{eq:def-Zn(m)^{q}}.  
These are required to generate the replica partition function $Z_n^{(\q)}$ for the R\'enyi $\q$-partite multi-entropy.

\begin{figure}[t]
    \centering
    \includegraphics[width=0.8\linewidth]{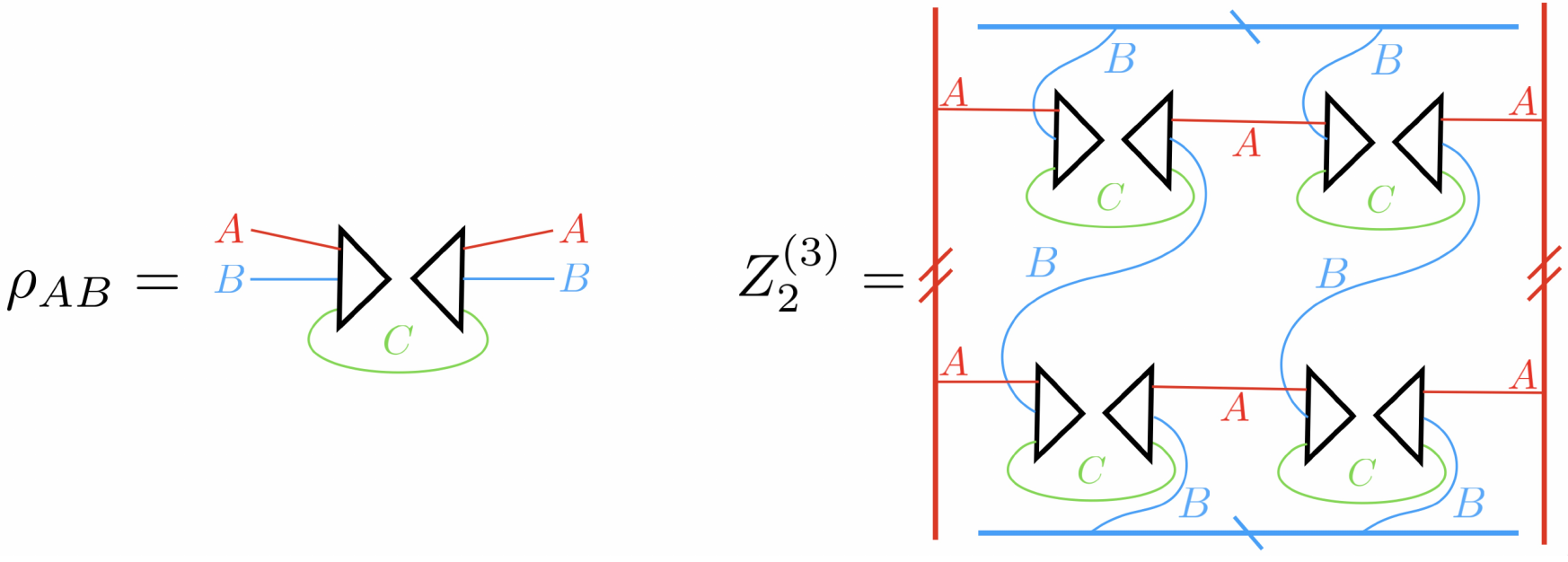}
    \caption{Contraction pattern of reduced density matrices $\rho_{AB}=\tr_{C} \left[ \ket{\psi}_{ABC} \bra{\psi}   \right]$ for $Z_{2}^{(3)}$.}
\label{fig:ContractionPattern}
\end{figure}

To see the contraction pattern explicitly, let us consider the contraction of the reduced density matrices for the $n=2$ R\'enyi $\q$-partite multi-entropy. For $\q=2$, we arrange $n^{\q-1}=n$ reduced density matrices on a one-dimensional line and contract the indices of the neighboring reduced density matrices. This contraction for $n=2$ is explicitly represented by
\begin{align}
\tr_{A} \left[ \left(\rho_{A}\right)^{2} \right]=\left(\rho_{A}\right)_{\alpha_1}^{\alpha_2} \left(\rho_{A}\right)_{\alpha_2}^{\alpha_1},
\end{align}
where $\left(\rho_{A}\right)_{\alpha_i}^{\alpha_j}$ is the matrix representation of the reduced density matrix $\rho_A$, and $\alpha_i$ is an index for subsystem $A$. For $\q=3$, we arrange $n^{\q-1}=n^2$ reduced density matrices on a two-dimensional square lattice and contract the indices of the neighboring reduced density matrices. The contraction of the reduced density matrix $\rho_{AB}$ for $n=2$ is given by
\begin{align}
\left(\rho_{AB}\right)_{\alpha_1\beta_1}^{\alpha_2\beta_3} \left(\rho_{AB}\right)_{\alpha_2\beta_2}^{\alpha_1\beta_4}\left(\rho_{AB}\right)_{\alpha_3\beta_3}^{\alpha_4\beta_1}\left(\rho_{AB}\right)_{\alpha_4\beta_4}^{\alpha_3\beta_2},
\end{align}
where $\beta_i$ is an index for subsystem $B$. Figure \ref{fig:ContractionPattern} graphically represents the contraction pattern of reduced density matrices for $\q=3$, $n=2$, $Z_{2}^{(3)}$. The indices for $A$ are contracted horizontally, and the indices for $B$ are contracted vertically.

For general $\q$ and $n$, we arrange $n^{\q-1}$ reduced density matrices on a $(\q-1)$-dimensional lattice of length $n$ and contract the indices of the neighboring reduced density matrices. This contraction is symmetric under the cyclic permutation in $\q-1$ directions on the $(\q-1)$-dimensional lattice, where each direction corresponds to the contraction directions of indices for $\q-1$ subsystems. Thus, the $n$-th R\'enyi $\q$-partite multi-entropy has the replica symmetry $\underbrace{\mathbb{Z}_n\otimes\dots\otimes \mathbb{Z}_n}_{\q-1}$.
See \cite{Penington:2022dhr,Gadde:2023zni,Iizuka:2025caq} for more details of the property of the $n$-th R\'enyi $\q$-partite multi-entropy.

The $\q$-partite multi-entropy is defined via the $n\to 1$ limit,
\begin{equation}
	S^{(\q )}\left(A_1: A_2: \cdots: A_{\q}\right)\coloneqq \lim_{n\to 1} S_n^{(\q )}\left(A_1: A_2: \cdots: A_{\q}\right).
\end{equation}

It has been proposed~\cite{Gadde:2022cqi,Gadde:2023zzj} that the holographic dual of the $\q$-partite multi-entropy is given by the minimal multiway cut that divides the bulk into $\q$ subregions.  
However, we note that there are some subtleties in this regard \cite{Penington:2022dhr}.  
In this paper, we {\it assume} that the holographic dual of the $\q$-partite multi-entropy is indeed given by the minimal multiway cut separating the bulk into $\q$ subregions.

\subsection{Reflected multi-entropy}\label{subsec:reflected}
The reflected multi-entropy was introduced in~\cite{Yuan:2024yfg} as a generalization of the multi-entropy for a given mixed state.  
Let $\rho_{A_{1},A_{2},\cdots,A_{\q-1}}$ be a mixed state on a $(\q{-}1)$-partite subsystem, obtained by tracing out $A_{\q}$ from a pure state $\ket{\psi}_{A_{1},A_{2},\cdots,A_{\q}}$ on the full $\q$-partite system,
\begin{equation}\label{eq:def-(q-1)-partite-mixed-state}
	\rho_{A_{1},A_{2},\cdots,A_{\q-1}}=\tr_{A_{\q}} \left[ \ket{\psi}_{A_{1},A_{2},\cdots,A_{\q}} \bra{\psi}\right].
\end{equation}

\begin{figure}[t]
    \centering
    \begin{subfigure}[b]{0.46\linewidth}
        \includegraphics[width=\linewidth]{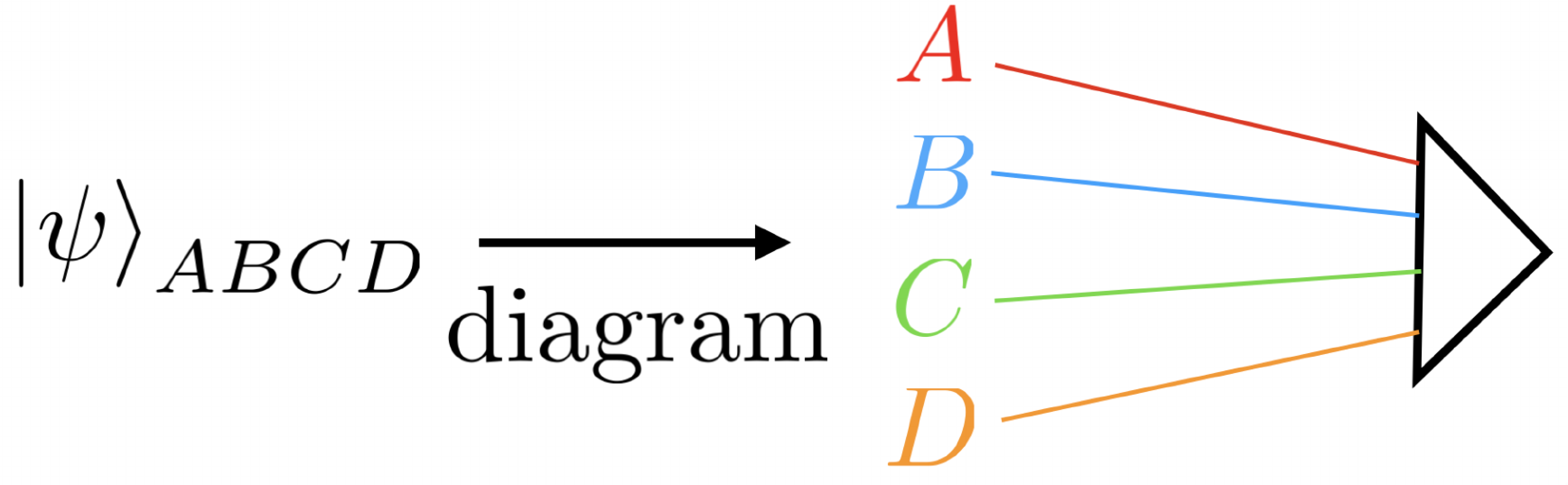}
        \caption{Diagram of the original $\q$-partite pure state ($\q=4$ case)}
        \label{fig:diagram-purified-state-q=4-1}
    \end{subfigure}
    \hfill
    \begin{subfigure}[b]{0.45\linewidth}
        \includegraphics[width=\linewidth]{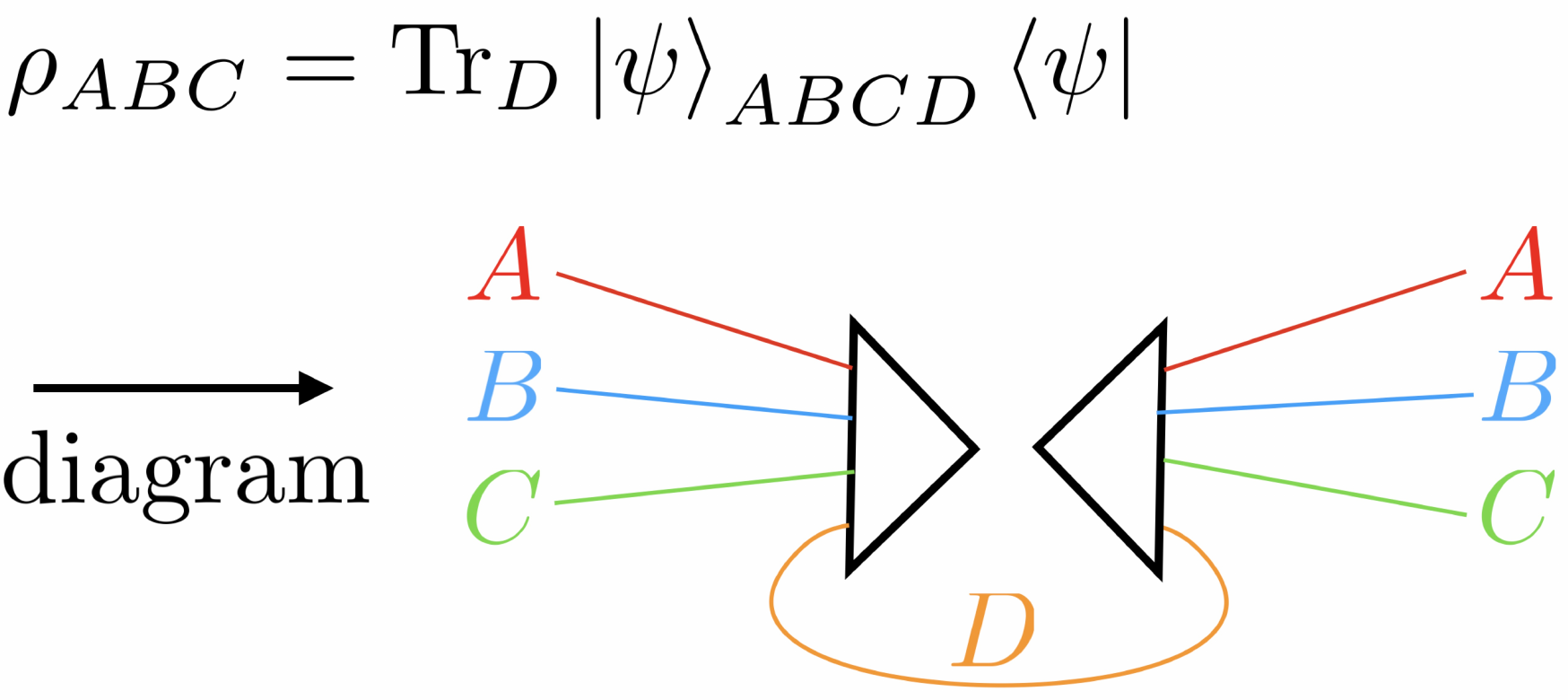}
        \caption{Diagram of the mixed state ($\q=4$ case)}
        \label{fig:diagram-purified-state-q=4-2}
    \end{subfigure}
    \vskip\baselineskip
    \begin{subfigure}[b]{1\linewidth}
        \includegraphics[width=0.9\linewidth]{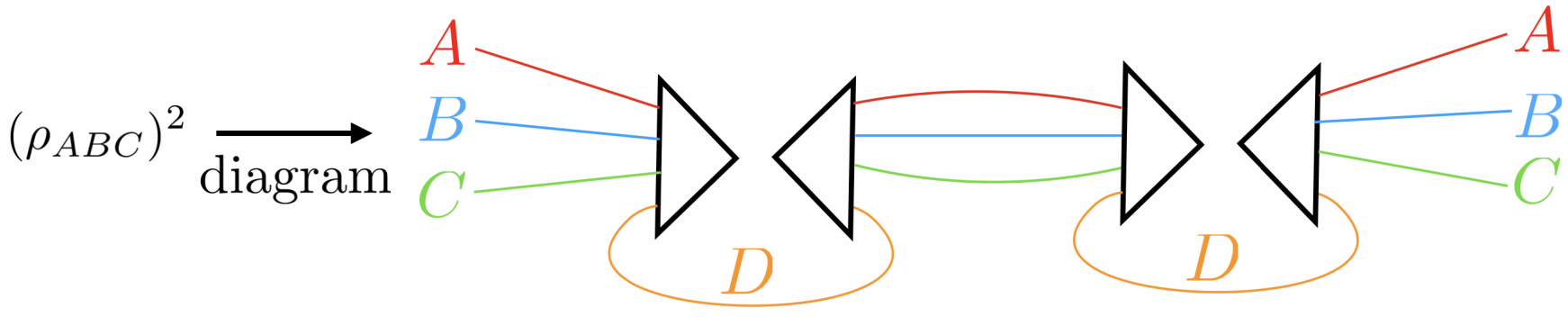}
        \caption{Diagram of the $m$-th power of the mixed state ($\q=4$, $m=2$ case)}
        \label{fig:diagram-purified-state-q=4-3}
        \vskip\baselineskip
        \includegraphics[width=0.75\linewidth]{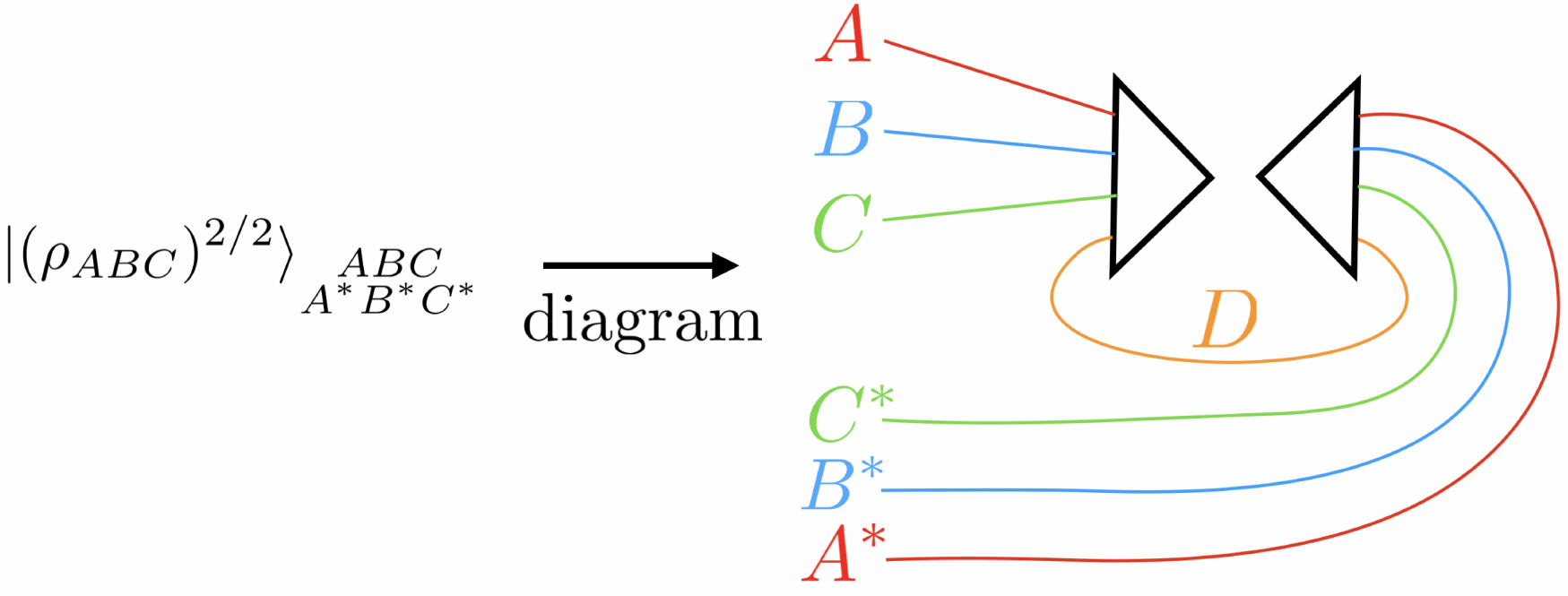}
        \caption{Diagram of the purification of the $m$-th power of the mixed state ($\q=4$, $m=2$ case)}
        \label{fig:diagram-purified-state-q=4-4}
    \end{subfigure}
    
    \caption{Diagrammatic understanding of the ($\q-1$)-partite purified state (\ref{eq:renyi-purified-state}) for $\q=4,m=2$.}
    \label{fig:diagram-purified-state-q=4}
\end{figure}

As in the previous section~\ref{subsec:reflected-entropy}, consider a mixed state $(\rho_{A_{1},A_{2},\cdots,A_{\q-1}})^{m}$.  
We introduce $A_{1}^{*}, A_{2}^{*}, \cdots, A_{\q-1}^{*}$ as canonical purifier subsystems to construct its canonical purification given by 
\begin{equation}\label{eq:renyi-purified-state}
	\ket{ (\rho_{A_{1},A_{2},\cdots,A_{\q-1}})^{m/2} }_{\substack{A_{1},A_{2},\cdots,A_{\q-1} \\ A_{1}^{*},A_{2}^{*},\cdots,A_{\q-1}^{*}}},
\end{equation}
with the condition
\begin{equation}\label{eq:requirement-for-purified-state}
	\tr_{A_{1}^{*},A_{2}^{*},\cdots,A_{\q-1}^{*}}\left[ \ket{ (\rho_{A_{1},A_{2},\cdots,A_{\q-1}})^{m/2} }_{\substack{A_{1},A_{2},\cdots,A_{\q-1} \\ A_{1}^{*},A_{2}^{*},\cdots,A_{\q-1}^{*}}} \bra{ (\rho_{A_{1},A_{2},\cdots,A_{\q-1}})^{m/2} } \right]=(\rho_{A_{1},A_{2},\cdots,A_{\q-1}})^{m}.
\end{equation}
Here, $m$ is again assumed to be an even number, $m\in 2\mathbb{N}$.
Similarly to the procedure as illustrated in Figure \ref{fig:diagram-purified-state-q=3}, the procedure for the $\q-1=3$ can also be visually understood as illustrated in Figure \ref{fig:diagram-purified-state-q=4}. 

Note that the state \eqref{eq:renyi-purified-state} is not normalized to be unity unless the state \eqref{eq:def-(q-1)-partite-mixed-state} is given by a pure state:
\begin{equation}\label{eq:normalization-purified-state}
    \hspace{-10mm}\tr_{\substack{A_{1},A_{2},\cdots,A_{\q-1} \\ A_{1}^{*},A_{2}^{*},\cdots,A_{\q-1}^{*}}}\left[ \ket{ (\rho_{A_{1},A_{2},\cdots,A_{\q-1}})^{m/2} }_{\substack{A_{1},A_{2},\cdots,A_{\q-1} \\ A_{1}^{*},A_{2}^{*},\cdots,A_{\q-1}^{*}}} \bra{ (\rho_{A_{1},A_{2},\cdots,A_{\q-1}})^{m/2} } \right] \\  =\tr_{A_{1},A_{2},\cdots,A_{\q-1}}\left[(\rho_{A_{1},A_{2},\cdots,A_{\q-1}})^{m}\right],
\end{equation}

For the $m$-dependent pure state given by  \eqref{eq:renyi-purified-state}, we define the following $n$-th R\'enyi multi-entropy: 
\begin{equation}\label{eq:renyi-reflected-multi-entropy-Original-form}
	\begin{aligned}
		&S_{n\,(m)}^{(\q-1 )}\left(A_{1}A_{1}^{*}: A_{2}A_{2}^{*}: \cdots: A_{\q-1}A_{\q-1}^{*}\right)\coloneqq \left. S_{n}^{(\q-1 )}\left(A_{1}A_{1}^{*}: A_{2}A_{2}^{*}: \cdots: A_{\q-1}A_{\q-1}^{*}\right) \right|_{\ket{ (\rho_{A_{1},A_{2},\cdots,A_{\q-1}})^{m/2} }}.
	\end{aligned}
\end{equation}
which depends on both $m$ and $n$. 

\begin{figure}[t]
    \centering
    \includegraphics[width=0.9\linewidth]{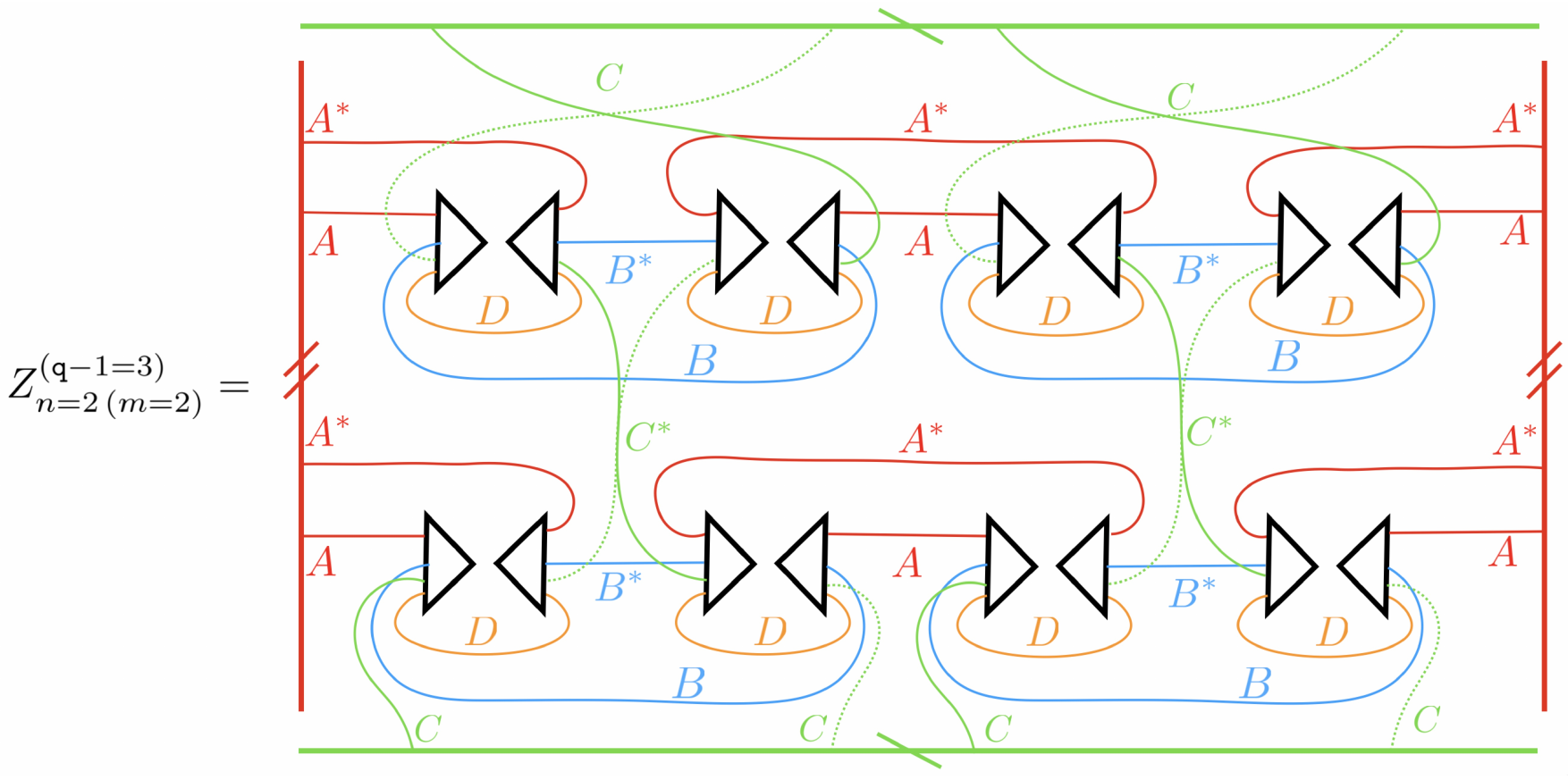}
    \caption{Diagrammatic representation of $Z_{n=2 \, (m=2)}^{(\q -1=3)}$ (\ref{eq:def-Zn(m)^{q-1}}) for $\q=4,m=n=2$ case. The top and bottom green lines are identified, and the left and right red lines are also identified. }
    \label{fig:Renyi-reflected-multi-q=4-n=2}
\end{figure}

 In this case, we also add the index $m$ to the partition functions appearing in the right hand side of \eqref{eq:def-renyi-multi-entropy} as follows,
\begin{equation}\label{eq:def-Zn(m)^{q-1}}
	Z_{n \, (m)}^{(\q -1)}\coloneqq \bra{ (\rho_{A_{1},A_{2},\cdots,A_{\q-1}})^{m/2} }^{\otimes n^{\q -2}} \Sigma_1\left(g_1\right) \Sigma_2\left(g_2\right) \ldots \Sigma_{\q-1 }\left(g_{\q-1}\right) \ket{ (\rho_{A_{1},A_{2},\cdots,A_{\q-1}})^{m/2} }^{\otimes n^{\q -2}}.
\end{equation}
since the purified state $\ket{ (\rho_{A_{1},A_{2},\cdots,A_{\q-1}})^{m/2}} $ depends on $m$.
Thus, the definition in equation~\eqref{eq:renyi-reflected-multi-entropy-Original-form} can equivalently be written as follows:
\begin{equation}\label{eq:def-renyi-multi-entropy-for-reflected}
	S_{n\,(m)}^{(\q-1 )}\left(A_{1}A_{1}^{*}: A_{2}A_{2}^{*}: \cdots: A_{\q-1}A_{\q-1}^{*}\right)\coloneqq\frac{1}{1-n}\cdot  \frac{1}{n^{\q-3}} \log \frac{Z_{n \, (m)}^{(\q -1)}}{\left(Z_{1 \, (m)}^{(\q -1)}\right)^{n^{\q-2}}},
\end{equation}
The normalization factor appearing in the denominator of the logarithm is given by
\eqref{eq:normalization-purified-state}, thereofore, 
    \begin{equation}
    \label{Z1mq-1norm}
Z_{1 \, (m)}^{(\q -1)}  =\tr_{A_{1},A_{2},\cdots,A_{\q-1}}\left[(\rho_{A_{1},A_{2},\cdots,A_{\q-1}})^{m}\right] \,.
    \end{equation}
The partition function $Z_{n \, (m)}^{(\q -1)}$ can be visualized using the same diagrammatic approach as in Figures~\ref{fig:diagram-purified-state-q=3} and~\ref{fig:diagram-purified-state-q=4}.  
As a concrete example, the case $Z_{n=2 \, (m=2)}^{(\q -1=3)}$ is illustrated in Figure~\ref{fig:Renyi-reflected-multi-q=4-n=2}.

Then, the $(m,n)$ R\'enyi reflected multi-entropy $S_{R\, (\q-1)}^{(m,n)}\left(A_1: A_2: \cdots: A_{\q-1}\right)$ of a mixed state on $(\q-1)$-partite subsystems \cite{Yuan:2024yfg} is defined by the $(m,n)$ R\'enyi multi-entropy \eqref{eq:renyi-reflected-multi-entropy-Original-form},
\begin{equation}\label{eq:def-renyi-reflected-multi-entropy}
	\begin{aligned}
		S_{R\, (\q-1)}^{(m,n)}\left(A_1: A_2: \cdots: A_{\q-1}\right)
\coloneqq  S_{n\,(m)}^{(\q-1 )}\left(A_{1}A_{1}^{*}: A_{2}A_{2}^{*}: \cdots: A_{\q-1}A_{\q-1}^{*}\right).
	\end{aligned}
\end{equation}
In short, the $(m,n)$ R\'enyi reflected multi-entropy is the $n$-th R\'enyi $(\q-1)$-partite multi-entropy of the canonical purified state (\ref{eq:renyi-purified-state}).
This definition is a natural generalization of the original R\'enyi reflected entropy \eqref{eq:def-renyi-reflected-entropy} for $\q=3$, to general $\q$. 
Indeed, by setting $\q-1=2$, we get the original R\'enyi reflected entropy,
\begin{equation}\label{eq:renyi-two-q=3-relation}
	\begin{aligned}
		S_{R\, (\q-1=2)}^{(m,n)}\left(A_1: A_2\right)&=  S_{n\,(m)}^{(\q-1=2)}\left(A_{1}A_{1}^{*}:A_{2}A_{2}^{*}\right)\\
		&=\left.S_{n}\left(A_{1}A_{1}^{*}\right)\right|_{\ket{ (\rho_{A_{1},A_{2}})^{m/2} }_{A_{1},A_{2}, A_{1}^{*},A_{2}^{*}}}\\
		&=S_{R}^{(m,n)}(A_1: A_2),
	\end{aligned}
\end{equation}
where in the second line we used the property that the (R\'enyi) multi-entropy for bi-partite subsystems is reduced to just the entanglement entropy between them, and in the final line, we used the definition of the original R\'enyi reflected entropy.

The reflected multi-entropy \cite{Yuan:2024yfg} is defined by the limit $m\to 1$, $n\to 1$ of \eqref{eq:def-renyi-reflected-multi-entropy},
\begin{equation}
	S_{R\, (q-1)}\left(A_1: A_2: \cdots: A_{\q-1}\right)\coloneqq \lim_{m\to 1, \\n\to 1}S_{R\, (q-1)}^{(m,n)}\left(A_1: A_2: \cdots: A_{\q-1}\right).
\end{equation}
Using the holographic dual of the multi-entropy, it has been proposed in~\cite{Yuan:2024yfg} that the reflected multi-entropy for a mixed state corresponds to the minimal multiway cut in the bulk, where the cut is anchored to the Ryu–Takayanagi surface of the combined region $A_1 A_2 \cdots A_{\q-1}$.

\section{Multipartite Markov Gap: Geometric obstruction to perfect recovery in multipartite systems}
\label{sec:MMG}

In this section, we assume that, in the AdS$_{d+1}$/CFT$_d$ correspondence ($d\geq 2$), the holographic dual of the $\q$-partite multi-entropy $S^{(\q )}\left(A_1: A_2: \cdots :A_{\q}\right)$ is given by the minimal multiway cut $\gamma_{A_{1}A_{2}\cdots A_{\q }}^{M}$ that divides the bulk into $\q$ subregions~\cite{Gadde:2022cqi,Gadde:2023zzj}\footnote{Note that the holographic proposal of the (reflected) multi-entropy has a subtle point regarding the analytic continuation $n\to1$. In particular, the replica symmetry breaking of the bulk geometry for $S_n^{(3 )}\left(A_1: A_2: A_{3}\right) \, $  in holographic states is argued by \cite{Penington:2022dhr}, and as a result, $S_n^{(3 )}\left(A_1: A_2: A_{3}\right)$ with $n\geq 3$ is not dual to the minimal triway cuts in the bulk. On the other hand, in random tensor network states \cite{Hayden:2016cfa,Pastawski:2015qua} and fixed area states \cite{Akers:2018fow,Dong:2018seb},\, and the holographic proposal at $n\to1$ would be valid at least in these states.}
\begin{equation}
	S^{(\q )}\left(A_1: A_2: \cdots :A_{\q}\right)=\frac{\mathrm{Area}(\gamma_{A_{1}A_{2}\cdots A_{\q }}^{M})}{4G_{N}}.
\end{equation}
By using holographic techniques with this assumption, we propose a method to generalize the Markov gap defined for the $\q=3$ case in \eqref{eq:def-Markov-gap} to more general $\q$-partite cases with $\q \geq 4$\footnote{If the assumption of the duality between multi-entropy   and the minimal multiway cut in the $n\to 1$ limit is not valid, we may need to modify the proposal described below since we construct the generalized Markov gap by combining minimal multiway cut and minimal surfaces such that it takes non-negative values. }.
To this end, we first review the holographic derivation of the tripartite Markov gap for $\q=3$, which provides the essential intuition and methodology for our generalization.

\subsection{Review of the holographic derivation of the lower bound of reflected entropy in \texorpdfstring{$\q$}{q}=3 case}
\label{subsec:Holographic-derivation-lower-bound}

In introducing the tripartite Markov gap, the existence of a lower bound of the reflected entropy as \eqref{eq:inequality-q=3-reflected-entropy-and-mutual-information} is essential for formulating the Markov gap combination.
For $\q=3$, a holographic analysis of the inequality \eqref{eq:SSA-q=3-purified-situation} was first presented explicitly in \cite{Hayden:2021gno}.
Through holographic methods, this lower bound \eqref{eq:inequality-q=3-reflected-entropy-and-mutual-information} can be understood more intuitively, and this intuition plays a crucial role in extending the tripartite Markov gap to general multipartite cases, as we discuss below.
We therefore follow \cite{Hayden:2021gno} and review the holographic derivation of inequality \eqref{eq:SSA-q=3-purified-situation}, which establishes the lower bound \eqref{eq:inequality-q=3-reflected-entropy-and-mutual-information}.  Readers familiar with this result may skip to subsection~\ref{HMGdef}.

\begin{figure}[t]
    \centering
    \begin{subfigure}[b]{0.43\linewidth}
        \centering
        \includegraphics[width=0.85\linewidth]{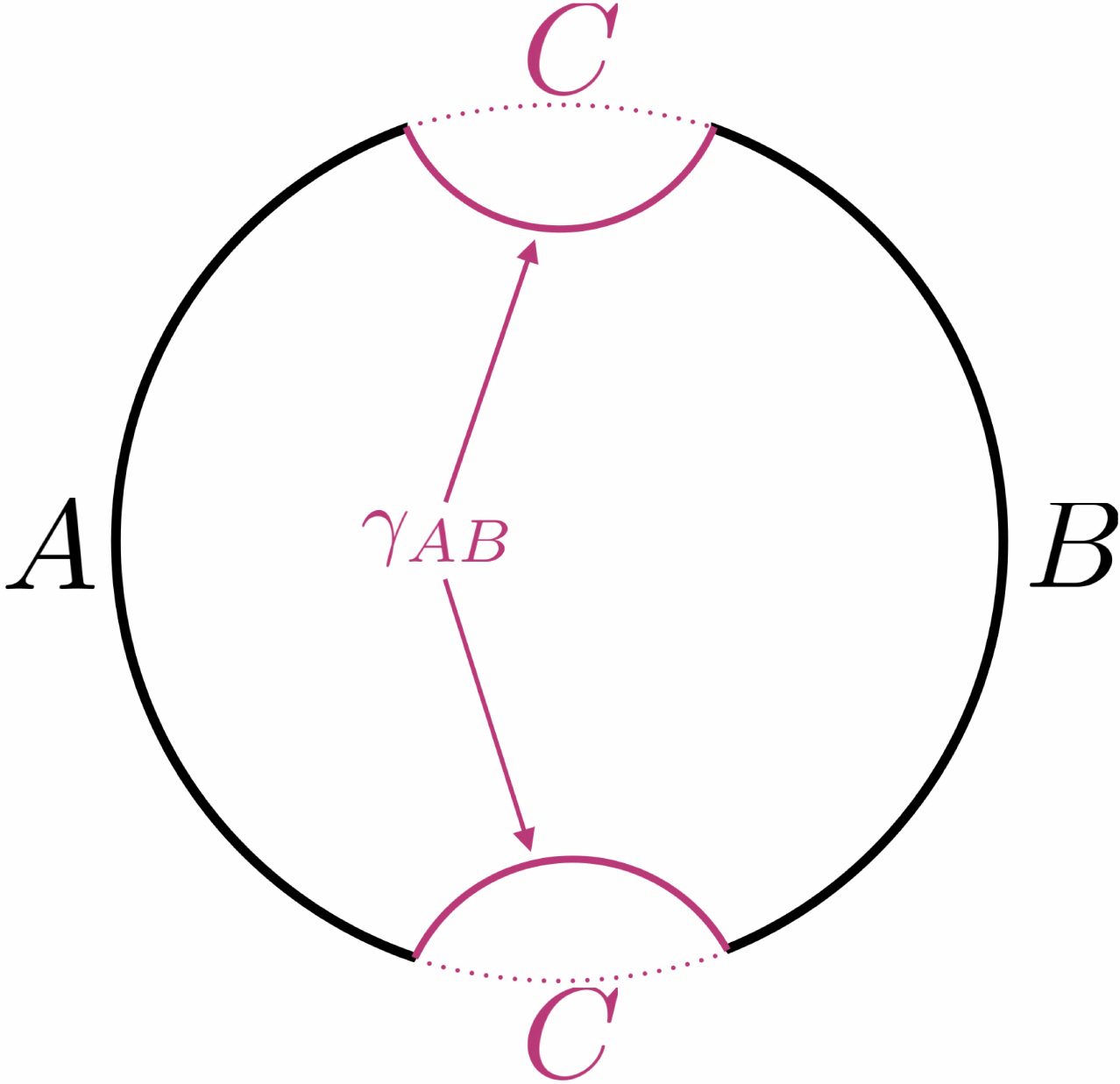}
        \caption{Connected entanglement wedge of subsystem $AB$.\\}
        \label{fig:Canonical-purified-geometry-1}
    \end{subfigure}
    \,\,\,\,\,
    \begin{subfigure}[b]{0.45\linewidth}
        \centering
        \includegraphics[width=\linewidth]{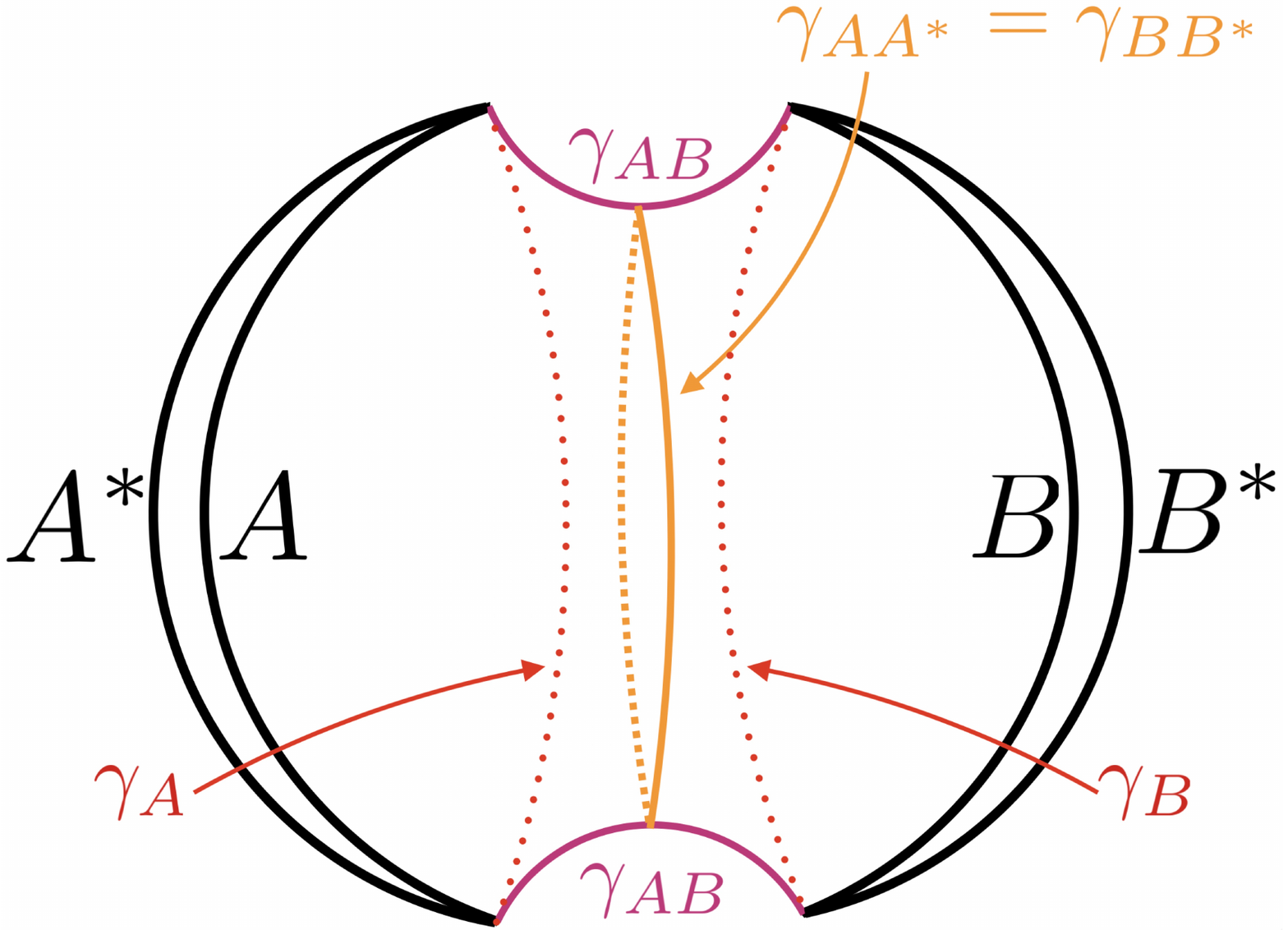}
        \caption{Bulk geometry for the canonical purification of subsystem $AB$ with reflected minimal surface $\gamma_{AA^{*}}=\gamma_{BB^{*}}$ and minimal surfaces $\gamma_{A},\gamma_{B}$.}
        \label{fig:Canonical-purified-geometry-2}
    \end{subfigure}
    \caption{Bulk geometry for the canonical purification of subsystem $AB$ with minimal surfaces.}
    \label{fig:Canonical-purified-geometry}
\end{figure}

We consider three subsystems $A,B,C$, which together form a pure state. We trace out subsystem $C$, and for the remaining subsystems $A,B$, we consider the canonical purification of their subsystems by introducing auxiliary copies of subsystems $A^*,B^*$. By this canonical purification, we need to consider a corresponding bulk geometry. As discussed in \cite{Dutta:2019gen}, this bulk geometry can be prepared by duplicating the entanglement wedge of subsystem $AB$, and gluing the original and the copy along the minimal surface $\gamma_{AB}$ \cite{Dutta:2019gen}\footnote{For simplicity, we consider minimal surfaces, but we can generalize our discussion to that using quantum extremal surfaces.}. In the case where subsystem $C$ is relatively small such that the entanglement wedge of $AB$ is connected, as shown in Figure \ref{fig:Canonical-purified-geometry-1}, the resulting purified geometry is illustrated in Figure \ref{fig:Canonical-purified-geometry-2}. On the other hand, if the entanglement wedge of $AB$ is disconnected, the bulk geometry consists simply of the entanglement wedges of $A$ and $B$ separately.

We focus on the situation where the entanglement wedge of $AB$ is connected, otherwise, the discussion below becomes trivial. 
\begin{figure}[h]
    \centering
    \includegraphics[width=0.5\linewidth]{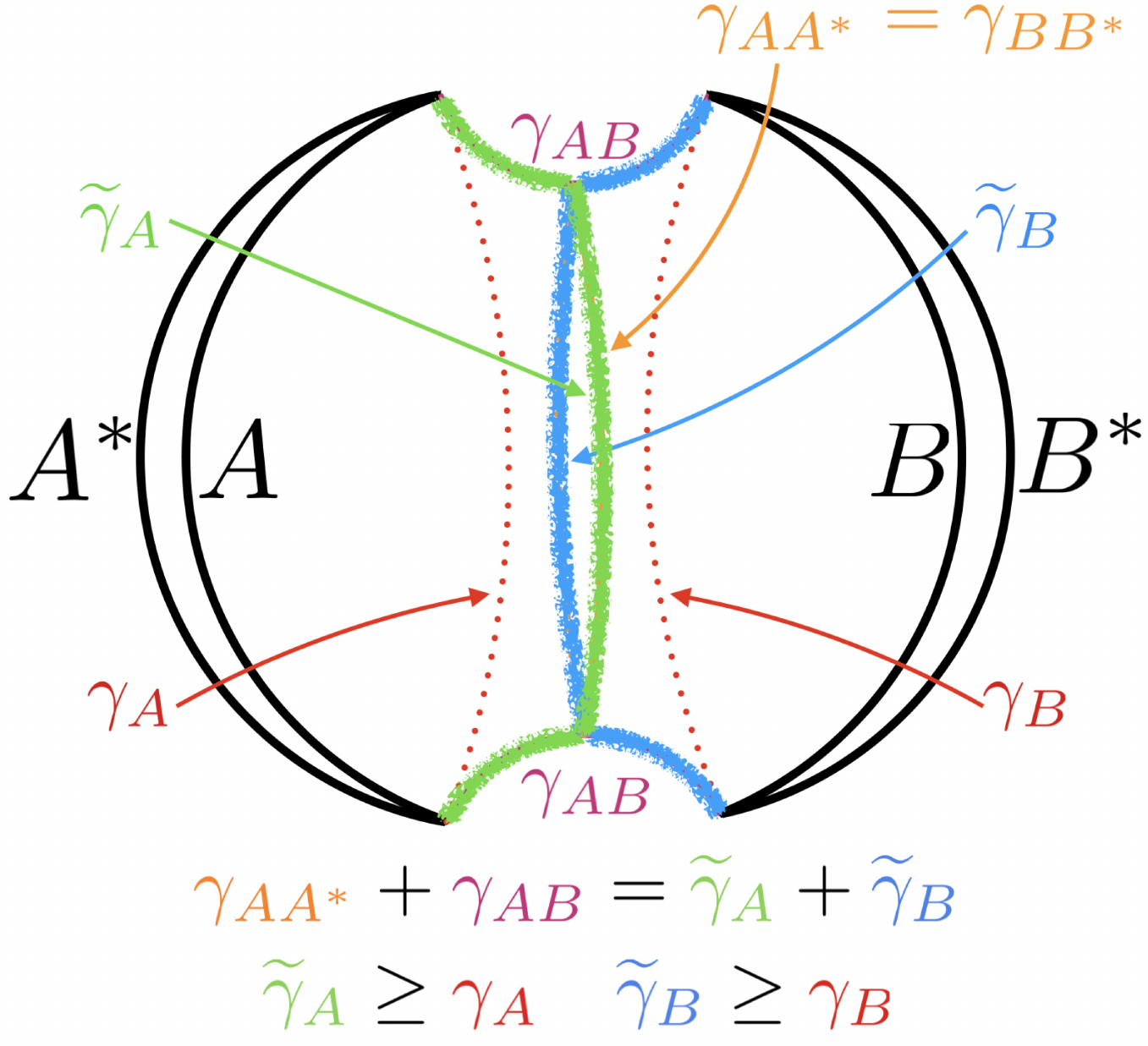}
    \caption{Rearrangement of minimal surfaces to introduce the lower bound of the reflected entropy ($\q=3$ case).}
    \label{fig:Holographic-inequality-q=3}
\end{figure}
In this case, the minimal surfaces corresponding to the reflected entropy $S_{R}(A)$, and the entanglement entropies $S(A),S(B),S(AB)$ (which define the mutual information), are denoted by 
\begin{equation}
\gamma_{AA^*} \left( = \gamma_{BB^*} \right), \quad
	\gamma_{A},\quad \gamma_{B}, \quad \gamma_{AB} \left( = \gamma_C \right).
\end{equation} 
See Figure \ref{fig:Canonical-purified-geometry-2} for an illustration of these minimal surfaces.
As can be clearly seen from  Figure \ref{fig:Holographic-inequality-q=3}, these minimal surfaces satisfy the inequality,
\begin{equation}
\label{q3casearearela}
	\mathrm{Area}(\gamma_{AA^{*}}) + \mathrm{Area}(\gamma_{AB}) \geq \mathrm{Area}(\gamma_{A}) + \mathrm{Area}(\gamma_{B}).
\end{equation}
This inequality arises because the union of $\gamma_{AB}$ and  $\gamma_{AA^*}$ gives rise to two non-minimal surfaces:  one connecting the endpoints of $\gamma_{A}$, and the other connecting the endpoints of $\gamma_{B}$.
This geometrical relation leads to the following inequality \eqref{eq:SSA-q=3-purified-situation} where  $S(B^*)$ is replaced by  $S(B)$,
\begin{equation}\label{eq:inequaitli-q=3}
	\begin{aligned}
		S(AA^{*}) + S(AB) \geq S(A) + S(B) \qquad \Longrightarrow \qquad S_{R}(A) + S(AB) \geq S(A) + S(B).
	\end{aligned}
\end{equation}
This provides a lower bound for the reflected entropy \eqref{eq:inequality-q=3-reflected-entropy-and-mutual-information}, and the Markov gap is defined as the left-hand side minus the right-hand side of the inequality.

From this holographic derivation, the Markov gap can be interpreted as the difference between the non-minimal surfaces consisting of $\gamma_{AB}, \gamma_{AA^*}$ and the minimal surfaces $\gamma_{A}, \gamma_{B}$. More intuitively, the Markov gap measures the portion of the bulk region contained in the entanglement wedge of $AB$ that is not captured by the union of the entanglement wedges of $A$ and $B$.
When the Markov gap vanishes, no such bulk region exists.

\subsection{Multipartite Markov Gaps}
\label{HMGdef}

Motivated by the holographic inequality derived for the $\q = 3$ case \eqref{eq:inequaitli-q=3}, which naturally leads to the Markov gap, we consider a generalization to multipartite settings with $\q \geq 4$.
Although deriving such inequalities from a purely information-theoretic perspective remains challenging, they arise more naturally in holographic frameworks.
While such a quantity can, in principle, be defined \emph{without reference to holography}, our construction is guided by holographic considerations.
We refer to it as the {\it Multipartite Markov gap}, denoted by $MG^{M(\q-1)}(A_{1} : A_{2} : \cdots : A_{\q-1})$. In what follows, we assume the validity of holographic duality in our discussion.

\subsubsection{\texorpdfstring{$\q$}{q}=4 case (Connected entanglement wedge)}

\begin{figure}[t]
    \centering
    \begin{subfigure}[b]{0.3\textwidth}
        \centering
        \includegraphics[width=\linewidth]{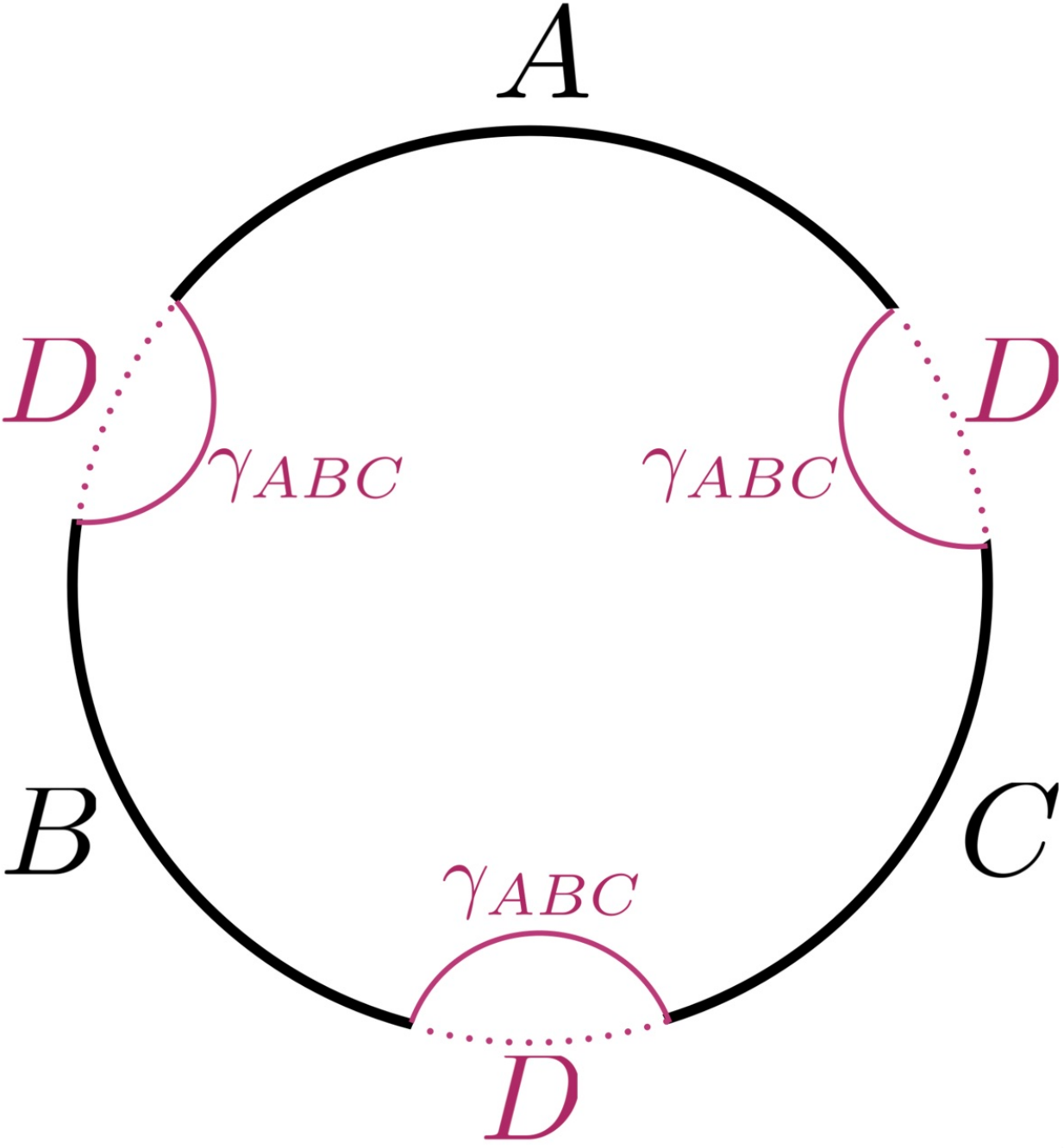}
        \caption{Fully connected entanglement wedge of $ABC$.}
        \label{fig:Assumption-of-subsystem-sizes-1}
    \end{subfigure}
    \hspace{0.03\textwidth} 
    \begin{subfigure}[b]{0.3\textwidth}
        \centering
        \includegraphics[width=\linewidth]{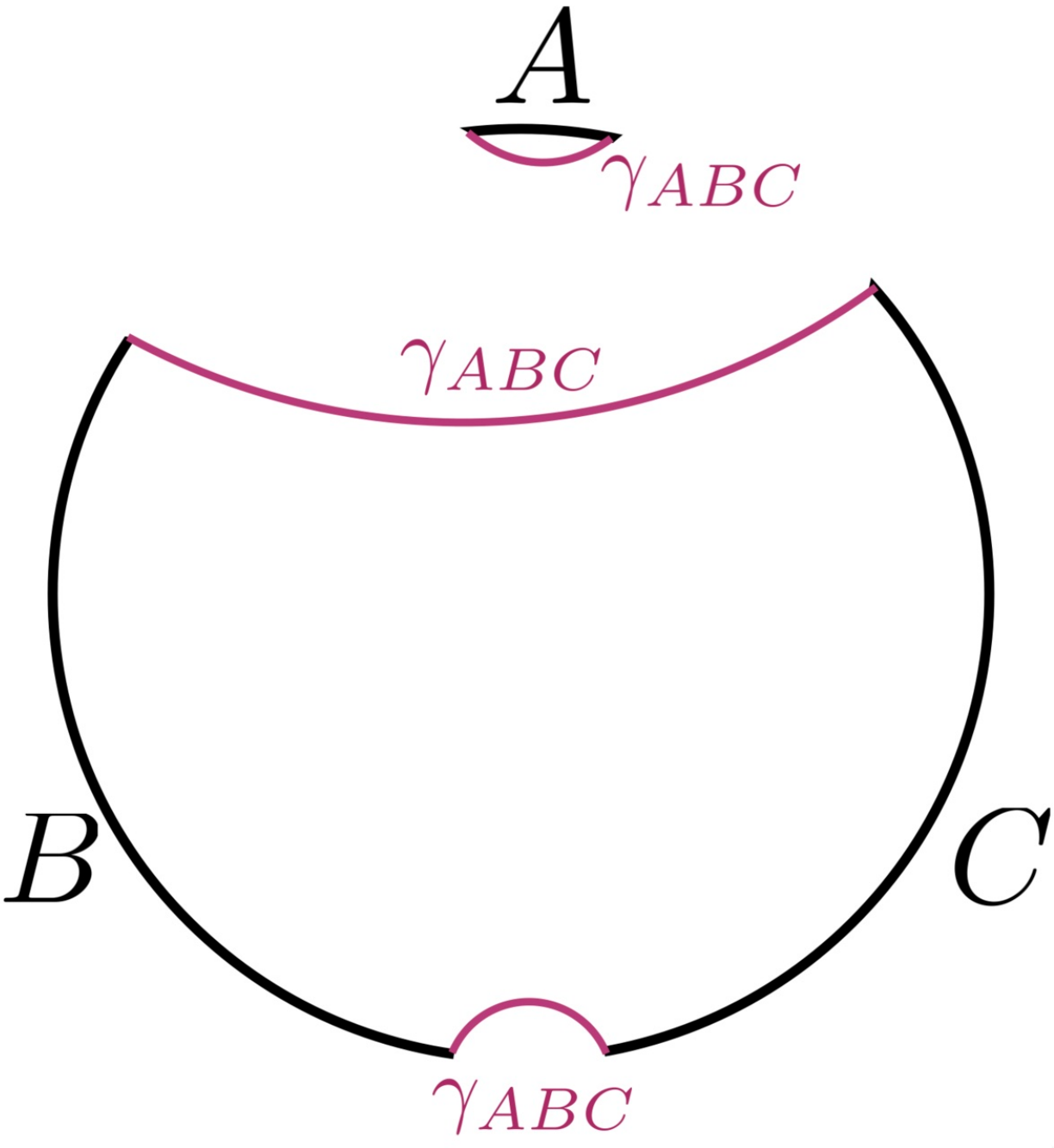}
        \caption{Partly connected entanglement wedge of $ABC$.}
        \label{fig:Assumption-of-subsystem-sizes-2}
    \end{subfigure}
    \hspace{0.03\textwidth}
    \begin{subfigure}[b]{0.3\textwidth}
        \centering
        \includegraphics[width=\linewidth]{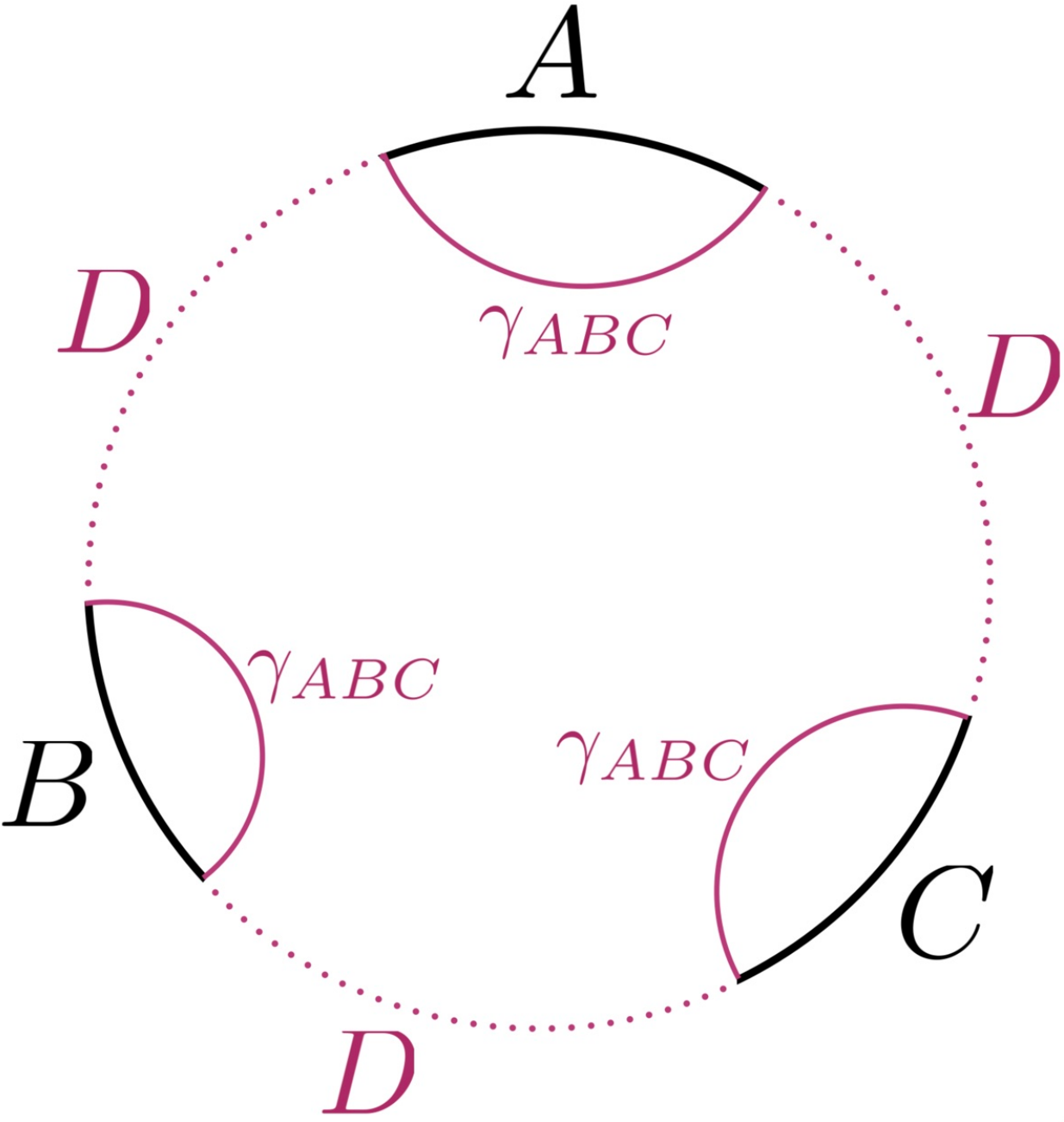}
        \caption{Fully disconnected entanglement wedge of $ABC$.}
        \label{fig:Assumption-of-subsystem-sizes-3}
    \end{subfigure}
    \caption{Three situations of the entanglement wedge of $ABC$ ($\q=4$ case). Since $ABCD$ forms a total pure state, the minimal surface of $ABC$ is equal to that of $D$ due to mutuality  $\gamma_{ABC} = \gamma_D$.}
    \label{fig:Assumption-of-subsystem-sizes}
\end{figure}

\begin{figure}[h]
    \centering
    \begin{subfigure}[b]{0.4\linewidth}
        \centering
        \includegraphics[width=1\linewidth]{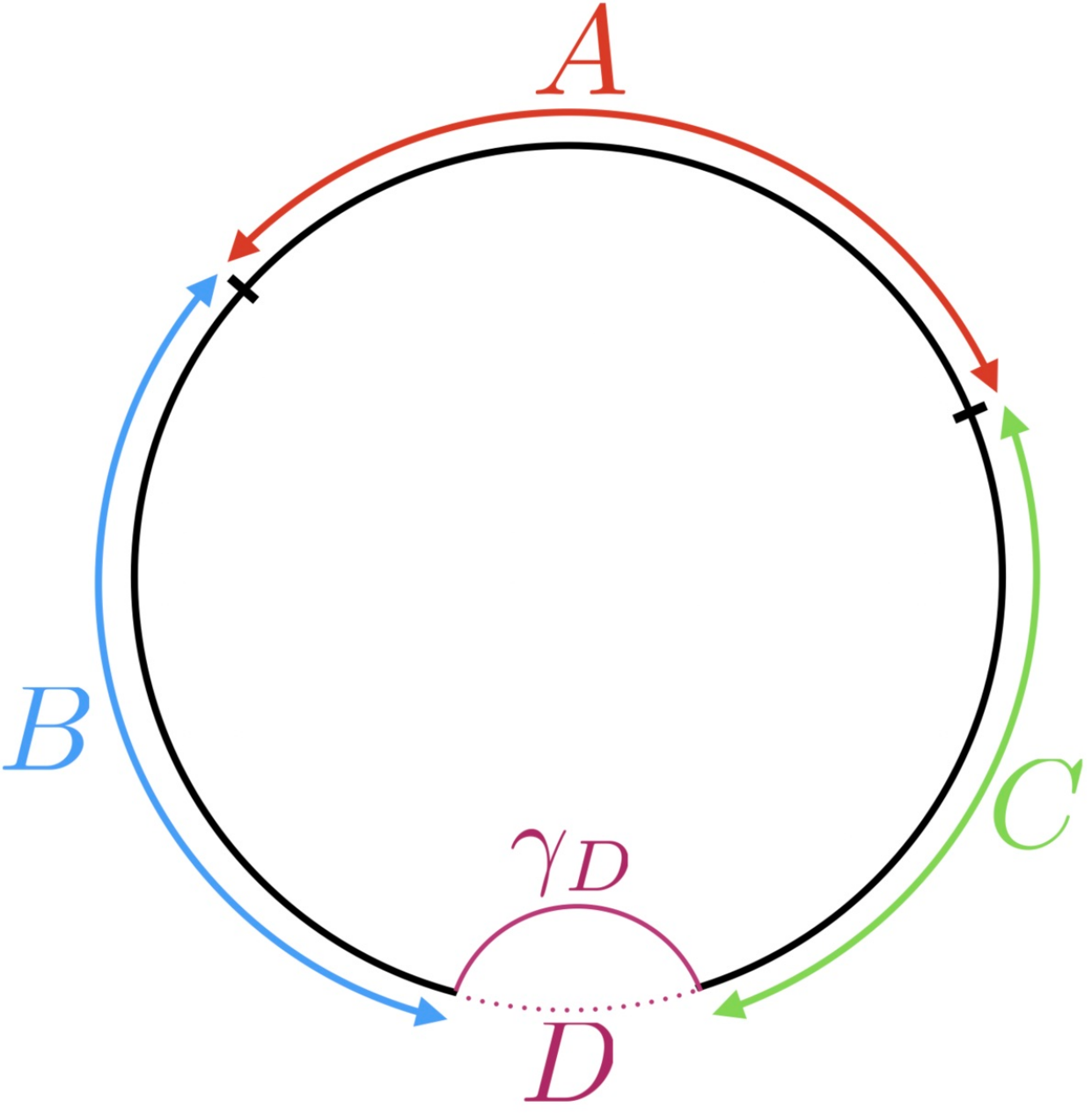}
        \caption{$A$ is adjacent to subsystems $B$ and $C$. \\}
        \label{subfig:Subsystem-Adjacent-1}
    \end{subfigure}
    \hspace{0.04\textwidth}
    \begin{subfigure}[b]{0.42\linewidth}
        \centering   \includegraphics[width=1.1\linewidth]{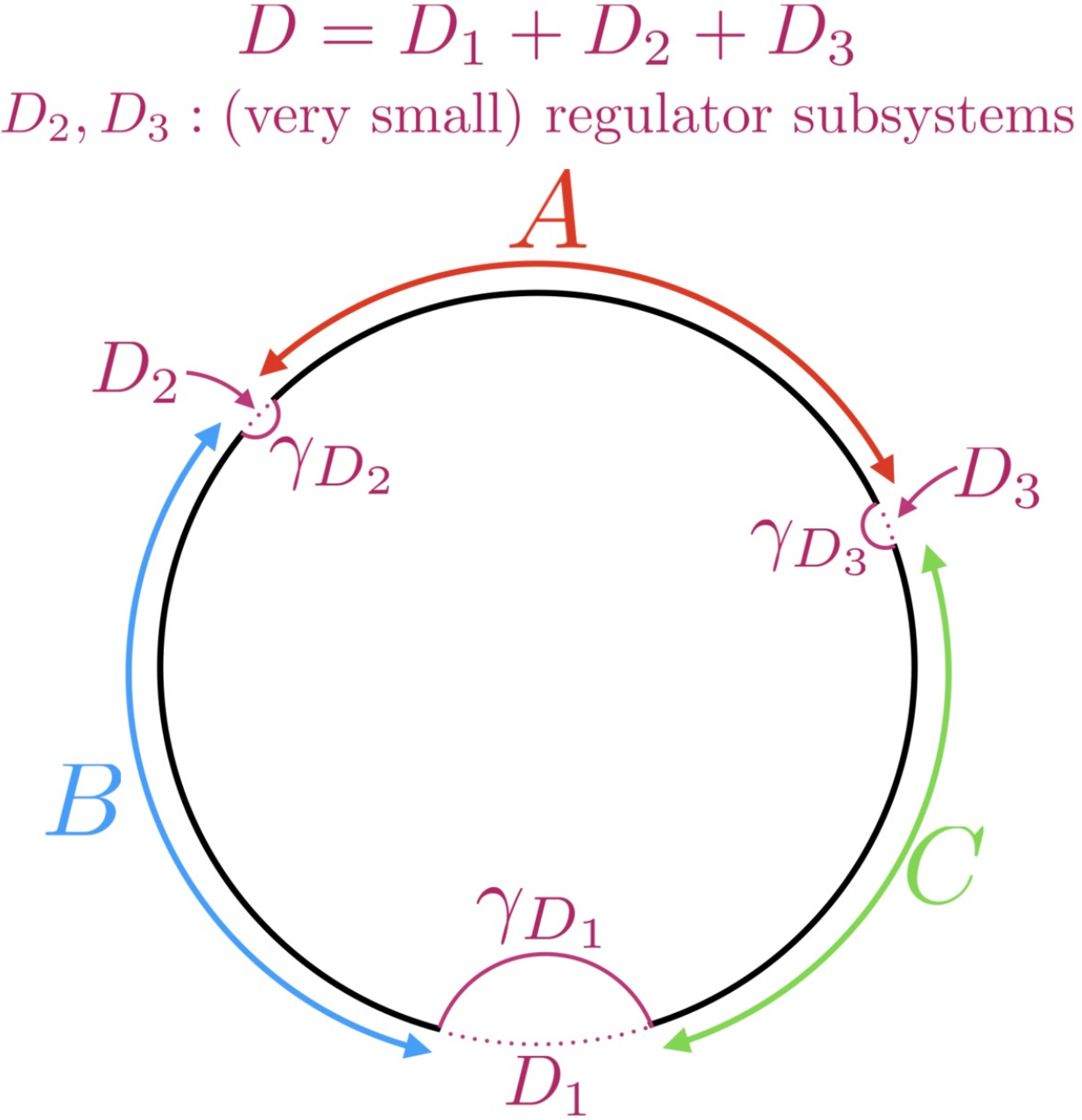}
        \caption{Very small regulator subsystems $D_{2},D_{3}$ are inserted.}
        \label{subfig:Subsystem-Adjacent-2}
    \end{subfigure}
    \caption{A situation that subsystem $A$ is adjacent to subsystems $B$ and $C$, and $D$ is adjacent to $B$ and $C$ ($\q=4$ case). The situation can be treated as the limit of the situation where subsystems $D=D_{1}+D_{2}+D_{3}$ are inserted between $A$, $B$ and $C$, and the subsystems $D_{2},D_{3}$ become very small.}
    \label{fig:Subsystem-Adjacent}
\end{figure}

We begin by considering a holographic setup with $\q = 4$, where the total system is in a pure state composed of four subsystems: $A$, $B$, $C$, and $D$.  
Let us assume that the subsystem $D$ is relatively small, such that the entanglement wedge of $ABC$ is connected, and that $D$ is geometrically located between the other three subsystems. See Figure~\ref{fig:Assumption-of-subsystem-sizes-1} for an illustration of this configuration.
One may alternatively consider cases where subsystem $A$ is adjacent to both subsystem $B$ and $C$, while $D$ is adjacent to $B$ and $C$, as shown in Figure \ref{subfig:Subsystem-Adjacent-1}. However, as illustrated in Figure \ref{subfig:Subsystem-Adjacent-2},
such a configuration can be viewed as a limiting case of the setup
in Figure \ref{fig:Assumption-of-subsystem-sizes-1}, where part of subsystem $D$ serves as a regulator for the UV divergences that arise when two subsystems, such as A and B or A and C, become adjacent.
Therefore, the configuration shown in Figure \ref{fig:Assumption-of-subsystem-sizes-1} provides the simplest non-trivial building block to analyze the $\q=4$ case. In this subsection, we focus exclusively on this configuration. Additional configurations, including those with disconnected entanglement wedges as illustrated in Figures \ref{fig:Assumption-of-subsystem-sizes-2} and \ref{fig:Assumption-of-subsystem-sizes-3}, will be discussed in \ref{sec:HolographicMarkovGapDisconnectedEW}.

\begin{figure}[h]
    \centering
    \begin{subfigure}[b]{\linewidth}
        \hspace{38mm} \includegraphics[width=0.46\linewidth]{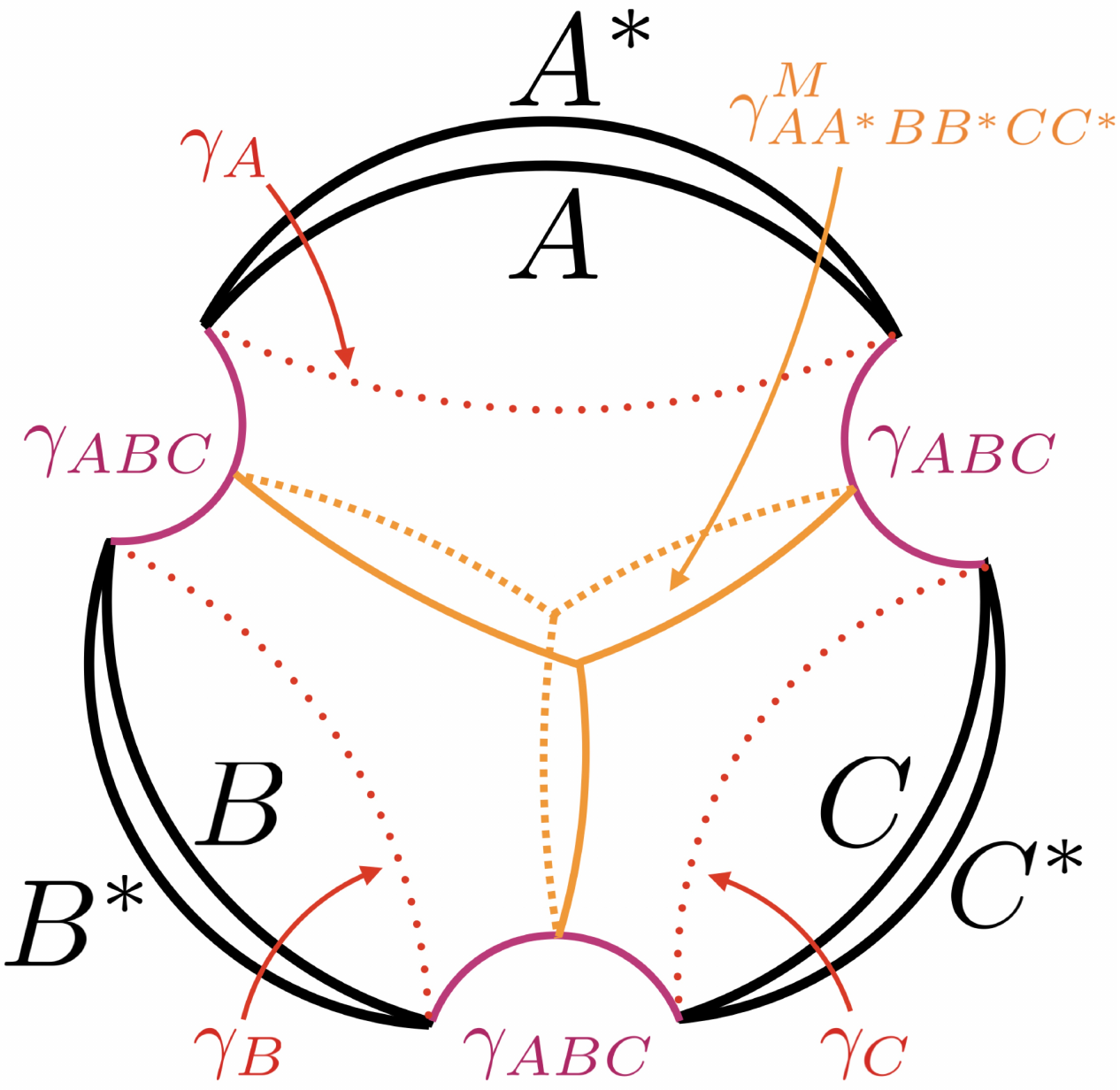}
        \caption{Bulk geometry for the canonical purification of subsystem $ABC$, having a connected entanglement wedge ($\q=4$ case) with the multiway cut $\gamma_{AA^{*}BB^{*}CC^{*}}^{M}$ and the minimal surfaces.}
        \label{fig:Holographic-inequality-q=4-1}
    \end{subfigure}
    \vspace{1em}
    \begin{subfigure}[b]{\linewidth}
        \centering  
\includegraphics[width=0.6\linewidth]{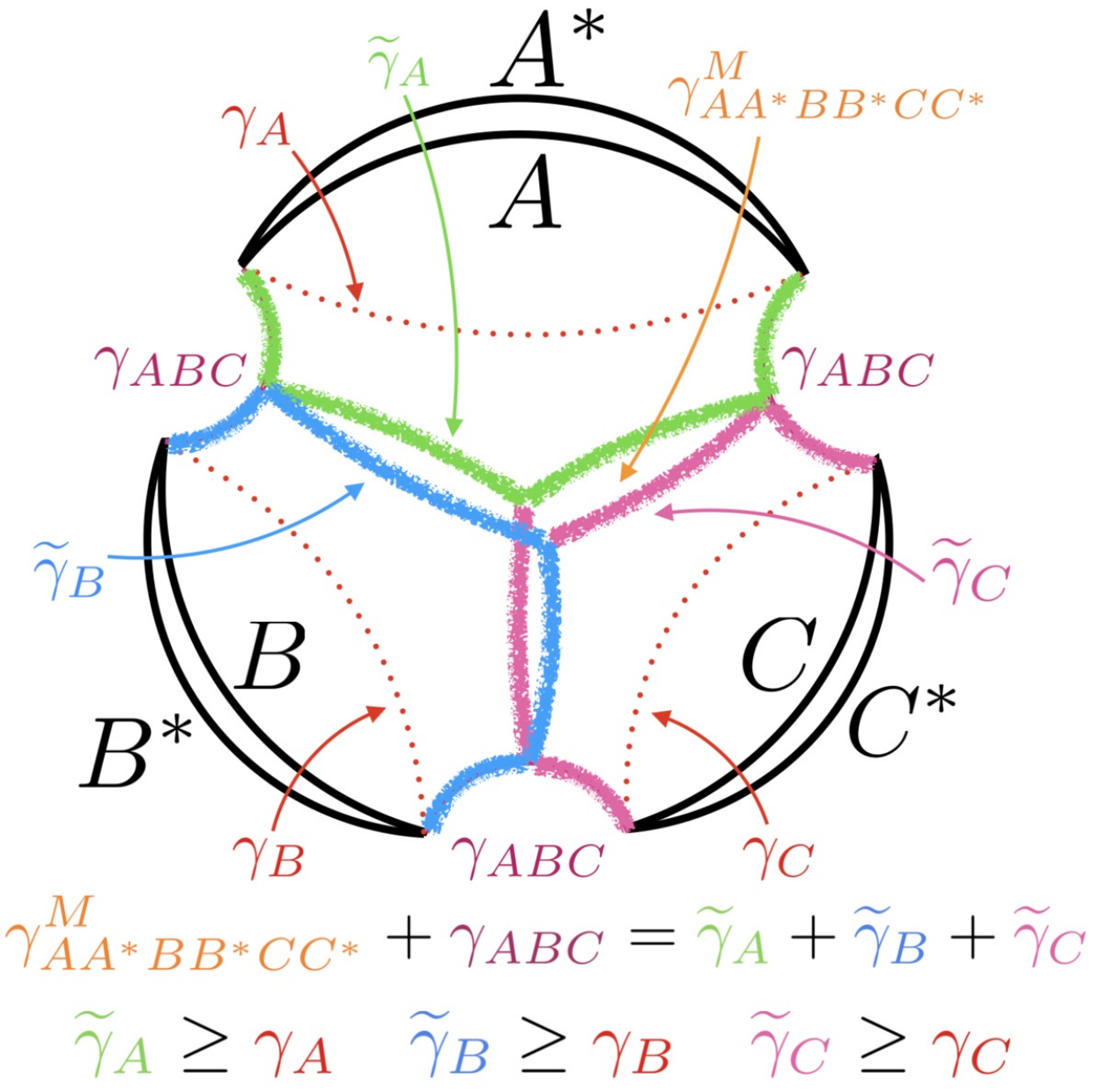}
        \caption{Rearranged non-minimal surfaces $\widetilde{\gamma}$ and minimal surfaces $\gamma$ ($\q=4$ case). The combined surfaces $\widetilde{\gamma}_{A},\widetilde{\gamma}_{B},\widetilde{\gamma}_{C}$ are not minimal ones.}
        \label{fig:Holographic-inequality-q=4-2}
    \end{subfigure}
    \caption{Bulk geometry with multiway cut and minimal surfaces in the inequality (\ref{eq:inequality-q=4-extremal-surfaces}) for the $\q=4$ case (top), and their rearrangements of the cut and surfaces (bottom). }
    \label{fig:Holographic-inequality-q=4}
\end{figure}

We begin by considering the minimal multiway cut $\gamma_{AA^* BB^* CC^*}^M$ in the bulk geometry obtained by the canonical purification of the subsystems $ABC$. 
This configuration is illustrated in Figure \ref{fig:Holographic-inequality-q=4-1}. The minimal multiway cut is defined as follows. Consider a time slice $\Sigma$ in the bulk that includes $\mathtt{p}$ boundary subregions $A_1,A_2,\dots,A_\mathtt{p}$. A $\mathtt{p}$-way cut is given by a bulk web $\mathcal{W}$ that divides the bulk $\Sigma$ into $\mathtt{p}$ bulk subregions such that $\mathcal{W}$ contains sub-webs that are homologous to the $\mathtt{p}$ boundary subregions $A_1,A_2,\dots,A_\mathtt{p}$. For given $A_1,A_2,\dots,A_\mathtt{p}$, the minimal multiway cut is defined by the web whose area is minimal over all possible $\mathtt{p}$-way cuts. For example, $\gamma_{AA^* BB^* CC^*}^M$ is the minimal multi-way cut for $\mathtt{p}=3$ boundary subregions $AA^*,BB^*,CC^*$ in the bulk geometry
obtained by the canonical purification of the subsystems $ABC$.

By mimicking the discussion of the $\q=3$ case in Section \ref{subsec:Holographic-derivation-lower-bound}, we also consider minimal surfaces of subsystems $A$, $B$, $C$, and $D$ or equivalently $ABC$.    These minimal surfaces, denoted by  $\gamma_{A}$, $\gamma_{B}$, $\gamma_{C}$, and $\gamma_{ABC}$ are also illustrated in Figure \ref{fig:Holographic-inequality-q=4-1}. 
To establish a lower bound of the reflected multipartite entropy, we combine the minimal multiway cut $\gamma_{AA^* BB^* CC^*}^M$ with the minimal surfaces $\gamma_{A}$, $\gamma_{B}$, $\gamma_{C}$, and $\gamma_{ABC}$. The strategy is straightforward: by rearranging segments of $\gamma_{AA^* BB^* CC^*}^M$ and $\gamma_{ABC}$ as illustrated in Figure~\ref{fig:Holographic-inequality-q=4-2}, we construct non-minimal surfaces $\widetilde{\gamma}_{A}$, $\widetilde{\gamma}_{B}$, and $\widetilde{\gamma}_{C}$.

Since $\widetilde{\gamma}_{A},\widetilde{\gamma}_{B},\widetilde{\gamma}_{C}$ are non-minimal surfaces, the areas of them are bounded from below by those of the minimal surfaces $\gamma_{A},\gamma_{B},\gamma_{C}$. Thus, we get the inequality:
\begin{equation}\label{eq:inequality-q=4-extremal-surfaces}
	\begin{aligned}
		\mathrm{Area}(\gamma_{AA^{*}BB^{*}CC^{*}}^{M}) + \mathrm{Area}(\gamma_{ABC} )&= \mathrm{Area}(\widetilde{\gamma}_{A}) + \mathrm{Area}(\widetilde{\gamma}_{B})+ \mathrm{Area}(\widetilde{\gamma}_{C})\\
		& \geq \mathrm{Area}(\gamma_{A}) + \mathrm{Area}(\gamma_{B})+ \mathrm{Area}(\gamma_{C}).
	\end{aligned}
\end{equation}
This leads to the following new inequality for the holographic entropies:
\begin{equation}\label{eq:ineq-q=4-reflected-multientropy}
	\begin{aligned}
		&S^{(\q-1=3)}(AA^{*}:BB^{*}:CC^{*}) + S(ABC) \geq S(A) + S(B)+ S(C)\\
		 &\qquad\Longrightarrow \qquad S_{R\, (\q-1=3)}(A:B:C) \geq \left[ S(A) + S(B)+ S(C) - S(ABC) \right] \, \left( \, \ge 0 \,  \right). \, 
	\end{aligned}
\end{equation}
where we used \eqref{eq:def-renyi-reflected-multi-entropy}. Indeed, the right hand side of this inequality is guaranteed to be non-negative, as it follows from \eqref{eq:q=4-identity-1} and the non-negativity of mutual information. 
Note that the lower bound of the reflected multi-entropy we obtained in \eqref{eq:ineq-q=4-reflected-multientropy} is different of the lower bound obtained in \cite{Yuan:2024yfg}. 
Note also that, unlike the $\q=3$ case, inequality \eqref{eq:ineq-q=4-reflected-multientropy} may not hold in non-holographic settings. It would be interesting to explore whether it remains valid in such contexts.

Thus, as in the $\q=3$ case, by subtracting the right-hand side from the left-hand side of the above inequality, we define the 4-partite Markov gap\footnote{Equivalent expressions for the 4-partite Markov gap can be given in terms of the reflected multi-entropy and mutual information, using the relations \eqref{eq:q=4-identity-1} and \eqref{eq:q=4-identity-2}.}:
\begin{equation}\label{eq:def-q=4-holographic-reflected-multi-entropy}
	\begin{aligned}
		MG^{M(\q-1=3)}(A: B: C)&\coloneqq S_{R\, (\q-1=3)}(A:B:C) - \left[ S(A) + S(B)+ S(C) - S(ABC)\right].
	\end{aligned}
\end{equation}
Note that this 4-partite Markov gap is again non-negative by construction in holographic setups, but it is unclear whether this inequality remains non-negative in non-holographic setups.

This 4-partite Markov gap can be generalized to the R\'enyi case by simply replacing each quantity with its R\'enyi counterpart:
\begin{equation}\label{eq:def-renyi-holographic-reflected-multi-entropy}
	\begin{aligned}
		MG^{M(\q-1=3)}_{m,n}(A: B: C)
		&\coloneqq S_{R\, (\q-1=3)}^{(m,n)}(A:B:C) - \left[ S_{n}(A) + S_{n}(B) + S_{n}(C) - S_{n}(ABC) \right]. 
	\end{aligned}
\end{equation}
By this R\'enyi generalization of the 4-partite Markov gap, it is not clear that this R\'enyi 4-partite Markov gap is non-negative even in holographic setups.

\subsubsection{\texorpdfstring{$\q$}{q}=4 case (Disconnected entanglement wedge)}\label{sec:HolographicMarkovGapDisconnectedEW}

\begin{figure}
    \centering
    \includegraphics[width=0.7\linewidth]{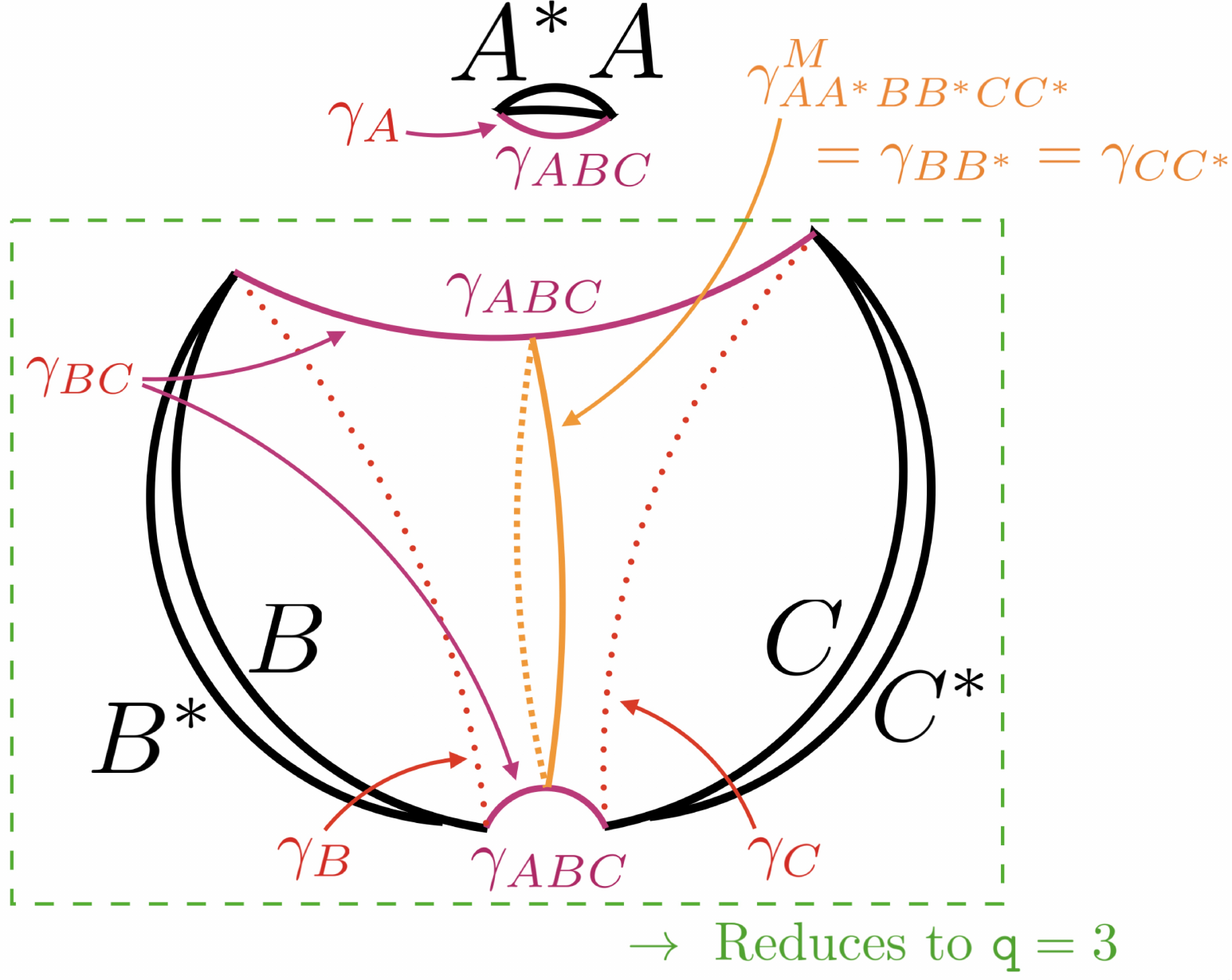}
    \caption{Bulk geometry for the canonical purification of subsystem $ABC$, having a connected entanglement wedge between $B$ and $C$, $\gamma_{ABC}=\gamma_{A}+\gamma_{BC}$, with multiway cut and minimal surfaces in the inequality (\ref{eq:inequality-q=4-extremal-surfaces}), ($\q=4$ case). The multiway cut $\gamma_{AA^* BB^* CC^*}^M$ is equal to the reflected minimal surface $\gamma_{BB^{*}}=\gamma_{CC^{*}}$. In this case, the situation is effectively reduced to the $\q=3$ case, which is boxed in a dotted rectangle. }
    \label{fig:Holographic-proof-positivity-for-biseparable-case-q=4}
\end{figure}

Next, we continue to focus on the $\q = 4$ case, but consider the limit in which one of the four subsystems becomes small, so that the entanglement wedge for $ABC$ connects two of the boundary regions while remaining one disconnected. For example, consider a situation where the subsystems $AD$ are small compared to $BC$, as illustrated in \ref{fig:Holographic-proof-positivity-for-biseparable-case-q=4}. In this case, the multiway cut $\gamma_{AA^* BB^* CC^*}^M$ reduces to the reflected minimal surface $\gamma_{BB^{*}}=\gamma_{CC^{*}}$, and the minimal surface $\gamma_{ABC}$ factorizes as  $\gamma_{A}+\gamma_{BC}$. As a result, the inequality \eqref{eq:inequality-q=4-extremal-surfaces} reduces to that of the $\q=3$ case;
\begin{equation}
	\begin{aligned}
		\mathrm{Area}(\gamma_{AA^{*}BB^{*}CC^{*}}^{M}) + \mathrm{Area}(\gamma_{ABC} )&= \mathrm{Area}(\gamma_{BB^{*}}) + \mathrm{Area}(\gamma_{BC})+ \mathrm{Area}(\gamma_{A})\\
		& \geq \mathrm{Area}(\gamma_{A}) + \mathrm{Area}(\gamma_{B})+ \mathrm{Area}(\gamma_{C}),\\
		\Longrightarrow \qquad S_{R\, (\q-1=3)}(A:B:C)=S_{R}(B:C) &\geq \left[ S(A) + S(B)+ S(C) - S(ABC) \right]\\
        &=\left[ S(B)+ S(C) - S(BC) \right]
	\end{aligned}
\end{equation}
where we used the result $\mathrm{Area}(\gamma_{BB^{*}}) + \mathrm{Area}(\gamma_{BC})\geq \mathrm{Area}(\gamma_{B})+ \mathrm{Area}(\gamma_{C})$ for the case $\q=3$, as given in \eqref{q3casearearela}.  Thus, the inequality \eqref{eq:ineq-q=4-reflected-multientropy} continues to hold even without saturation. This implies that the 4-partite Markov gap takes a positive value equal to the Markov gap between subsystems $B$ and $C$.
\begin{equation}
	MG^{M(\q-1=3)}(A: B: C)=MG(B: C) >0.
\end{equation}

\begin{figure}
    \centering
    \includegraphics[width=0.5\linewidth]{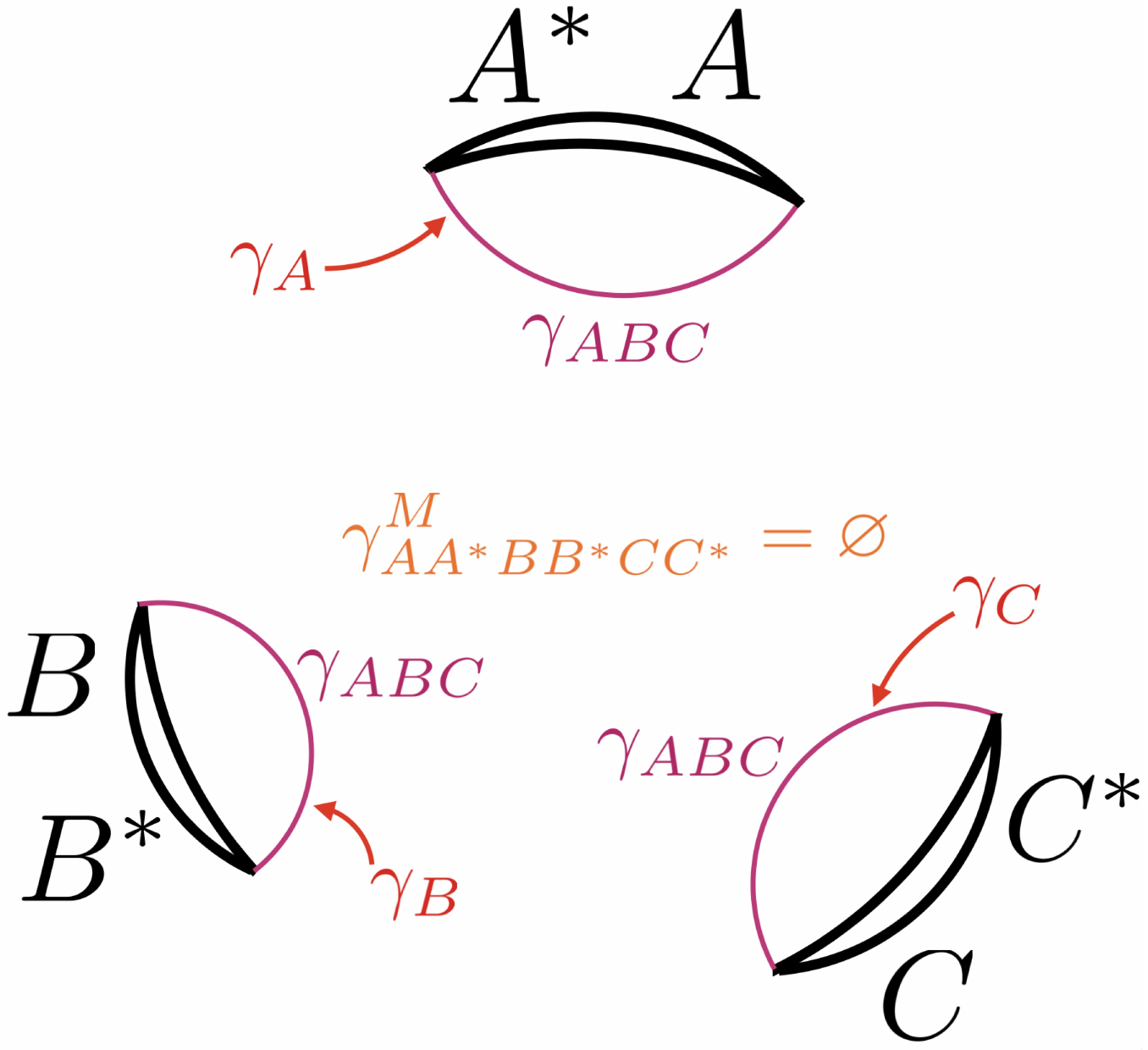}
    \caption{Bulk geometry for the canonical purification of subsystem $ABC$, having a disconnected entanglement wedge, $\gamma_{ABC}=\gamma_{A}+\gamma_{B}+\gamma_{C}$, with empty multiway cut $\gamma_{AA^{*}BB^{*}CC^{*}}^{M}=\varnothing$ and minimal surfaces in the inequality (\ref{eq:inequality-q=4-extremal-surfaces}), ($\q=4$ case). In this case, the inequality (\ref{eq:inequality-q=4-extremal-surfaces}) is saturated.}
    \label{fig:Holographic-vanishing-Markov-gap-q=4}
\end{figure}
Also, in Figure~\ref{fig:Holographic-vanishing-Markov-gap-q=4}, we show a configuration in which the entanglement wedge is entirely disconnected. In this case, the multiway cut $\gamma_{AA^* BB^* CC^*}^M$ is empty, and the minimal surface $\gamma_{ABC}$ splits into three components: $\gamma_{ABC} = \gamma_{A} + \gamma_{B} + \gamma_{C}$. Consequently, the areas of the multiway cut and the minimal surfaces satisfy the equality
\begin{equation}
	\begin{aligned}
		\mathrm{Area}(\underset{=\varnothing}{\gamma_{AA^{*}BB^{*}CC^{*}}^{M}}) + \mathrm{Area}(\gamma_{ABC} )&= \mathrm{Area}(\gamma_{A}) + \mathrm{Area}(\gamma_{B})+ \mathrm{Area}(\gamma_{C}),\\		\Longrightarrow \qquad 0=S_{R\, (\q-1=3)}(A:B:C) &= \left[ S(A) + S(B)+ S(C) - S(ABC) \right].
	\end{aligned}
\end{equation}
Therefore, both sides of the inequality \eqref{eq:ineq-q=4-reflected-multientropy} become zero, saturating the inequality.
This means that the 4-partite Markov gap vanishes in this case;
\begin{equation}
	MG^{M(\q-1=3)}(A: B: C)=0.
\end{equation}
In this way, the 4-partite Markov gap \eqref{eq:def-q=4-holographic-reflected-multi-entropy} takes non-negative values in holographic setups.

\subsubsection{\texorpdfstring{$\q=5,6$}{q=5,6} and beyond}
\begin{figure}[h]
    \centering
    \begin{subfigure}[b]{\linewidth}
        \centering
        \includegraphics[width=0.6\linewidth]{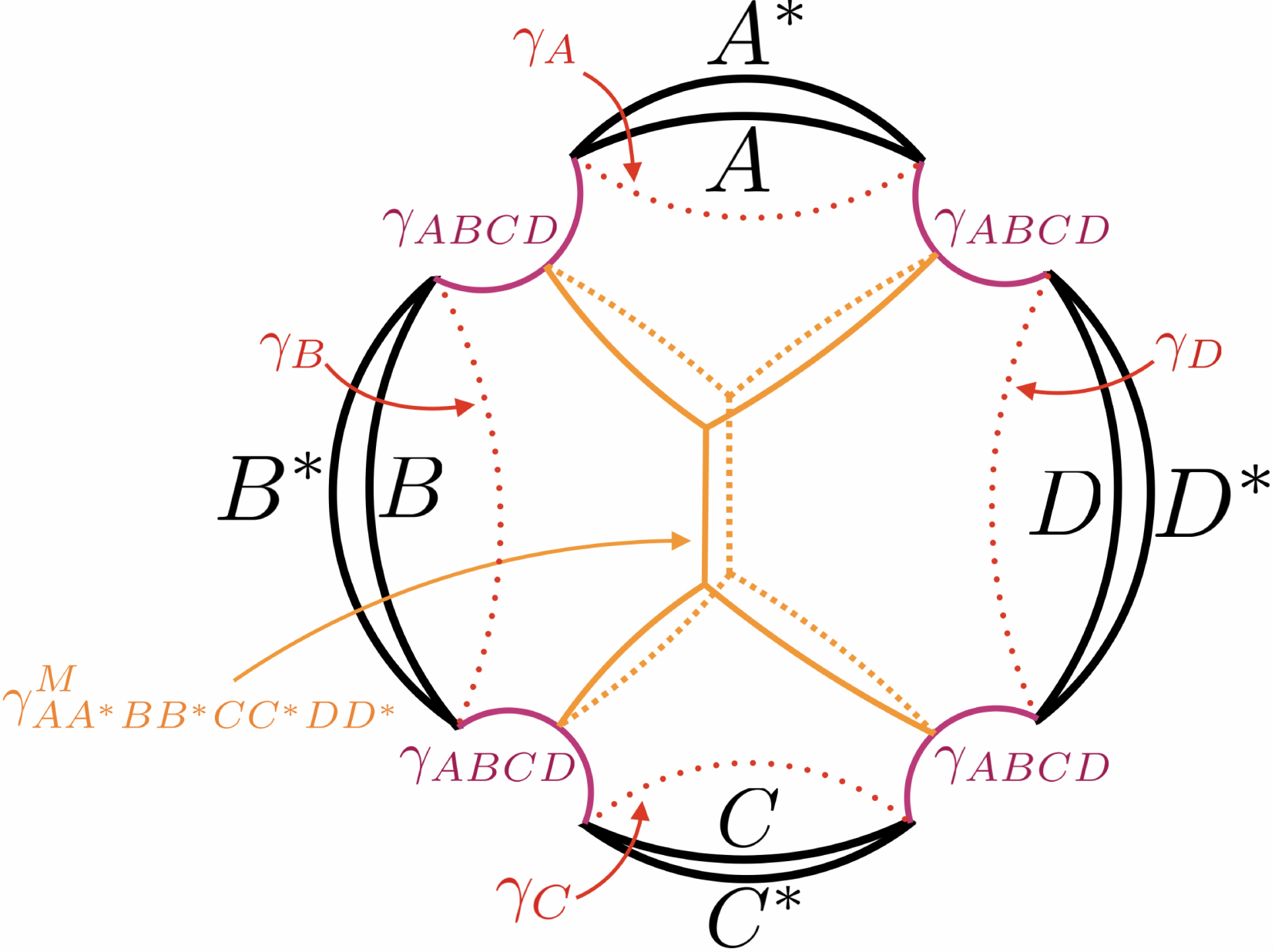}
        \caption{One possible bulk geometry for the canonical purification of subsystem $ABCD$ ($\q=5$ case) with the multiway cut $\gamma_{AA^{*}BB^{*}CC^{*}DD^{*}}^{M}$ and the minimal surfaces.}
    \end{subfigure}
    \vspace{1em}
    \begin{subfigure}[b]{\linewidth}    
$\hspace{20mm}$
\includegraphics[width=0.8\linewidth]{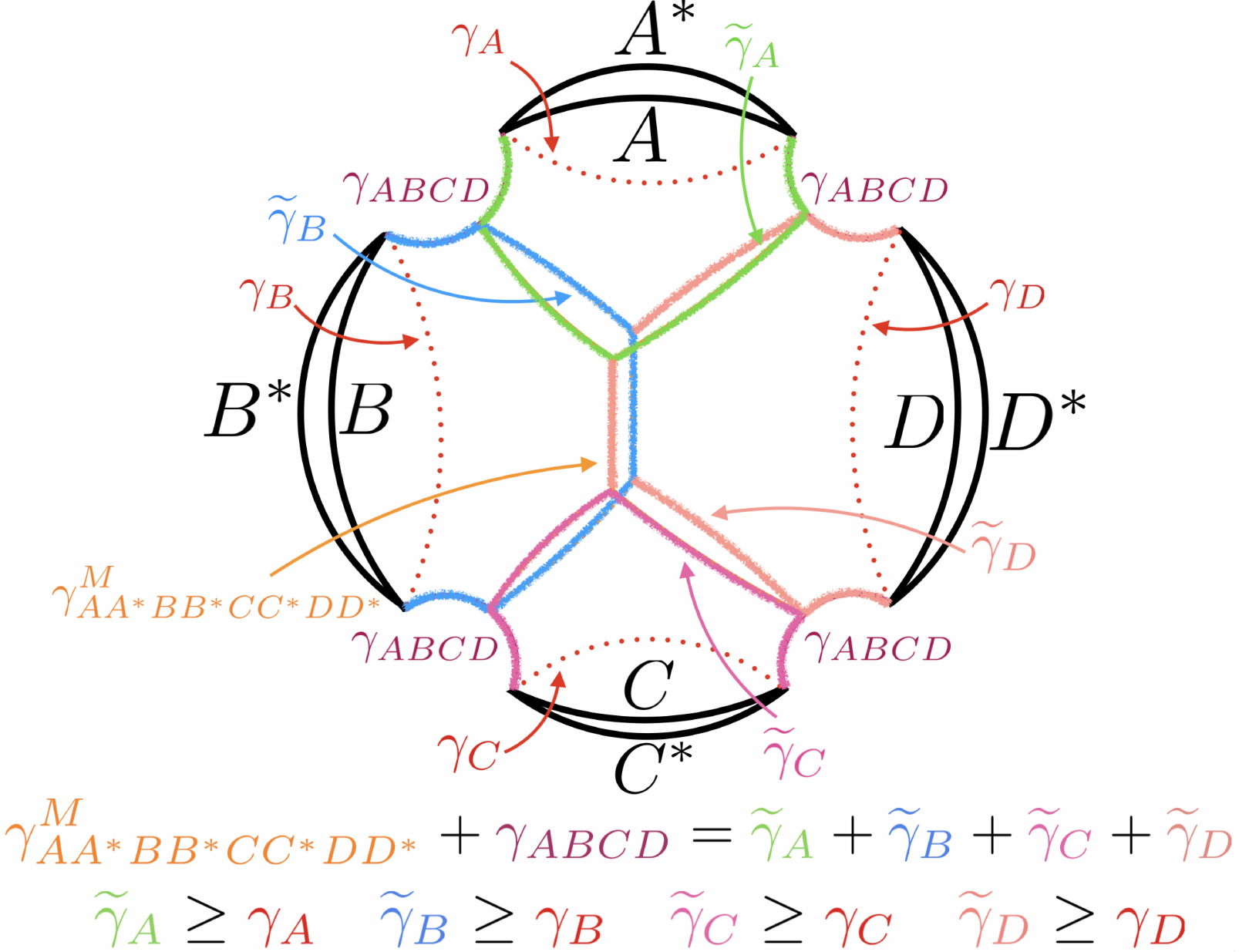}
        \caption{Rearranged non-minimal surfaces $\widetilde{\gamma}$ and minimal surfaces $\gamma$ ($\q=5$ case).}
    \end{subfigure}
    \caption{Multiway cut and minimal surfaces in the inequality (\ref{eq:inequality-q=5-extremal-surfaces}) for the $\q=5$ case.}
    \label{fig:Holographic-inequality-q=5}
\end{figure}

\begin{figure}[h]
    \centering
    \begin{subfigure}[b]{\linewidth}
        \centering
        \includegraphics[width=0.5\linewidth]{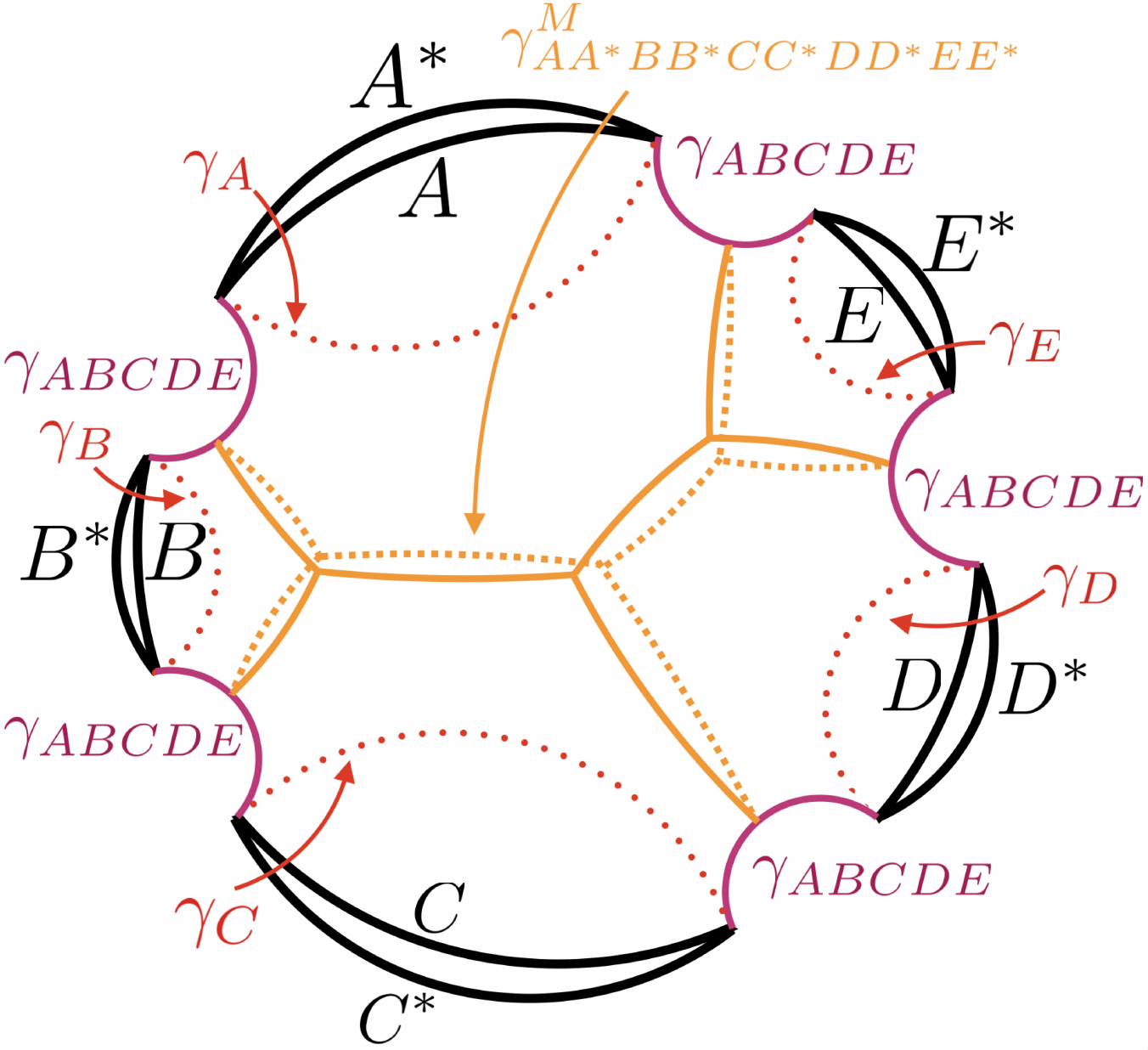}
        \caption{One possible bulk geometry for the canonical purification of subsystem $ABCDE$ ($\q=6$ case) with the multiway cut $\gamma_{AA^{*}BB^{*}CC^{*}DD^{*}EE^{*}}^{M}$ and the minimal surfaces.}
    \end{subfigure}
    \vspace{1em}
    \begin{subfigure}[b]{\linewidth}
        \centering
        \includegraphics[width=0.96\linewidth]{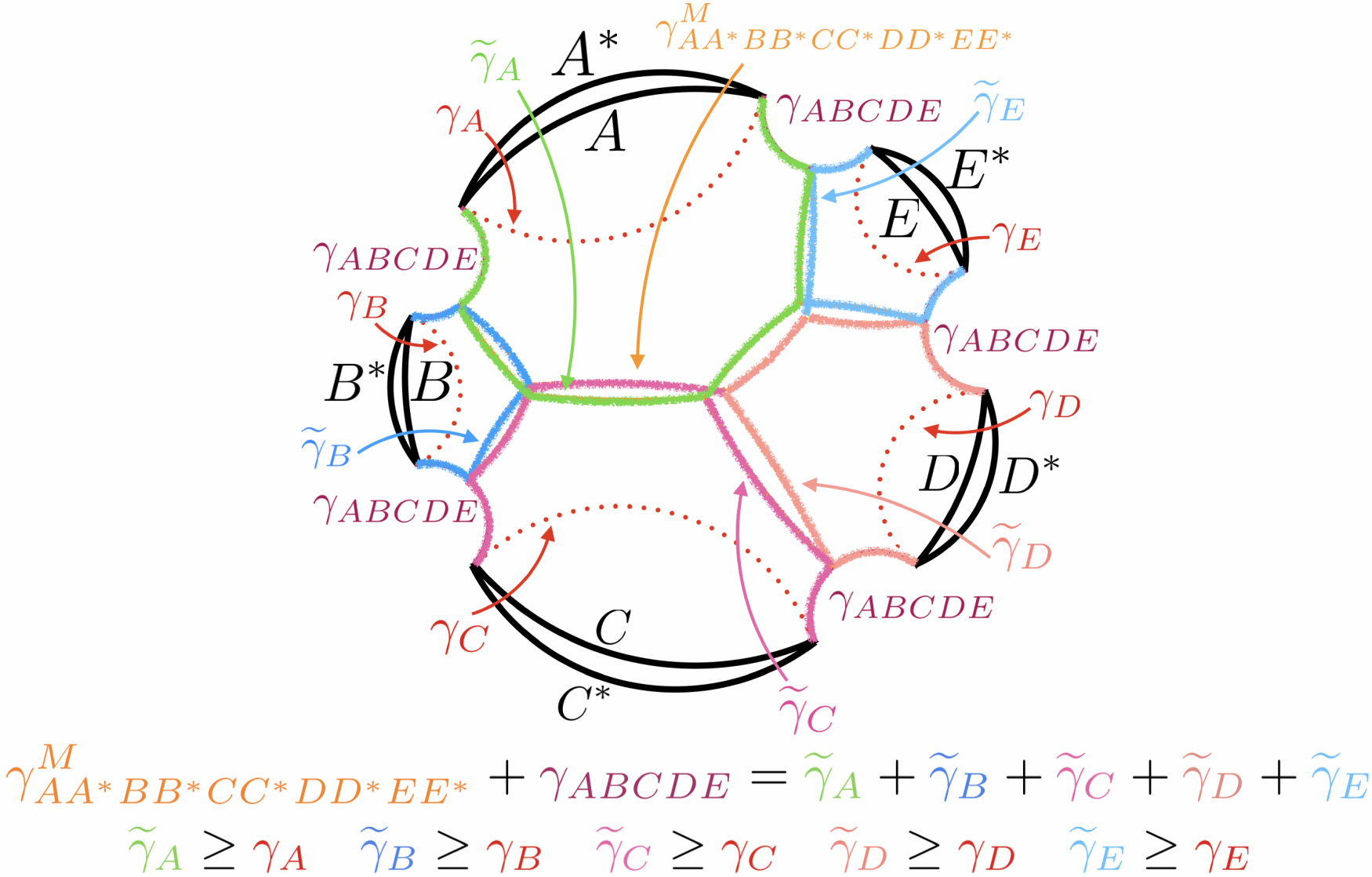}
        \caption{Rearranged non-minimal surfaces $\widetilde{\gamma}$ and minimal surfaces $\gamma$ ($\q=6$ case).}
    \end{subfigure}
    \caption{Multiway cut and minimal surfaces in the inequality (\ref{eq:inequality-q=6-extremal-surfaces}) for the $\q=6$ case.}
    \label{fig:Holographic-inequality-q=6}
\end{figure}

Similar to the $\q=4$ case, the $\q=5$ and $\q=6$ cases can be treated in the same manner. By rearranging the minimal multiway cut and the minimal surfaces, we again introduce non-minimal surfaces and compare them with the corresponding minimal ones. The relevant surfaces are illustrated in Figures \ref{fig:Holographic-inequality-q=5} and \ref{fig:Holographic-inequality-q=6}. From these configurations, the following inequalities follow:
\begin{equation}
\label{eq:inequality-q=5-extremal-surfaces}
	\begin{aligned}
		\mathrm{Area}(\gamma_{AA^{*}BB^{*}CC^{*}DD^{*}}^{M}) + \mathrm{Area}(\gamma_{ABCD} )&= \mathrm{Area}(\widetilde{\gamma}_{A}) + \mathrm{Area}(\widetilde{\gamma}_{B})+ \mathrm{Area}(\widetilde{\gamma}_{C})+ \mathrm{Area}(\widetilde{\gamma}_{D})\\
		& \geq \mathrm{Area}(\gamma_{A}) + \mathrm{Area}(\gamma_{B})+ \mathrm{Area}(\gamma_{C})+ \mathrm{Area}(\gamma_{D}),
	\end{aligned}
\end{equation}
for $\mathtt{q} = 5$ case, and
\begin{equation}
\label{eq:inequality-q=6-extremal-surfaces}
	\begin{aligned}
		\mathrm{Area}(\gamma_{AA^{*}BB^{*}CC^{*}DD^{*}EE^{*}}^{M}) +& \mathrm{Area}(\gamma_{ABCDE} )\\
		&= \mathrm{Area}(\widetilde{\gamma}_{A}) + \mathrm{Area}(\widetilde{\gamma}_{B})+ \mathrm{Area}(\widetilde{\gamma}_{C})+ \mathrm{Area}(\widetilde{\gamma}_{D})+ \mathrm{Area}(\widetilde{\gamma}_{E})\\
		& \geq \mathrm{Area}(\gamma_{A}) + \mathrm{Area}(\gamma_{B})+ \mathrm{Area}(\gamma_{C})+ \mathrm{Area}(\gamma_{D})+ \mathrm{Area}(\gamma_{E}).
	\end{aligned}
\end{equation}
for $\mathtt{q} = 6$ case. 

Thus, these inequalities lead to the lower bound of the reflected multi-entropy for $\q=5,6$;
\begin{equation}
	\begin{aligned}
		S_{R}^{(\q-1=4)}(A:B:C:D)&\geq S(A)+S(B)+S(C)+S(D)-S(ABCD) (\geq 0),\\
	\end{aligned}
\end{equation}
and 
\begin{equation}
	\begin{aligned}
	&S_{R}^{(\q-1=5)}(A:B:C:D:E)\geq S(A)+S(B)+S(C)+S(D)+S(E)-S(ABCDE) (\geq 0).
	\end{aligned}
\end{equation}
Using the relations \eqref{eq:q=5-identity-1} and \eqref{eq:q=6-identity-1}, along with the non-negativity of mutual information, the right hand sides of the above two inequalities are manifestly non-negative.
Again, note that these inequalities may fail to hold in non-holographic setups. It would be interesting to investigate whether this is indeed the case.

Based on these inequalities, 5- and 6-partite Markov gaps are defined as the left hand side minus the right hand side of the corresponding inequalities,
 \begin{equation}
 	\begin{aligned}
 	&MG^{M(\q-1=4)}(A: B: C:D)\\
 		&\coloneqq S_{R}^{(\q-1=4)}(A:B:C:D)- \left[ S(A)+S(B)+S(C)+S(D)-S(ABCD)\right].\\
 	\end{aligned}
    	\label{5markov}
 \end{equation}
 and 
  \begin{equation}
 	\begin{aligned}
 	&MG^{M(\q-1=5)}(A: B: C: D: E)\\
 		&\coloneqq S_{R}^{(\q-1=5)}(A:B:C:D:E)- \left[ S(A)+S(B)+S(C)+S(D)+S(E)-S(ABCDE)\right].\\
 	\end{aligned}
    	\label{6markov}
 \end{equation}
 These 5- and 6-partite Markov gaps are non-negative in holographic setups by construction.

These results can be generalized to the general $\q$ case.
Again, let us assume a situation where subsystem $A_{\q}$ is relatively small such that the entanglement wedge of $A_{1}A_{2}\cdots A_{\q-1}$ is connected. Then, as in \eqref{eq:inequality-q=4-extremal-surfaces}, areas of minimal surfaces and a minimal multiway satisfy the inequality
\begin{equation}
	\mathrm{Area}(\gamma_{A_{1}A_{1}^{*}A_{2}A_{2}^{*}\cdots A_{\q-1}A_{\q-1}^{*}}^{M}) + \mathrm{Area}(\gamma_{A_{1}A_{2}\cdots A_{\q-1}}) \geq \mathrm{Area}(\gamma_{A_{1}}) + \mathrm{Area}(\gamma_{A_{2}})+ \cdots + \mathrm{Area}(\gamma_{A_{\q-1}}).
\end{equation}
This leads to the following new inequality:
\begin{equation}\label{eq:general-q-lower-bound-reflected}
	\begin{aligned}
		&S^{(\q-1)}(A_{1}A_{1}^{*}:A_{2}A_{2}^{*}:\cdots:A_{\q-1}A_{\q-1}^{*}) + S(A_{1}A_{2}\cdots A_{\q-1}) \geq S(A_{1}) + S(A_{2})+ \cdots + S(A_{\q-1})\\
		&\hspace{-5mm}\Longrightarrow S_{R\, (\q-1)}(A_{1}:A_{2}:\cdots:A_{\q-1}) \geq  \left[ S(A_{1}) + S(A_{2})+ \cdots + S(A_{\q-1}) - S(A_{1}A_{2}\cdots A_{\q-1}) \right] (\geq 0). 
	\end{aligned}
\end{equation}
As in the previous cases, the right hand side of the inequality is non-negative from the relation \eqref{eq:q-identity-1} and the non-negativity of mutual information.

From this inequality, one can define a $\q$-partite Markov gap as follows:
\begin{equation}\label{eq:def-general-q-holographic-reflected-multi-entropy}
	\begin{aligned}
		&MG^{M(\q-1)}(A_{1}:A_{2}:\cdots:A_{\q-1})\\
		&\coloneqq S_{R\, (\q-1)}(A_{1}:A_{2}:\cdots:A_{\q-1}) - \left[ S(A_{1}) + S(A_{2})+ \cdots + S(A_{\q-1}) - S(A_{1}A_{2}\cdots A_{\q-1}) \right].\\
	\end{aligned}
\end{equation}
Again, the $\q$-partite Markov gap is non-negative in holographic setups by construction.

The R\'enyi generalization of this quantity is given by
\begin{equation}\label{eq:def-general-q-reyi-holographic-reflected-multi-entropy}
	\begin{aligned}
		&MG^{M(\q-1)}_{m,n}(A_{1}:A_{2}:\cdots:A_{\q-1})\\
		&\coloneqq S_{R\, (\q-1)}^{(m,n)}(A_{1}:A_{2}:\cdots:A_{\q-1}) - \left[ S_{n}(A_{1}) + S_{n}(A_{2})+ \cdots + S_{n}(A_{\q-1}) - S_{n}(A_{1}A_{2}\cdots A_{\q-1}) \right].
	\end{aligned}
\end{equation}

\subsection{Bounding the $\q$-partite Markov Gap from below by the $(\q{-}1)$-partite Markov Gap}\label{subsub:lower-bound-q}
\subsubsection{$\q=4$ case}\label{subsub:q=4-bound}

\begin{figure}
\centering
\includegraphics[width=0.97\linewidth]{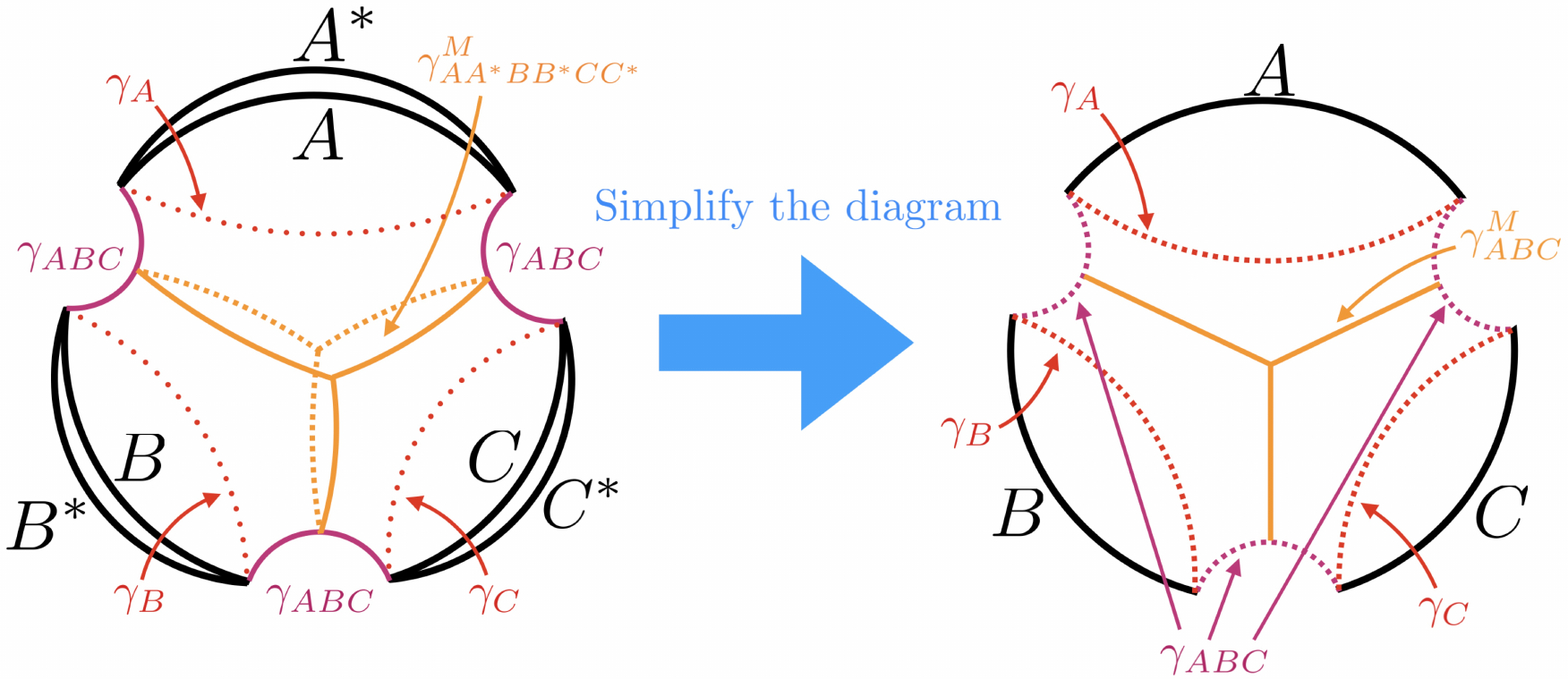}	
    \caption{An example using the single-copy notation. For simplicity, the second copy ($A^{*}B^{*}C^{*}$) is omitted from the illustration.}
\label{fig:Simplified-diagram}
\end{figure}

In this subsection, assuming the holographic duality, we establish the following inequalities between the $\q=4$ and conventional $\q=3$ Markov gaps:
\begin{align}
MG^{M(\q-1=3)}(A: B: C)&\ge MG(A: BC),\label{inequality1}\\
MG^{M(\q-1=3)}(A: B: C)&\ge MG(B: CA),\label{inequality12}\\
MG^{M(\q-1=3)}(A: B: C)&\ge MG(C: AB).\label{inequality13}
\end{align}
Using \eqref{eq:def-Markov-gap} and \eqref{eq:def-q=4-holographic-reflected-multi-entropy}, these inequalities are rewritten in terms of the reflected multi-entropy and the mutual information as
\begin{align}
S_{R\, (\q-1=3)}(A:B:C)&\ge S_{R}(A:BC)+I^{(2)}(B:C),\label{inequality2}\\
S_{R\, (\q-1=3)}(A:B:C)&\ge S_{R}(B:CA)+I^{(2)}(C:A),\\
S_{R\, (\q-1=3)}(A:B:C)&\ge S_{R}(C:AB)+I^{(2)}(A:B).
\end{align}
Below, we derive the inequality (\ref{inequality2}), which is equivalent to (\ref{inequality1}). The other inequalities can be obtained in a similar manner by permuting the subsystems.
In our derivation, we assume that the holographic dual of the reflected multi-entropy is given by the minimal multiway cut in the bulk for a mixed state, as proposed in \cite{Yuan:2024yfg}.
To simplify the bulk illustration, we use the single-copy picture of the bulk, as shown in Figure~\ref{fig:Simplified-diagram}, in which the reflected multi-entropy is proportional to twice the minimal area of the multiway cut.
\begin{enumerate} 
\item[\bf Case 1:]  \underline{Connected entanglement wedge with $I^{(2)}(B:C)=0$}

First, let us consider a bulk configuration with a connected entanglement wedge and vanishing mutual information $I^{(2)}(B:C) = 0$, as illustrated in Figure~\ref{fig:BulkConnectedEWzeroI}. 
\begin{figure}[t]
    \centering
    \includegraphics[width=0.47\linewidth]{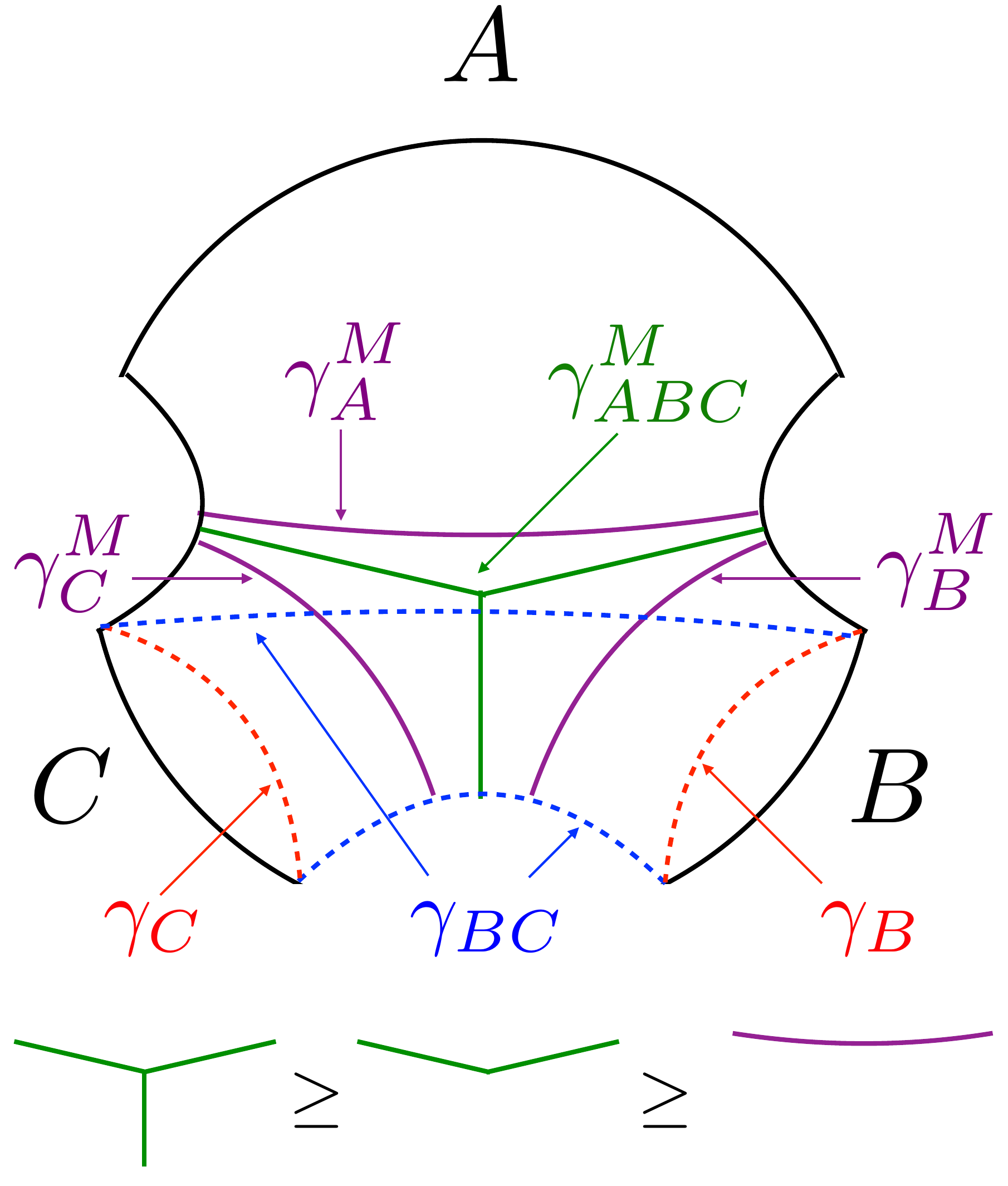}
    \caption{Bulk configuration for the connected entanglement wedge with $I^{(2)}(B:C)=0$ and a geometrical inequality between area of the minimal cuts.}
\label{fig:BulkConnectedEWzeroI}
\end{figure}
Here, $I^{(2)}(B:C) = 0$ corresponds to the condition $S(B) + S(C) = S(BC)$, or equivalently, $\mathrm{Area}(\gamma_{BC}) \ge \mathrm{Area}(\gamma_{B}) + \mathrm{Area}(\gamma_{C})$.
In this setup, $\gamma_{ABC}^M$ denotes the minimal triway cut, and $\gamma_{A}^M,\gamma_{B}^M,\gamma_{C}^M$ are half of the reflected minimal surfaces to exclude single subsystem $AA^{*}$, $BB^{*}$, $CC^{*}$, respectively. We also draw the minimal surfaces $\gamma_B,\gamma_C,\gamma_{BC}$ anchored to the boundary, where $I^{(2)}(B:C)=0$ means $\text{Area}(\gamma_{BC})\ge\text{Area}(\gamma_{B}+\gamma_{C})$.\footnote{As shown in the figure, $\gamma_{BC}$ is the minimal surface that connects $B$ and $C$ through the bulk. However, it does not necessarily provide the dominant contribution to $S_{BC}$. It becomes the dominant contribution only when its area is smaller than that of the disconnected surface $\gamma_B + \gamma_C$.} Using the holographic duality and the geometrical inequality shown in Figure \ref{fig:BulkConnectedEWzeroI}, we obtain the following.
\begin{align}
\begin{aligned}
& S_{R\, (\q-1=3)}(A:B:C)=\frac{2}{4G_N}\text{Area}(\gamma_{ABC}^M)\ge \frac{2}{4G_N}\text{Area}(\gamma_{A}^M)\\
&\, \, \, \ge \frac{2}{4G_N}\text{Min} \{\text{Area}(\gamma_{A}^M), \text{Area}(\gamma_{B}^M+\gamma_{C}^M) \}=S_{R}(A:BC) \,,
\end{aligned}
\end{align}
which gives a geometrical proof of the inequality (\ref{inequality2}) when $I^{(2)}(B:C)=0$.

\begin{figure}[t]
    \centering
    \includegraphics[width=0.44\linewidth]{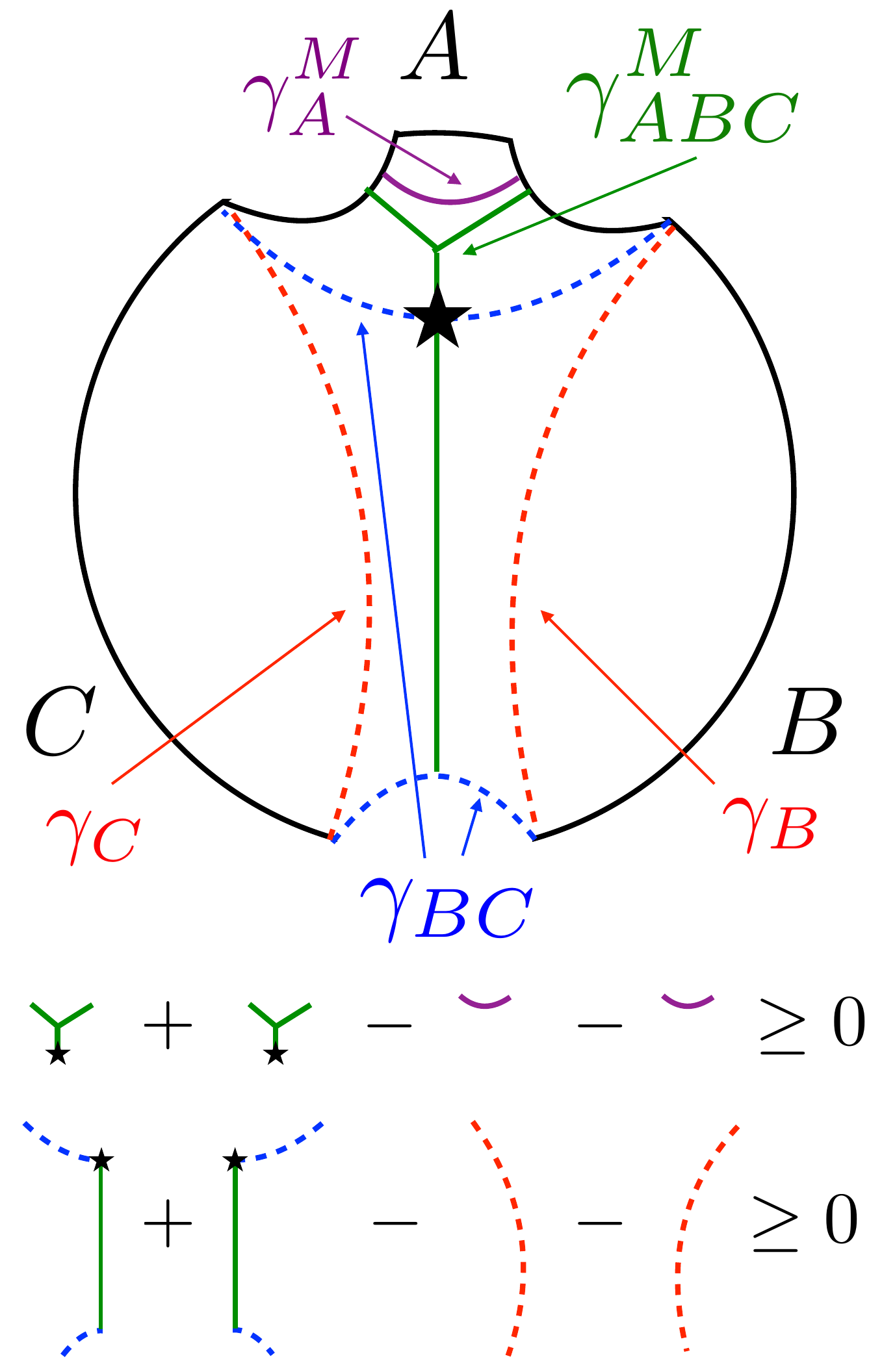}
    \caption{Bulk configuration for the connected entanglement wedge with $I^{(2)}(B:C)>0$ when the center of the triway cut is outside the entanglement wedge of $BC$, and  geometrical inequalities between area of the minimal cuts.}
\label{fig:BulkConnectedEWOutside}
\end{figure}

\begin{figure}[t]
    \centering
    \includegraphics[width=0.44\linewidth]{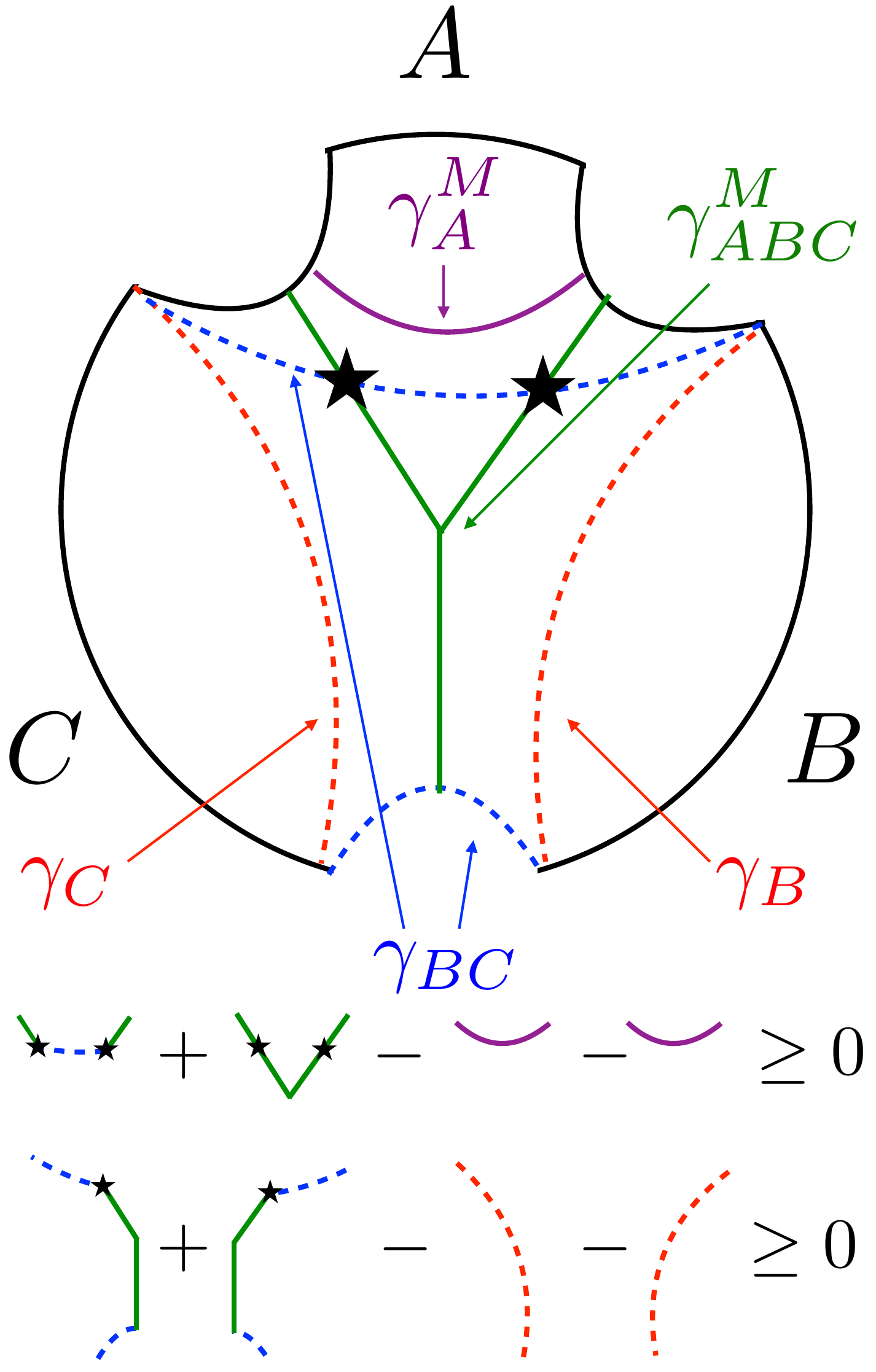}
    \caption{Bulk configuration for the connected entanglement wedge with $I^{(2)}(B:C)>0$ when the center of the triway cut is inside the entanglement wedge of $BC$, and geometrical inequalities between the area of the minimal cuts.}
\label{fig:BulkConnectedEWInside}
\end{figure}

\item[\bf Case 2:]  \underline{Connected entanglement wedge with $I^{(2)}(B:C)>0$}

Next, let us consider a bulk configuration with a connected entanglement wedge and mutual information that is non-zero, $I^{(2)}(B:C)>0$. This means $S(B)+S(C)>S(BC)$, or equivalently, $\mathrm{Area}(\gamma_{BC}) < \mathrm{Area}(\gamma_{B}) + \mathrm{Area}(\gamma_{C})$. In this case, we further classify the configurations into two types: (1) the center of the triway cut lies outside the entanglement wedge of $BC$ as shown in Figure \ref{fig:BulkConnectedEWOutside}, and (2) the center of the triway cut lies inside the entanglement wedge of $BC$ as shown in Figure \ref{fig:BulkConnectedEWInside}. In both cases, using the geometrical inequalities shown in Figures \ref{fig:BulkConnectedEWOutside} and \ref{fig:BulkConnectedEWInside}, one can show that
\begin{align}
\label{geometricalinequalityfor332}
\frac{2}{4G_N}\text{Area}(\gamma_{ABC}^M) +\frac{1}{4G_N}\text{Area}(\gamma_{BC}) - \frac{2}{4G_N}\text{Area}(\gamma_{A}^M)-\frac{1}{4G_N}\text{Area}(\gamma_{B}+\gamma_{C})
\ge 0 \,,
\end{align}
where $\gamma_B,\gamma_C,\gamma_{BC}$ are the minimal surfaces corresponding to $S(B),S(C),S(BC)$, respectively, and $I^{(2)}(B:C)>0$ means $\text{Area}(\gamma_{B}+\gamma_{C}) > \text{Area}(\gamma_{BC})$. 

Furthermore, using the following holographic relationship,
\begin{align}
S_{R}(A:BC)=\frac{2}{4G_N}\text{Min} \{\text{Area}(\gamma_{A}^M), \text{Area}(\gamma_{B}^M+\gamma_{C}^M) \}\le\frac{2}{4G_N}\text{Area}(\gamma_{A}^M) 
\end{align}
and 
\begin{align}
\begin{aligned}
   S_{R\, (\q-1=3)}(A:B:C) &= \frac{2}{4G_N}\text{Area}(\gamma_{ABC}^M) \,, \\
   I^{(2)}(B:C) &= \frac{1}{4G_N}\text{Area}(\gamma_{B}+\gamma_{C}) - \frac{1}{4G_N}\text{Area}(\gamma_{BC}) \,,
   \end{aligned}
\end{align}
we thus obtain the desired inequality as follows:
\begin{align}
\begin{aligned}
&\;\;\;\;\;S_{R\, (\q-1=3)}(A:B:C)-S_{R}(A:BC)-I^{(2)}(B:C)\\
&\ge \frac{2}{4G_N}\text{Area}(\gamma_{ABC}^M)- \frac{2}{4G_N}\text{Area}(\gamma_{A}^M)-\frac{1}{4G_N}\text{Area}(\gamma_{B}+\gamma_{C})+\frac{1}{4G_N}\text{Area}(\gamma_{BC}) \, \ge \, 0 \,,
\end{aligned}
\end{align}
where the final inequality follows from
\eqref{geometricalinequalityfor332}.

\item[\bf Case 3:]  \underline{Disconnected entanglement wedge}

Finally, let us consider bulk configurations in which the entanglement wedge is disconnected. We examine two cases: a disconnected entanglement wedge with bipartite entanglement between B and C, as shown in Figure~\ref{fig:BulkDisConnectedEWBipartite}, and a fully disconnected entanglement wedge, as shown in Figure~\ref{fig:BulkDisConnectedEWEntire}. 

As shown in Section~\ref{sec:HolographicMarkovGapDisconnectedEW}, for the configuration in Figure~\ref{fig:BulkDisConnectedEWBipartite}, we obtain:
\begin{align}
MG^{M(\q-1=3)}(A: B: C)=MG(B: CA)=MG(C: AB)>MG(A: BC)=0,
\end{align}
and for the configuration in Figure~\ref{fig:BulkDisConnectedEWEntire}, we find:
\begin{align}
MG^{M(\q-1=3)}(A: B: C)=MG(A: BC)=MG(B: CA)=MG(C: AB)=0.
\end{align}
These relations constitute special cases of the inequalities (\ref{inequality1}), (\ref{inequality12}), and (\ref{inequality13}).

\begin{figure}[t]
\centering
\begin{minipage}[b]{0.54\columnwidth}
    \centering
    \includegraphics[width=0.47\columnwidth]{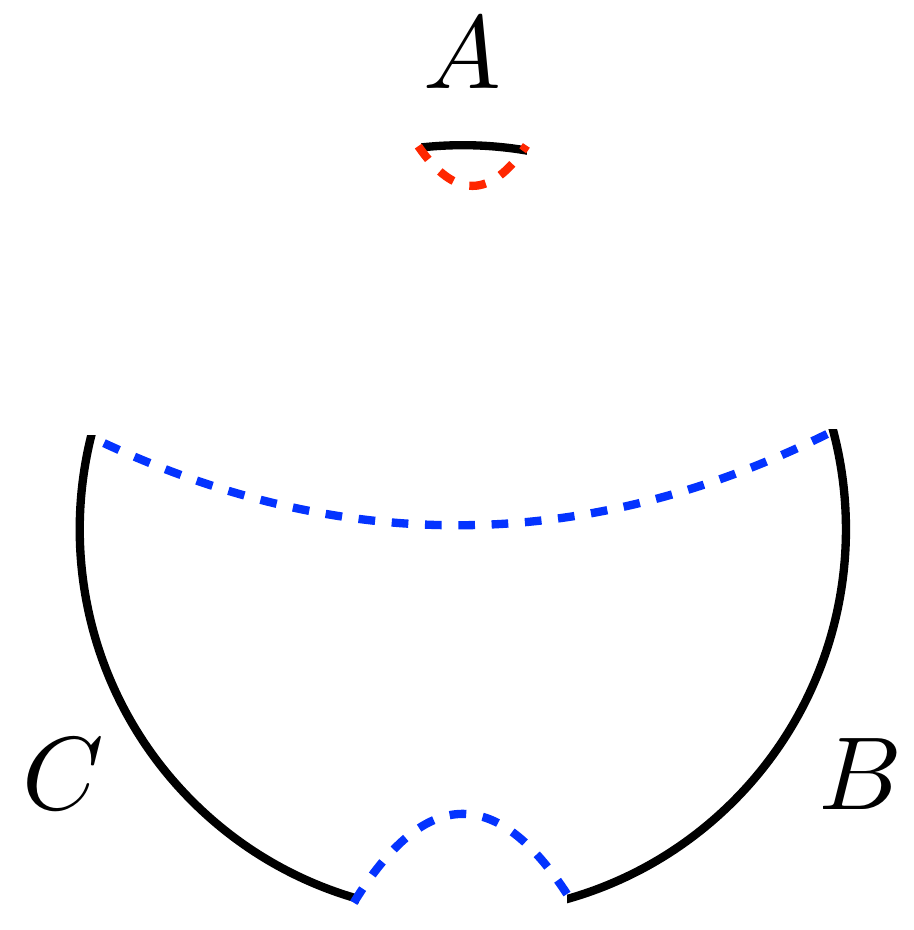}
    \caption{Disconnected entanglement wedge with bipartite entanglement between $B$ and $C$.}
    \label{fig:BulkDisConnectedEWBipartite}
\end{minipage}
\begin{minipage}[b]{0.44\columnwidth}
    \centering
    \includegraphics[width=0.65\columnwidth]{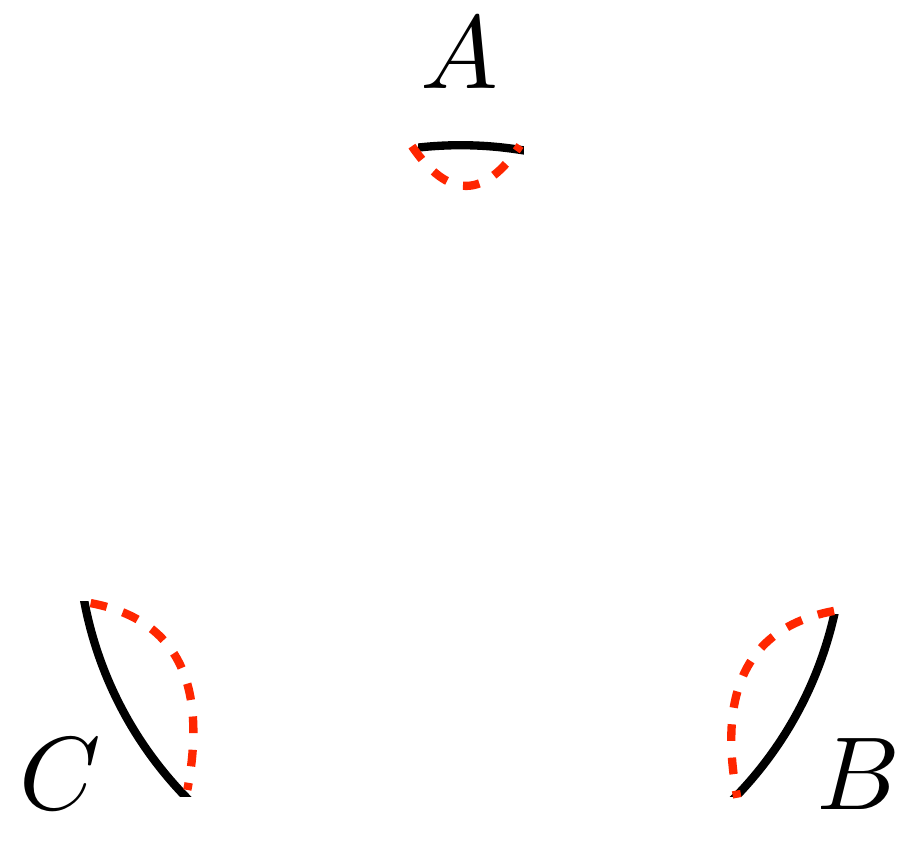}
    \caption{Entirely disconnected entanglement wedge.}
    \label{fig:BulkDisConnectedEWEntire}
\end{minipage}
\end{figure}

\end{enumerate}
\subsubsection{$\q \ge 5$ cases}

As an extension of the lower bounds on the Multipartite Markov gap observed in \eqref{inequality1} for $\q=4$ case, we conjecture that the following inequality holds for higher values of $\q$:
\begin{equation}
\label{genericqbounds}
MG^{M(\q-1)}(A_1: A_2: \cdots : A_{\q-1}) \geq MG^{M(\q-2)} (A_1A_2: \cdots : A_{\q-1}) \,,
\end{equation}
which is expected to hold for all integers $\q \ge 5$.

We provide partial evidence for this conjecture by analyzing several explicit examples of bulk configurations and their corresponding geometrical inequalities.

As the simplest nontrivial case, we begin by examining $\q = 5$.
In this case, the inequality \eqref{genericqbounds} becomes:
\begin{align}
MG^{M(\q-1=4)}(A: B: C:D)&\ge MG^{M(\q-2=3)}(AB: C: D),
\end{align}
 which can be equivalently rewritten as
\begin{align}\label{inequalityq5q4}
S_{R\, (\q-1=4)}(A:B:C:D)&\ge S_{R\, (\q-2=3)}(AB:C:D)+I^{(2)}(A:B).
\end{align}

\begin{figure}[t]
    \centering
    \includegraphics[width=0.5\linewidth]{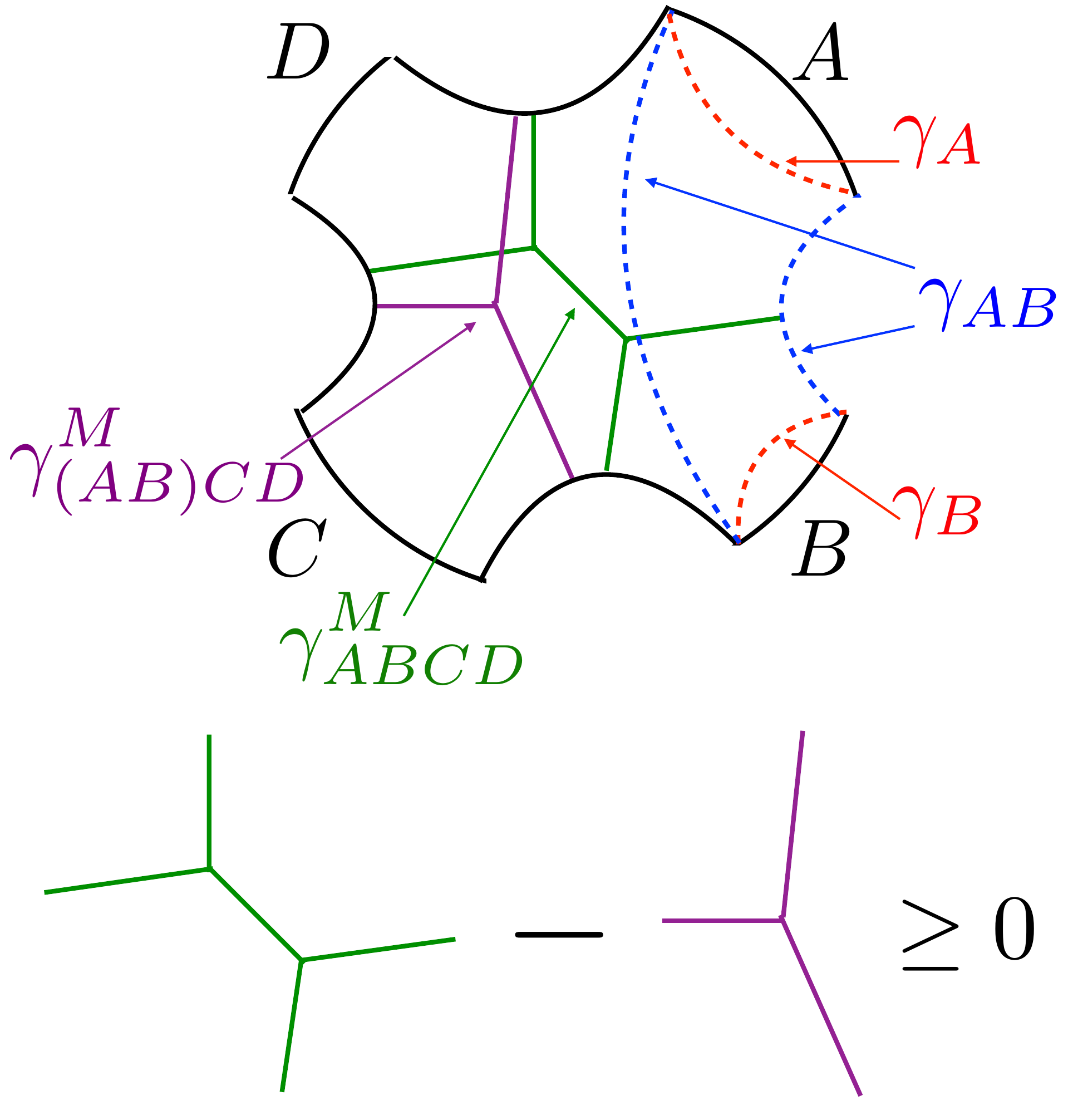}
    \caption{Bulk configuration for the connected entanglement wedge with $I^{(2)}(A:B)=0$, which means $\text{Area}(\gamma_{AB})\ge\text{Area}(\gamma_{A}+\gamma_{B})$, and a geometrical inequality between area of the minimal cuts $\text{Area}(\gamma_{ABCD}^M)-\text{Area}(\gamma_{(AB)CD}^M)\ge0$.}
\label{fig:BulkConnectedEWzeroIq5q4}
\end{figure}

\begin{figure}[t]
    \centering
    \includegraphics[width=0.43\linewidth]{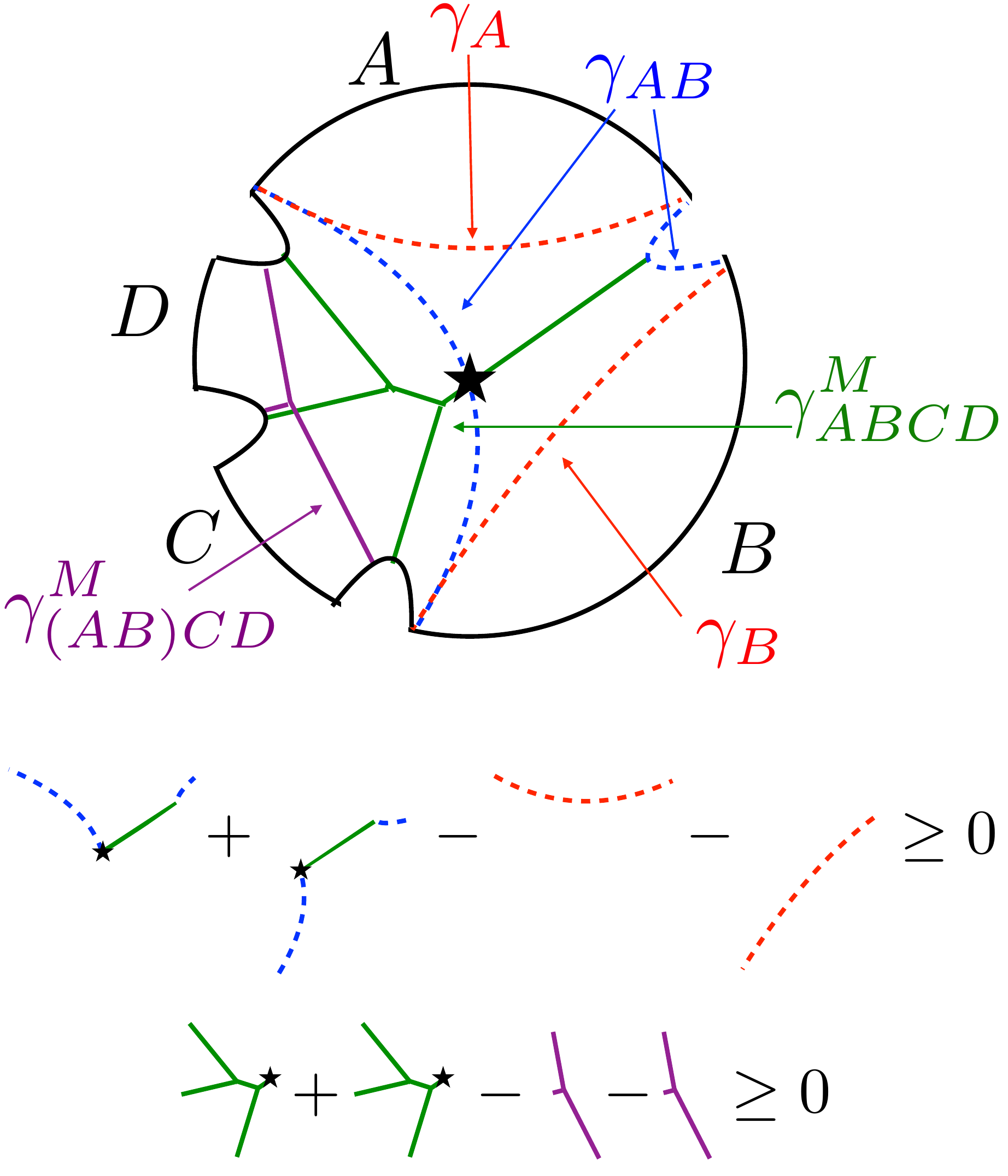}
    \caption{Bulk configuration for the connected entanglement wedge with $I^{(2)}(A:B)>0$, which means $\text{Area}(\gamma_{A}+\gamma_{B})>\text{Area}(\gamma_{AB})$, when the intersection points of the minimal cut $\gamma_{ABCD}^M$ is outside the entanglement wedge of $AB$, and  geometrical inequalities between area of the minimal cuts for $2\text{Area}(\gamma_{ABCD}^M)-2\text{Area}(\gamma_{(AB)CD}^M)-\text{Area}(\gamma_{A}+\gamma_{B}) +\text{Area}(\gamma_{AB})\ge0$.}
\label{fig:BulkConnectedEWOutsideq5q4}
\end{figure}

\begin{figure}[t]
    \centering
    \includegraphics[width=0.43\linewidth]{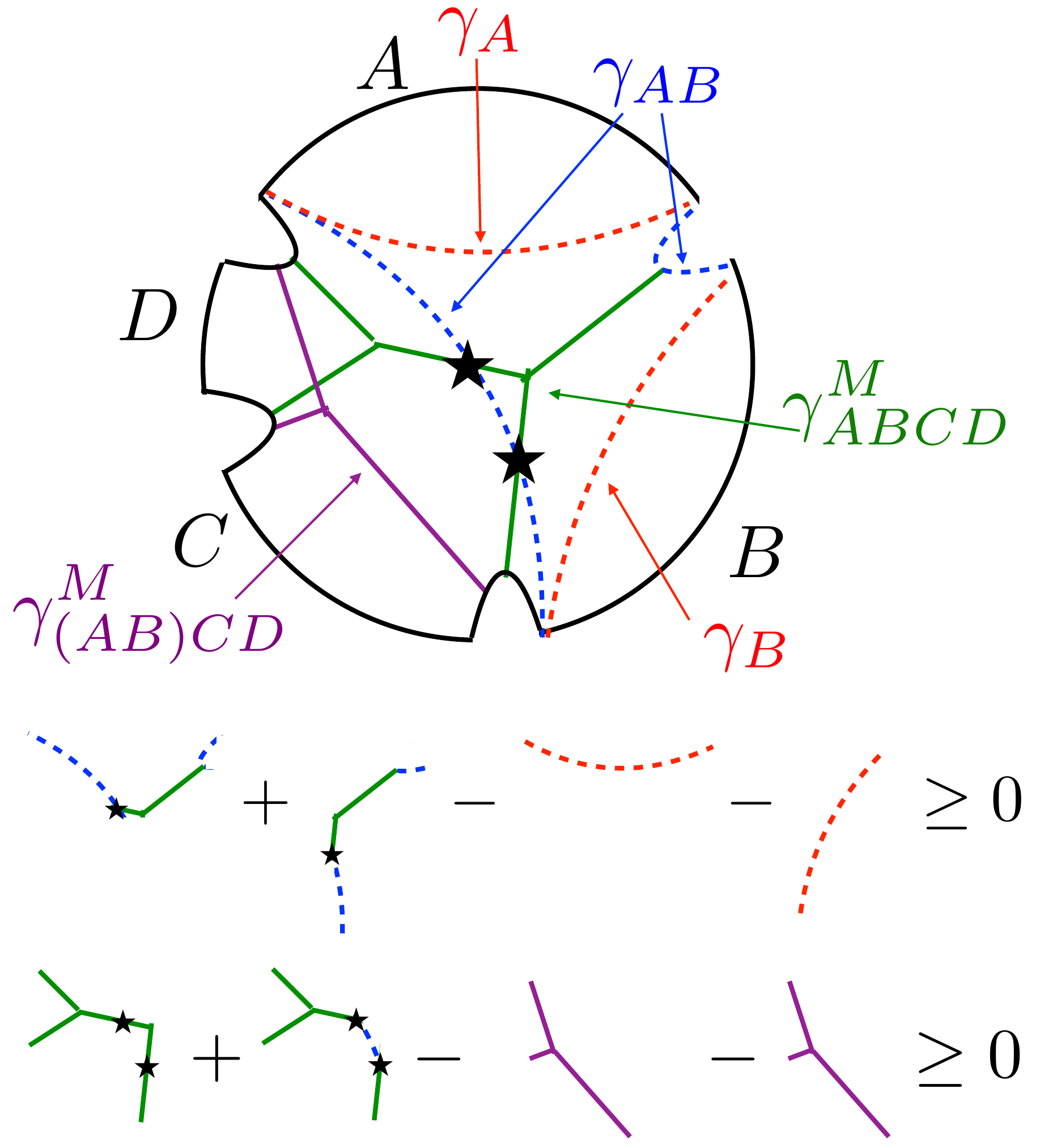}
    \caption{Bulk configuration for the connected entanglement wedge with $I^{(2)}(A:B)>0$, which means $\text{Area}(\gamma_{A}+\gamma_{B})>\text{Area}(\gamma_{AB})$, when one of the intersection points of the minimal cut $\gamma_{ABCD}^M$ is inside the entanglement wedge of $AB$, and  geometrical inequalities between area of the minimal cuts for $2\text{Area}(\gamma_{ABCD}^M)-2\text{Area}(\gamma_{(AB)CD}^M)-\text{Area}(\gamma_{A}+\gamma_{B}) +\text{Area}(\gamma_{AB})\ge0$.}
\label{fig:BulkConnectedEWInsideq5q4}
\end{figure}

To see how this inequality holds, we consider three examples of bulk configurations, as analyzed in the previous subsection and illustrated in Figures~\ref{fig:BulkConnectedEWzeroIq5q4}, \ref{fig:BulkConnectedEWOutsideq5q4}, and \ref{fig:BulkConnectedEWInsideq5q4}.
Here, $\gamma_{A}, \gamma_{B},$ and $\gamma_{AB}$ are minimal surfaces anchored to the boundary subsystems $A$, $B$, and $AB$, respectively.
The surface $\gamma_{ABCD}^M$ denotes the minimal cut that partitions the bulk into four subregions associated with boundary subsystems $A, B, C, D$, while $\gamma_{(AB)CD}^M$ is the minimal cut dividing the bulk into three subregions corresponding to $AB, C, D$.

In all three cases, applying the holographic dictionary gives 
\begin{align}
\begin{aligned}
S_{R\, (\q-1=3)}(AB:C:D)&=\frac{2}{4G_N}\text{Area}(\gamma_{(AB)CD}^M)\,,\\ 
   S_{R\, (\q-1=4)}(A:B:C:D) &= \frac{2}{4G_N}\text{Area}(\gamma_{ABCD}^M) \,, \\
   I^{(2)}(A:B) &= 
   \begin{cases}
   0 & (\text{Figure}~\ref{fig:BulkConnectedEWzeroIq5q4} )\\
   \frac{1}{4G_N}\text{Area}(\gamma_{A}+\gamma_{B}) - \frac{1}{4G_N}\text{Area}(\gamma_{AB}) & (\text{Figures}~\ref{fig:BulkConnectedEWOutsideq5q4}, \ref{fig:BulkConnectedEWInsideq5q4} ),
   \end{cases}
   \end{aligned}
\end{align}
Substituting these expressions into the geometric inequalities obtained from the three bulk configurations shown in Figures~\ref{fig:BulkConnectedEWzeroIq5q4}, \ref{fig:BulkConnectedEWOutsideq5q4}, and \ref{fig:BulkConnectedEWInsideq5q4}, we find 
\begin{align}
\begin{aligned}
&S_{R\, (\q-1=4)}(A:B:C:D)-\left(S_{R\, (\q-1=3)}(AB:C:D)+I^{(2)}(A:B)\right)\\
&\ge\frac{2}{4G_N}\text{Area}(\gamma_{ABCD}^M)-\frac{2}{4G_N}\text{Area}(\gamma_{(AB)CD}^M)\\
&\, \ge 0 \,,
\end{aligned}
\end{align}
for Figure \ref{fig:BulkConnectedEWzeroIq5q4}, and
\begin{align}
\begin{aligned}
&S_{R\, (\q-1=4)}(A:B:C:D)-\left(S_{R\, (\q-1=3)}(AB:C:D)+I^{(2)}(A:B)\right)\\
&\ge\frac{2}{4G_N}\text{Area}(\gamma_{ABCD}^M)-\frac{2}{4G_N}\text{Area}(\gamma_{(AB)CD}^M)-\frac{1}{4G_N}\text{Area}(\gamma_{A}+\gamma_{B}) + \frac{1}{4G_N}\text{Area}(\gamma_{AB})\\
&\, \ge 0 \,,
\end{aligned}
\end{align}
for Figures \ref{fig:BulkConnectedEWOutsideq5q4} and \ref{fig:BulkConnectedEWInsideq5q4}. Therefore, we confirm the inequality~\eqref{inequalityq5q4}.

A similar argument can be applied to higher values of $\q \ge 6$ in holographic settings, making the generalization conceptually straightforward. In this paper, we treat subsystems such as $A$ as a single connected domain, but one can generalize this even when $A$ is a disconnected domain.

\begin{figure}[t]
    \centering
\includegraphics[width=0.45\linewidth]{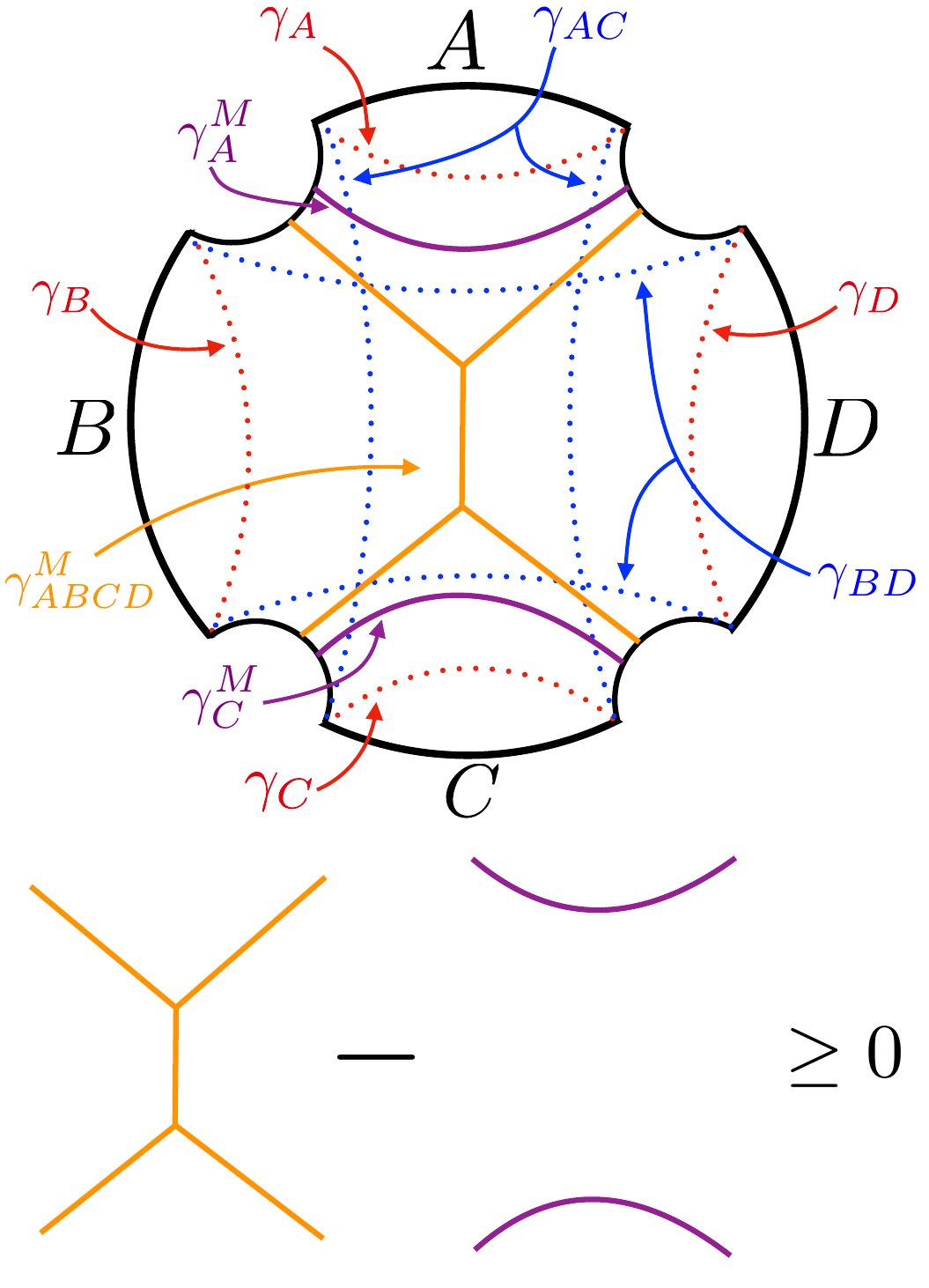}
    \caption{Bulk configuration for the connected entanglement wedge of four-partite subsystems $A,B,C,D$ with $I^{(2)}(A:C)=I^{(2)}(B:D)=0$, and a geometrical inequality between area of the minimal cuts.}
\label{fig:BulkConnectedEWq5}
\end{figure}

Note that the generic inequality \eqref{genericqbounds} implies the following sequence of lower bounds:
\begin{align}
MG^{M(\q-1)} \ge MG^{M(\q-2)} \ge MG^{M(\q-3)} \ge \dots \ge MG^{M(3)} \ge MG^{M(2)}.
\end{align}
For example, this leads to the inequality
\begin{align}
MG^{M(4)}(A: B: C: D) \ge MG(AC: BD), \label{inequalityq5}
\end{align}
for the case between $\q=5$ and $\q=3$, as illustrated in Figure~\ref{fig:BulkConnectedEWq5}, assuming $I^{(2)}(A:C) = I^{(2)}(B:D) = 0$.
Here, $\gamma_{ABCD}^M$, $\gamma_{A}^M$, and $\gamma_{C}^M$ denote the minimal cuts, while $\gamma_{AC}$, $\gamma_{BD}$, $\gamma_{A}$, $\gamma_{B}$, $\gamma_{C}$, and $\gamma_{D}$ represent minimal surfaces anchored to the boundary.
Here, the condition $I^{(2)}(A:C) = 0$ means that $\text{Area}(\gamma_{AC}) \ge \text{Area}(\gamma_{A}) + \text{Area}(\gamma_{C})$, and similarly, $I^{(2)}(B:D) = 0$ implies $\text{Area}(\gamma_{BD}) \ge \text{Area}(\gamma_{B}) + \text{Area}(\gamma_{D})$.

In terms of the reflected multi-entropy and mutual information, the inequality \eqref{inequalityq5} can be rewritten as
\begin{align}
S_{R (\q-1=4)}(A:B:C:D) \ge S_{R}(AC:BD) + I^{(2)}(A:C) + I^{(2)}(B:D).
\end{align}
For the bulk configuration shown in Figure~\ref{fig:BulkConnectedEWq5}, under  $I^{(2)}(A:C) = I^{(2)}(B:D) = 0$, we obtain
\begin{align}
S_{R (\q-1=4)}(A:B:C:D) &= \frac{2}{4G_N} \text{Area}(\gamma_{ABCD}^M)\notag
\\
&\ge \frac{2}{4G_N} \left(\text{Area}(\gamma_{A}^M) +  \text{Area}(\gamma_{C}^M)\right) 
\ge S_{R}(AC:BD) \,,
\end{align}
where we have used the holographic duality together with the geometric inequality illustrated in Figure~\ref{fig:BulkConnectedEWq5}.

\begin{figure}[t]
    \centering
\includegraphics[width=0.5\linewidth]{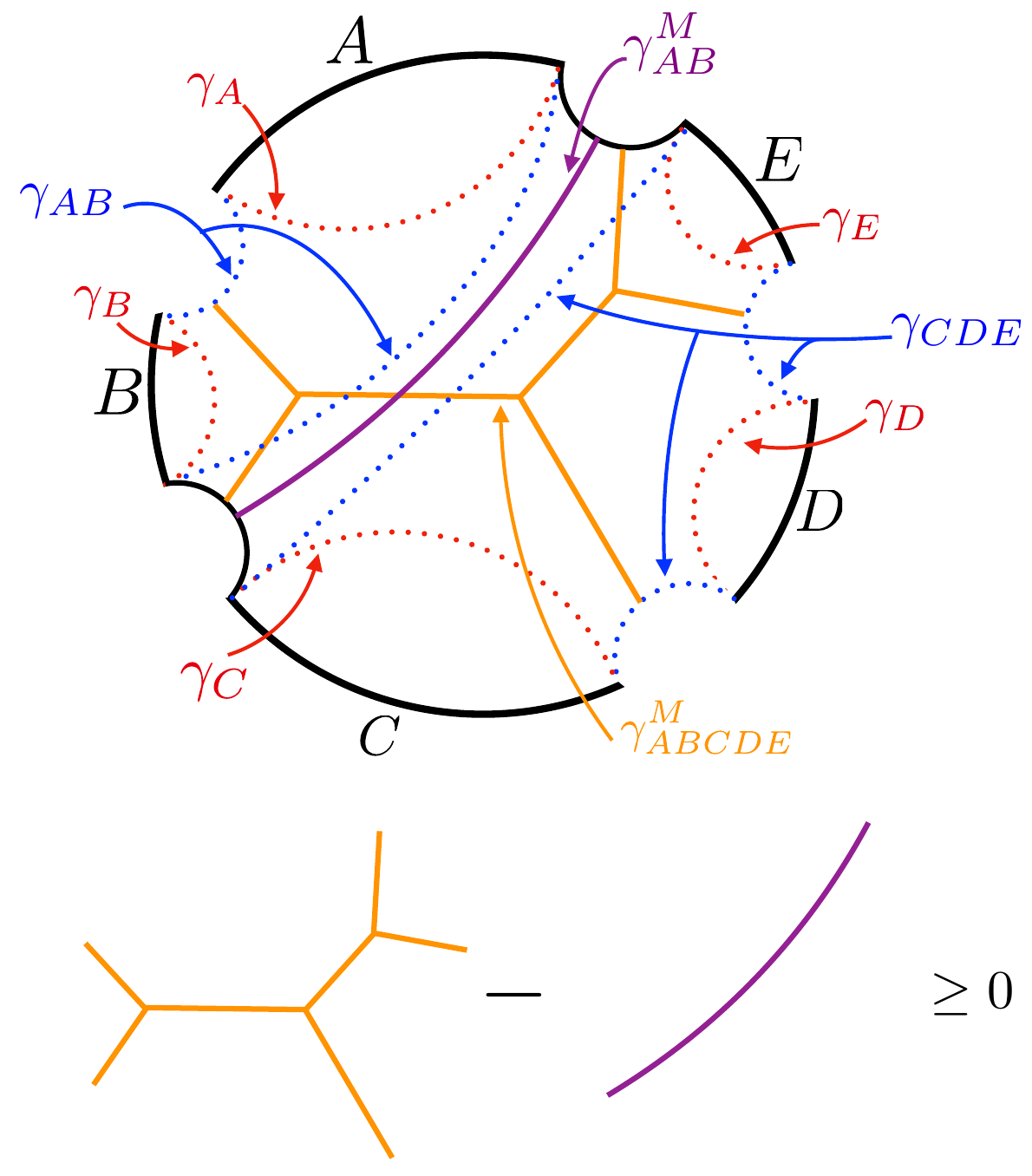}
    \caption{Bulk configuration for the connected entanglement wedge of five-partite subsystems $A,B,C,D,E$ with $I^{(2)}(A:B)=S(C)+S(D)+S(E)-S(CDE)=0$, and a geometrical inequality between area of the minimal cuts.}
\label{fig:BulkConnectedEWq6}
\end{figure}

Another example is the following inequality:
\begin{align}
MG^{M(\q-1=5)}(A: B: C: D: E) \ge MG(AB: CDE), \label{inequalityq6}
\end{align}
for the case between $\q=6$ and $\q=3$, as shown in Figure~\ref{fig:BulkConnectedEWq6}, under the assumption that
$I^{(2)}(A:B) = 0$ and $S(C) + S(D) + S(E) - S(CDE) = 0$.
Here, $\gamma_{ABCDE}^M$ and $\gamma_{AB}^M$ are the minimal cuts, while $\gamma_{AB}$, $\gamma_{CDE}$, $\gamma_{A}$, $\gamma_{B}$, $\gamma_{C}$, $\gamma_{D}$, and $\gamma_{E}$ are minimal surfaces anchored to the boundary.
The condition $I^{(2)}(A:B) = 0$ implies $\text{Area}(\gamma_{AB}) \ge \text{Area}(\gamma_{A}) + \text{Area}(\gamma_{B})$, and the condition $S(C) + S(D) + S(E) - S(CDE) = 0$ implies $\text{Area}(\gamma_{CDE}) \ge \text{Area}(\gamma_{C}) + \text{Area}(\gamma_{D}) + \text{Area}(\gamma_{E})$.

In terms of the reflected multi-entropy and mutual information, inequality \eqref{inequalityq6} can be rewritten as:
\begin{align}
S_{R (\q-1=5)}(A:B:C:D:E)
&\ge S_{R}(AB:CDE)  \nonumber \\ &\quad + I^{(2)}(A:B) + S(C) + S(D) + S(E)
 - S(CDE).
\end{align}

Under the assumptions $I^{(2)}(A:B) = 0$ and $S(C) + S(D) + S(E) - S(CDE) = 0$, for the bulk configuration shown in Figure~\ref{fig:BulkConnectedEWq6}, we obtain:
\begin{align}
S_{R (\q-1=5)}(A:B:C:D:E)
= \frac{2}{4G_N} \text{Area}(\gamma_{ABCDE}^M)
\ge \frac{2}{4G_N} \text{Area}(\gamma_{AB}^M)
\ge S_{R}(AB:CDE),
\end{align}
where we use the holographic duality together with the geometric inequalities illustrated in Figure~\ref{fig:BulkConnectedEWq6}.

It is also possible to investigate the case without these assumptions, but we will not pursue it further here.

\subsection{Multipartite Markov gap for \texorpdfstring{$\q=4$}{q=4} and the Markov Recovery Problem}\label{subsec:Markov-recovery}

Thus far, we have introduced the Multipartite Markov gap and investigated its properties using holographic methods. Although our construction relies heavily on holographic intuition, we now turn to a deeper connection between the Multipartite Markov gap and quantum recovery processes, one that holds independently of holography. This generalizes the known relation between the conventional (tripartite) Markov gap and quantum Markov recovery, as first explored in~\cite{Hayden:2021gno}\footnote{One can consider another generalization of a similar recovery problem in multipartite systems \cite{Wilde_2015,Wilde:2015xoa}. However, the recovery operation is different from ours. It would be interesting to find the relation between their discussion and ours.}.

In this subsection, we present this connection in detail and illustrate it explicitly in qubit systems. For simplicity, we focus on the $\q=4$ case, although the extension to general $\q$ is straightforward. To set the stage, we begin with a brief review of the standard $\q=3$ case. Readers already familiar with this material may proceed directly to subsection~\ref{subsec:Markov-recovery-q=4}.
\subsubsection{Review of the Tripartite Markov Gap and Markov Recovery in the \texorpdfstring{$\q=3$}{q=3} Case} \label{subsub:review-Markov-recovery}

We begin by reviewing the quantum information-theoretic interpretation of the Markov gap, along with its geometric manifestation in holographic settings.
For a detailed discussion of the Markov gap and its role in quantum recovery, we refer the reader to~\cite{Hayden:2021gno}.
For foundational results on recovery maps and the quantum Markov condition, see also~\cite{Wilde:2015xoa, Wilde:2011npi}. This background will serve as a useful reference point for our generalization to multipartite settings in the following sections.

We start with the fact that the Markov gap is equal to the conditional mutual information \cite{Hayden:2021gno}:
\begin{equation}\label{eq:MG=CMI}
	MG(A:B)=S_{R}(A)-I(A:B)=I(A:B^{*}|B)=I(B:A^{*}|A).
\end{equation}
Here, the conditional mutual information $I(A:C|B)$ is defined by
\begin{equation}\label{eq:conditiona-MI}
	I(A:C|B)=S(A B)+S(B C)-S(A B C)-S(B)= I(A:BC)-I(A:B),
\end{equation}
and it takes non-zero values due to the strong sub-additivity. 
We can show the second equality of \eqref{eq:MG=CMI} by noting the relation 
\begin{equation}
	\begin{aligned}
		S_{R}(A)= S(AA^*)=S(BB^*)=S(BB^{*})+S(A)-S(ABB^{*})=I(A:BB^{*}),
	\end{aligned}
\end{equation}
where we use the fact that the state on $ABA^{*}B^{*}$ is pure, and we add the identity $0 = S(A^{*}) - S(ABB^{*})$, noting that $S(A) = S(A^{*})$.
Similarly, we can show the relation 
\begin{equation}
	\begin{aligned}
		S(AA^*)=S(AA^{*})+S(B)-S(AA^{*}B)=I(AA^{*}:B),
	\end{aligned}
\end{equation}
From this equality, we get the third equality of \eqref{eq:MG=CMI}.

We also observe that, based on the expression of the mutual information~\eqref{eq:conditiona-MI} in terms of entanglement entropies, the roles of subsystems $A$ and $C$ are symmetric
\begin{equation}
  I(A:C|B) = I(C:A|B) \,.
\end{equation}
Hence, the subsystems $A$ and $C$ can be interchanged throughout the subsequent analysis due to their symmetric appearance.

Using the relation~\eqref{eq:MG=CMI}, we now interpret the Markov gap in terms of the conditional mutual information.  
From~\eqref{eq:conditiona-MI}, the conditional mutual information \( I(A:C|B) \) quantifies the additional correlation between \( A \) and the composite system \( BC \), beyond the correlation between \( A \) and \( B \) alone.  
In other words, it measures how much the correlation between \( A \) and \( B \) increases when subsystem \( C \) is included.

When \( I(A:C|B) = 0 \), all correlations between \( A \) and \( C \) are mediated through \( B \), and the information of subsystem \( A \) can be perfectly recovered from \( B \) alone, without access to \( C \).  
By contrast, when \( I(A:C|B) > 0 \), some information about \( A \) is inaccessible from \( B \) alone, and access to \( C \) is necessary for perfect recovery.

These observations can be reformulated in terms of the existence and performance of a quantum recovery map.  
To make this connection explicit, let us recall that the conditional mutual information can be expressed as a difference of relative entropies~\cite{Hayden_2004}:
\begin{equation}
	\begin{aligned}
		I(A:C|B)
        &= I(A:BC)-I(A:B)\\
		&=S(A B)+S(B C)-S(A B C)-S(B)\\
		&=S\left(\rho_{A B C} \| \rho_{AB} \otimes \rho_{C}\right)-S\left(\rho_{B C} \| \rho_{B} \otimes \rho_{C}\right),
	\end{aligned}
    	\label{conditionalmutual}
\end{equation}
where $S(\rho||\sigma)$ denotes the relative entropy, defined by
\begin{equation}
	S(\rho||\sigma)=\tr\left[\rho\left( \log \rho-\log \sigma \right)\right] \,. 
\end{equation} 
Here we have used the relations
\begin{equation}
	I(AB:C)=S(AB)+S(C)-S(ABC)=S\left(\rho_{A B C} \| \rho_{AB} \otimes \rho_{ C}\right),
\end{equation}
and
\begin{equation}
	I(B:C)=S(B)+S(C)-S(BC)=S\left(\rho_{B C} \| \rho_{B} \otimes \rho_{C}\right).
\end{equation}
and the combination $S\left(\rho_{A B C} \| \rho_{AB} \otimes \rho_{C}\right)-S\left(\rho_{B C} \| \rho_{B} \otimes \rho_{C}\right)$ is called the relative entropy difference \cite{Seshadreesan:2014ddz}.

The relative entropy measures the distinguishability between two states $\rho$ and $\sigma$. It takes non-negative values and vanishes if and only if $\rho=\sigma$. 
Due to the monotonicity of relative entropy\footnote{See \textit{e.g.,} \cite{Witten:2018zxz} for the review on the derivation of the monotonicity of relative entropy.}, it can not increase under trace operation,
\begin{equation}
	S\left(\rho_{A B C} \| \rho_{AB} \otimes \rho_{C}\right)\geq S\left(\rho_{B C} \| \rho_{B} \otimes \rho_{C}\right).
\end{equation}
This means that, under the trace operation of $A$, it might become difficult to distinguish the two states. In the case where $I(A:C|B)=0$,  the above inequality is saturated:
\begin{equation} \label{eq:saturated-relative-entropy}
	S\left(\rho_{A B C} \| \rho_{AB} \otimes \rho_{C}\right)= S\left(\rho_{B C} \| \rho_{B} \otimes \rho_{C}\right).
\end{equation}
 In this case, as shown in \cite{Petz:2002eql}, we have the equality\footnote{By taking the derivative the equality with respect to $z$, sending $z\to 1$ and taking the expectation value by $\rho_{ABC}$, we get \eqref{eq:saturated-relative-entropy}. }
 \begin{equation}
 	(\rho_{ABC})^{z}(\rho_{AB}\otimes \rho_{C} )^{-z}= (\rho_{BC})^{z}(\rho_{B}\otimes \rho_{C} )^{-z}  \qquad \forall z\in \mathbb{C},
 \end{equation}
or more simply
 \begin{equation}
 	(\rho_{ABC})^{z}(\rho_{AB} )^{-z}= (\rho_{BC})^{z}(\rho_{B})^{-z}  \qquad \forall z\in \mathbb{C}.
 \end{equation}
 By taking $z=1/2-it/2$ ($t\in \mathbb{R}$), we have
 \begin{equation}
 \label{Petzid}
 	(\rho_{ABC})^{\frac{1}{2}-\frac{it}{2}}(\rho_{AB} )^{-\frac{1}{2}+\frac{it}{2}}= (\rho_{BC})^{\frac{1}{2}-\frac{it}{2}}(\rho_{B})^{-\frac{1}{2}+\frac{it}{2}},
 \end{equation}
 and its Hermit conjugation 
 \begin{equation}
 \label{HCPetzid}
 	(\rho_{AB} )^{-\frac{1}{2}-\frac{it}{2}} (\rho_{ABC})^{\frac{1}{2}+\frac{it}{2}}= (\rho_{B})^{-\frac{1}{2}-\frac{it}{2}} (\rho_{BC})^{\frac{1}{2}+\frac{it}{2}}.
 \end{equation}
By multiplying both sides of equation~\eqref{HCPetzid} on the right by equation~\eqref{Petzid}, and multiplying by $(\rho_{AB} )^{1/2 + it/2}$ from left and its Hermit conjugation from right, we obtain\footnote{The recovery map $\mathcal{R}_{B\to AB}^{P,t}$ is non-trivial due to the constraint that it can only act on $B$ and we can use only $\rho_{AB}$ and $\rho_{B}$. Otherwise, if we can use $C$ in addition to $B$, \textit{i.e.,} $\rho_{AB}\to \rho_{ABC}$ and $\rho_{B}\to \rho_{BC}$, then this recovery problem and the recovery map become trivial:
\begin{equation}
	\begin{aligned}
		(\rho_{ABC} )^{\frac{1}{2}+\frac{it}{2}}(\rho_{BC})^{-\frac{1}{2}-\frac{it}{2}} \, \rho_{BC}\, (\rho_{BC})^{-\frac{1}{2}+\frac{it}{2}} (\rho_{ABC} )^{\frac{1}{2}-\frac{it}{2}}&=(\rho_{ABC} )^{\frac{1}{2}+\frac{it}{2}} \, I_{BC}\,(\rho_{ABC} )^{\frac{1}{2}-\frac{it}{2}}\\
		&=\rho_{ABC}.
	\end{aligned}
\end{equation}
}
 \begin{equation}\label{eq:recovery-map}
 \begin{aligned}
 	\rho_{ABC}&=(\rho_{AB} )^{\frac{1}{2}+\frac{it}{2}}(\rho_{B})^{-\frac{1}{2}-\frac{it}{2}} \, \rho_{BC}\, (\rho_{B})^{-\frac{1}{2}+\frac{it}{2}} (\rho_{AB} )^{\frac{1}{2}-\frac{it}{2}}\\
 	& \eqqcolon \mathcal{R}_{B\to AB}^{P,t}\left[ \rho_{BC} \right],
 \end{aligned} \qquad \forall t\in \mathbb{R}
 \end{equation}
where $\mathcal{R}_{B\to AB}^{P,t}$ is the recovery map acting on subsystem $B$. 

This map is known to be a special case of the rotated Petz map, and in the $t=0$ case, it reduces to the conventional Petz map~\cite{Petz:1986tvy,Petz:1988usv}.  
Moreover, in the case where there is no correlation between $A$ and $B$, \textit{i.e.}, $\rho_{AB} = \rho_A \otimes \rho_B$, the recovery map further simplifies to the following form:
\begin{equation}\label{eq:recovery-map-special}
 \begin{aligned}
 	\rho_{ABC}&=(\rho_{A}\otimes \rho_{B} )^{\frac{1}{2}+\frac{it}{2}}(\rho_{B})^{-\frac{1}{2}-\frac{it}{2}} (\rho_{BC})(\rho_{B})^{-\frac{1}{2}+\frac{it}{2}} (\rho_{A}\otimes \rho_{B} )^{\frac{1}{2}-\frac{it}{2}}\\
 	&=\rho_{A}\otimes \rho_{BC}\\
 	&\eqqcolon\mathcal{R}_{B\to AB}^{P}\left[ \rho_{BC} \right].
 \end{aligned}
 \end{equation}
Therefore, as long as $I(A:C|B) = 0$, the above recovery map functions correctly, enabling the reconstruction of information in subsystem $A$ from $B$ alone, without accessing $C$. This setup is known as the Markov recovery problem~\cite{Hayden_2004}. A key feature of this problem is that the recovery map is restricted to act only on subsystem $B$, rather than on the joint system $BC$. This constraint renders the recovery task significantly more challenging compared to the situation in which operations are allowed on the full $BC$ system.

When $I(A:C|B) > 0$, the equality in \eqref{eq:saturated-relative-entropy} no longer holds, and thus it becomes impossible to construct a recovery map that perfectly reconstructs the original information solely from subsystem $B$.  
Nevertheless, in such cases, one can approximately recover the information of subsystem $A$ from $B$ using the rotated Petz map $\mathcal{R}_{B \to AB}^{P, t}$.  
It is known that the fidelity between the original state $\rho_{ABC}$ and the approximately recovered state $\mathcal{R}_{B \to AB}^{P, t}[\rho_{BC}]$ is bounded below by a function of the conditional mutual information $I(A:C|B)$\footnote{One can also consider the bound by the \textit{twirled Petz map} \cite{Junge:2015lmb}. However, since later we numerically evaluate the fidelity between the original state and the recovered state, it is easier to evaluate the fidelity when we consider the rotated Petz map. Thus, we continue to focus on the rotated Petz map for simplicity.} \cite{Wilde:2015xoa}:
\begin{equation}\label{FidelityBound}
	I(A:C|B) = S\left(\rho_{A B C} \| \rho_{AB} \otimes \rho_{C}\right)-S\left(\rho_{B C} \| \rho_{B} \otimes \rho_{C}\right) \geq -\sup_{t \in \mathbb{R}} \left\{  \log \left[ F\left(\rho_{ABC}, \mathcal{R}_{B\to AB}^{P,t}\left[ \rho_{BC} \right]\right) \right]\right\},
\end{equation}
where $F(\rho,\sigma)$ is the fidelity defined by
\begin{equation}
\label{fidelitydef}
	F(\rho, \sigma)=\left[\tr \sqrt{\sqrt{\rho} \sigma \sqrt{\rho}}\right]^2 \qquad 0\leq F(\rho, \sigma) \leq 1.
\end{equation}
The fidelity is a measure of the closeness or similarity between two quantum states $\rho,\sigma$, and it takes $1$ when  the two states are identical, and 0 when they have  orthogonal support\footnote{Let us briefly explain the fidelity by using a simple example. We take $\ket{\psi},\ket{\psi'}$ to be orthogonal normalized states, $\braket{\psi|\psi'}=0$, and consider the states $\sigma=\ket{\psi}\bra{\psi}$ and $\rho=p\ket{\psi'}\bra{\psi'} + (1-p) \sigma $ ($0\leq p \leq 1$). Then, we can see that
\begin{equation}
	\sqrt{\rho} \sigma \sqrt{\rho}= \left( \sqrt{p}\ket{\psi'}\bra{\psi'} + \sqrt{1-p} \, \sigma  \right)\sigma \left( \sqrt{p}\ket{\psi'}\bra{\psi'} + \sqrt{1-p} \,  \sigma  \right)= (1-p)\sigma 
\end{equation}
and 
\begin{equation}
	\tr \sqrt{\sqrt{\rho} \sigma \sqrt{\rho}}= \sqrt{1-p}\,  \tr \left[ \sigma  \right]= \sqrt{1-p} \qquad \Longrightarrow \qquad F(\rho, \sigma)=1-p.
\end{equation}
  }.

Returning to our setting, we have the equality \eqref{eq:MG=CMI} for the Markov gap, and the general relation of the fidelity and the conditional mutual information \eqref{FidelityBound}. Combining them, we obtain the following bounds~\cite{Hayden:2021gno,Wilde:2015xoa}\footnote{One can consider a general recovery map instead of the rotated Petz map by simply replacing the rotated Petz map $ \mathcal{R}_{B \rightarrow BB^{*}}^{P,t}$ with a general recovery map $ \mathcal{R}_{B \rightarrow BB^{*}}$ and the operation $\sup\limits_{t \in \mathbb{R}}$ with $\max\limits_{\mathcal{R}_{B \rightarrow BB^{*}}}$. However, in this paper, for the purpose of numerical calculations, we focus on the rotated Petz map case. }
\begin{equation}\label{eq:q=3-fidelity-bound-1}
	MG(A:B)=I(A:B^{*}|B) \geq -\sup_{t\in \mathbb{R}} \log F\left(\rho_{ABB^{*}}, \mathcal{R}_{B \rightarrow BB^{*}}^{P,t}\left(\rho_{AB}\right)\right) \geq0,
\end{equation}
\begin{equation}\label{eq:q=3-fidelity-bound-2}
	MG(A:B)=I(B:A^{*}|A) \geq -\sup_{t\in \mathbb{R}} \log F\left(\rho_{AA^{*}B}, \mathcal{R}_{A \rightarrow AA^{*}}^{P,t}\left(\rho_{AB}\right)\right)\geq 0,
\end{equation}
where, in giving $\geq 0$, we used the fact that the fidelity is less than or equal to $1$.
Here, we note the relation to the relative entropy difference
\begin{equation}
	\begin{aligned}
		MG(A:B)&=I(A:B^{*}|B)=S(\rho_{ABB^{*}}||\rho_{A}\otimes \rho_{BB^{*}})-S(\rho_{AB}||\rho_{A}\otimes \rho_{B})\\
		&=I(B:A^{*}|A)=S(\rho_{AA^{*}B}||\rho_{AA^{*}}\otimes \rho_{B})-S(\rho_{AB}||\rho_{A}\otimes \rho_{B}).
	\end{aligned}
\end{equation}

Thus, from the inequalities \eqref{eq:q=3-fidelity-bound-1} and \eqref{eq:q=3-fidelity-bound-2}, when the Markov gap vanishes $MG(A:B)=0$, the fidelities become 1. Therefore, this implies the following exact recovery of the states:
\begin{equation}
	\rho_{ABB^{*}}=\mathcal{R}^{P,t}_{B\to BB^{*}}\left[ \rho_{AB} \right]\qquad \forall t\in \mathbb{R},
\end{equation}
and
\begin{equation}
	\rho_{AA^{*}B}=\mathcal{R}_{A\to AA^{*}}^{P,t}\left[ \rho_{AB} \right] \qquad \forall t\in \mathbb{R}.
\end{equation}
We again emphasize that these recovery maps are constrained to act only on a single subsystem, either $A$ or $B$, and not on both. This restriction is a crucial and nontrivial feature of the Markov recovery problem. Without this constraint, one could trivially reconstruct the original tripartite state by considering the canonical purification of the input state $\rho_{AB}$ and tracing out the auxiliary subsystem $B^*$ (or $A^*$).  
However, under the constraint of acting on $A$ or $B$ alone, the success of the recovery map depends sensitively on the correlation structure between the two subsystems.

From a holographic perspective, this recovery process can be interpreted as follows: starting from the entanglement wedge of $AB$, one performs a suitable bulk recovery operation supported entirely within the entanglement wedge of $B$ (or $A$), thereby reconstructing the entanglement wedge corresponding to $ABB^*$ (or $AA^*B$).

\paragraph{\underline{Examples in the Qubit Case: GHZ and W States}}

$\vspace{-7mm}$\\

In the above discussion, we provided an abstract explanation of the Markov gap and its information-theoretic significance.  
Here, we complement that discussion with explicit numerical evaluations for two canonical examples in a tripartite qubit system: the GHZ state~\eqref{eq:def-tripartie-GHZ} and the W state\footnote{In tripartite qubit systems, it is well known that the GHZ and W states represent two inequivalent classes of genuine tripartite entanglement~\cite{Dur:2000zz}.}.  
The W state is defined as
\begin{equation}
	\ket{\mathrm{W}_{3}}_{ABC}=\frac{1}{\sqrt{2}} \left( \ket{001}_{ABC}+ \ket{010}_{ABC}+\ket{100}_{ABC} \right).
\end{equation}

For these states, we compute the following two quantities to examine their relationship:
\begin{itemize}
	\item The Markov gap, which is equivalent to the difference of relative entropies:
		\begin{equation}
			\begin{aligned}
				MG(A:B)&=S(\rho_{ABB^{*}}||\rho_{A}\otimes \rho_{BB^{*}})-S(\rho_{AB}||\rho_{A}\otimes \rho_{B})\\
			&=S(\rho_{AA^{*}B}||\rho_{AA^{*}}\otimes \rho_{B})-S(\rho_{AB}||\rho_{A}\otimes \rho_{B}),
			\end{aligned}
		\end{equation}
	\item The fidelities, which measure how accurately the original state can be recovered:
	\begin{equation}
		\sup_{t\in \mathbb{R}} F\left(\rho_{ABB^{*}}, \mathcal{R}_{B \rightarrow BB^{*}}^{P,t}\left(\rho_{AB}\right)\right),  \quad \sup_{t\in \mathbb{R}} F\left(\rho_{AA^{*}B}, \mathcal{R}_{A \rightarrow AA^{*}}^{P,t}\left(\rho_{AB}\right)\right).
	\end{equation}
\end{itemize}

Due to computational resource constraints, we are unable to explore all possible values of $t \in \mathbb{R}$. Therefore, in our concrete computations, we impose a cutoff on the range of $t$. As a result, we cannot determine the exact value. However, we are able to obtain reasonable numerical estimates.

We summarize their values in Table \ref{tab:Summary-q=3-result}. As one can see that the Markov gap is closely tied to the fidelity of quantum recovery.

\begin{table}[t]
\centering
\begin{tabular}{|c|c|c|}
\hline
 & $MG(A:B)$ & Fidelity \\
\hline
$\ket{\mathrm{GHZ}_{3}}$ & $0$ & $1$ \\
\hline
$\ket{\mathrm{W}_{3}}$ & $\frac{5 \log 2}{3}-\frac{\log \left(\sqrt{3}+2\right)}{\sqrt{3}}\approx 0.394899$ & $\frac{1}{108} \left(8 \sqrt{3}+2 \sqrt{116 \sqrt{3}+202}+29\right) \approx 0.768537$ \\
\hline
\end{tabular}
\caption{Values of the Markov gap and the fidelity for the GHZ state and W state in a tripartite qubit system. The fidelities $ F\left(\rho_{ABB^{*}}, \mathcal{R}_{B \rightarrow BB^{*}}^{P,t}\left(\rho_{AB}\right)\right)$, $ F\left(\rho_{AA^{*}B}, \mathcal{R}_{A \rightarrow AA^{*}}^{P,t}\left(\rho_{AB}\right)\right)$ take the same maximal value within the range $t\in [-300,300]$. Here, the value 300 is cutoff for $t$.}
\label{tab:Summary-q=3-result}
\end{table}

\subsubsection{Non-holographic recoverability through a sequential Petz bound}
\label{subsec:nonholo-seqpetz}

Let us consider a sequential (rotated) Petz procedure to reconstruct the global state from $B$ step by step. In that case, the total recoverability error is controlled by the sum of the conditional correlations revealed at each step. Consequently, we can obtain a holography-independent operational lower bound on the fidelity of recovery. In this subsection, we consider this \emph{holography-independent} recovery bound. 

Let $\rho_{A_1\cdots A_{\q} B}$ be a state and $\pi$ a permutation of $\{1,\dots,\q\}$.
For $i=1,\dots,\q$ set
\begin{align}
A_{\pi(<i)}&:=\{A_{\pi(1)},\ldots,A_{\pi(i-1)}\}\quad\text{(already-recovered past)},\\
A_{\pi(>i)}&:=\{A_{\pi(i+1)},\ldots,A_{\pi(\q)}\}\quad\text{(not-yet-recovered remainder)}.
\end{align}
We use the conditional mutual information $I(A{:}C \mid B)$ as in Eq.~\eqref{conditionalmutual}, the fidelity $F(\cdot,\cdot)$ from Eq.~\eqref{fidelitydef}, and the rotated-Petz inequality Eq.~\eqref{FidelityBound}.

In earlier sections, we use the unconditional multi-information
\begin{equation}\label{eq:I-uncond-def}
I^{(0)}(A_1{:}\cdots{:}A_\q) :=\ \sum_{i=1}^{\q} S(A_i)\ -\ S(A_1\cdots A_\q) \,, 
\end{equation}
to define the multipartite Markov gap 
\begin{align}
MG^{M(\q-1)} = S_{R\,(\q-1)} - I^{(0)}(A_1{:}\cdots{:}A_\q) \,,
\end{align}
since it gives the minimal value for the reflected multi-entropy in holography, 
see \eqref{eq:def-q=4-holographic-reflected-multi-entropy}, \eqref{5markov}, \eqref{6markov}. 
Here, because we study recovery from $B$, we use the conditional version, conditional multi-information, 
\begin{equation}\label{eq:I-cond-def}
I_{\mathrm{multi}}(A_1{:}\cdots{:}A_\q \mid B)\ :=\ \sum_{i=1}^{\q} S(A_i \mid B)\ -\ S(A_1\cdots A_\q \mid B),
\end{equation}
where the conditional entropy is defined by
\begin{equation}\label{eq:cond-entropy-def}
S(A \mid B)\ :=\ S(AB)\ -\ S(B) \,.
\end{equation}

The conditional mutual information given by \eqref{conditionalmutual} can be written in terms of the conditional entropy as 
\begin{align}
I(A{:} C \mid B ) &=\ S(AB)+S(BC)-S(B)-S(ABC) \;=\;  I(C{:} A \mid B )\nonumber \\
&= S(A\mid B) - S(A\mid BC) \,,
\label{eq:CMI-alt-def}
\end{align}
which satisfies the chain rule,
\begin{align}
I(A{:} CD \mid B ) & = S(A\mid B) - S(A\mid BC) + S(A\mid BC)  - S(A\mid BCD)  \nonumber \\ 
&= I(A{:} C \mid B ) + I(A{:} D \mid BC ) \,.
\label{eq:CMI-chain}
\end{align}

Thus, the conditional multi-information is
\begin{equation}\label{eq:I-cond-uncond-relation}
I_{\mathrm{multi}}(A_1{:}\cdots{:}A_\q \mid B)
=\Big[\sum_{i=1}^{\q} S(A_iB)\ -\ S(A_1\cdots A_\q B)\Big]\ -\ (\q-1)\,S(B),
\end{equation}
and it coincides with the unconditional multi-information when $B=\varnothing$.

The key point is that the conditional multi-information defined by \eqref{eq:I-cond-def} can be written as 
\begin{align}
I_{\mathrm{multi}}(A_1{:}\cdots{:}A_{\q} \mid B)
&= \sum_{i=1}^{\q} S(A_i \mid B) \;-\; S(A_1\cdots A_{\q} \mid B) \\
&= \sum_{i=1}^{\q} \Bigl[ S(A_{\pi(i)} \mid B) \; -  S( A_{\pi(i)} \mid B A_{\pi(<i)}) \Bigr] \\
&= \sum_{i=1}^{\q} I\!\big(A_{\pi(i)}{:}A_{\pi(<i)} \,\big|\, B \big) \,,
\label{eq:def-Imulti-342}
\end{align}
for any permutation $\pi$ (order-independence follows from the chain rule). 
The quantity $I_{\mathrm{multi}}$ measures the redundancy among the $A_i$ that remains even after conditioning on $B$.

Define
\begin{equation}\label{eq:Lambda-pi-def}
\Lambda_\pi(\rho)\ :=\ \sum_{i=1}^{\q} I\!\big(A_{\pi(i)}{:}A_{\pi(>i)} \,\big|\, B A_{\pi(<i)}\big).
\end{equation}
By the conditional mutual information chain rule \eqref{eq:CMI-chain}, for each $i$ we have
\begin{equation}\label{eq:per-i-chain}
I\!\big(A_{\pi(i)}{:}A_{\pi(<i)}A_{\pi(>i)} \,\big|\, B\big)
= I\!\big(A_{\pi(i)}{:}A_{\pi(<i)} \,\big|\, B\big)
+ I\!\big(A_{\pi(i)}{:}A_{\pi(>i)} \,\big|\, B A_{\pi(<i)}\big).
\end{equation}
Since $A_{\pi(<i)}A_{\pi(>i)}=A_{\neq \pi(i)}$, summing \eqref{eq:per-i-chain} over $i=1,\dots,\q$ yields
\begin{equation}\label{eq:Lambda-identity}
\sum_{i=1}^{\q} I\!\big(A_i{:}A_{\neq i}\mid B\big)
= \underbrace{\sum_{i=1}^{\q} I\!\big(A_{\pi(i)}{:}A_{\pi(<i)} \,\big|\, B \big)}_{=\,I_{\mathrm{multi}}(A_1{:}\cdots{:}A_\q\mid B)\ \text{by }\eqref{eq:def-Imulti-342}}
\;+\; \Lambda_\pi(\rho).
\end{equation}
Rearranging gives the order-independent identity
\begin{equation}\label{eq:Lambda-closed}
\Lambda_\pi(\rho)\ =\ \sum_{i=1}^{\q} I\!\big(A_i{:}A_{\neq i}\mid B\big)\ -\ I_{\mathrm{multi}}(A_1{:}\cdots{:}A_\q\mid B)
\;=:\; \Lambda(\rho).
\end{equation}
Since $\Lambda_\pi$ does not depend on the permutation, we simply write $\Lambda(\rho)$. Moreover, each summand in \eqref{eq:Lambda-pi-def} is a conditional mutual information and thus nonnegative, so $\Lambda(\rho)\ge 0$.

\medskip

Let $\mathcal{R}^{(\pi)}=\mathcal{R}_{\pi(\q)}\circ\cdots\circ\mathcal{R}_{\pi(1)}$ be the composition of (rotated) Petz maps where, at step $i$, $\mathcal{R}_{\pi(i)}$ attempts to recover $A_{\pi(i)}$ from $B A_{\pi(<i)}$.
In our sequential setting at step $i$, set
\begin{equation}
X:=A_{\pi(i)},\qquad
Y:= B\,A_{\pi(<i)},\qquad
Z:= A_{\pi(>i)}.
\end{equation}
Thus,
\begin{equation}
\rho_{XYZ}=\rho_{A_1\cdots A_\q B}\,,\qquad
\rho_{YZ}=\rho_{B A_{\pi(<i)} A_{\pi(>i)}}=\tr_{A_{\pi(i)}} \rho \, .
\end{equation}
Applying the one–step rotated Petz bound \eqref{FidelityBound} to the tripartition
\(\big(A_{\pi(i)}\big):\big(A_{\pi(>i)}\big)\mid\big(B A_{\pi(<i)}\big)\),
there exists a parameter \(t_i\in\mathbb{R}\) such that
\begin{align}\label{eq:FR-step-fidelity}
&F \Big(\rho_{A_1\cdots A_\q B},\;
(\mathrm{id}_{A_{\pi(>i)}} \otimes \mathcal{R}^{P,t_i}_{\,B A_{\pi(<i)}\to B A_{\pi(<i)}A_{\pi(i)}})\big[\rho_{B A_{\pi(<i)} A_{\pi(>i)}}\big]\Big)
\ \nonumber \\
&\qquad \qquad \qquad \ge\ \exp\!\Big(-\,I\!\big(A_{\pi(i)}{:}A_{\pi(>i)}\,\big|\,B A_{\pi(<i)}\big)\Big).
\end{align}
Here $F(\cdot,\cdot)$ is the fidelity defined in \eqref{fidelitydef}.
Let $\widetilde F(\sigma,\tau):=\sqrt{F(\sigma,\tau)}$ denote the root fidelity and define the Bures angle
$A(\sigma,\tau):=\arccos \widetilde F(\sigma,\tau)$.
From \eqref{eq:FR-step-fidelity} we obtain the angle form
\begin{align}\label{eq:FR-step-angle}
&A \Big(\rho_{A_1\cdots A_\q B},\;
(\mathrm{id}_{A_{\pi(>i)}}\!\otimes \mathcal{R}^{P,t_i}_{\,B A_{\pi(<i)}\to B A_{\pi(<i)}A_{\pi(i)}})\big[\rho_{B A_{\pi(<i)} A_{\pi(>i)}}\big]\Big)
\ \nonumber \\ 
& \qquad \qquad \qquad \le\ \arccos\!\Big(e^{-\tfrac12\,I(A_{\pi(i)}{:}A_{\pi(>i)}\mid B A_{\pi(<i)})}\Big).
\end{align}
Using $e^{-u}\ge 1-u$ and $\arccos(1-x)\le \sqrt{2x}$ for $x\in[0,1]$, we further get the convenient bound
\begin{equation}\label{eq:FR-step-angle-sqrt}
A \Big(\rho_{A_1\cdots A_\q B},\;
(\mathrm{id}_{A_{\pi(>i)}}\!\otimes \mathcal{R}^{P,t_i}_{\,B A_{\pi(<i)}\to B A_{\pi(<i)}A_{\pi(i)}})\big[\rho_{B A_{\pi(<i)} A_{\pi(>i)}}\big]\Big)
\ \le\ \sqrt{\,I\!\big(A_{\pi(i)}{:}A_{\pi(>i)}\mid B A_{\pi(<i)}\big)}.
\end{equation}
Summing \eqref{eq:FR-step-angle-sqrt} over $i$ and using the triangle inequality of the Bures angle, we get
\begin{equation}\label{eq:angle-sum-synced}
A \big(\rho,\ \mathcal{R}^{(\pi)}(\rho_B)\big)
\ \le\ \sum_{i=1}^{\q}\sqrt{\,I\!\big(A_{\pi(i)}{:}A_{\pi(>i)}\mid B A_{\pi(<i)}\big)}
\ \le\ \sqrt{\ \q\,\Lambda(\rho)\,},
\end{equation}
where $\mathcal{R}^{(\pi)}:=\mathcal{R}^{P,t_\q}_{\,B A_{\pi(<\q)}\to\cdots}\circ\cdots\circ
\mathcal{R}^{P,t_1}_{\,B\to B A_{\pi(1)}}$ is the composed rotated Petz map
and $\Lambda(\rho)=\sum_{i=1}^{\q} I(A_{\pi(i)}{:}A_{\pi(>i)}\mid BA_{\pi(<i)})$ is the order–independent sum from \eqref{eq:Lambda-identity}.
Since $\widetilde F(\sigma,\tau)=\cos A(\sigma,\tau)$, the inequality \eqref{eq:angle-sum-synced} is equivalent to the fidelity bound
\begin{align}\label{eq:F-bound-synced}
\sqrt{F \big(\rho,\ \mathcal{R}^{(\pi)}(\rho_B)\big)}\ &\ge\ \cos\!\Big(\sqrt{\ \q\,\Lambda(\rho)\,}\Big),
\\
-\,\log F \big(\rho,\ \mathcal{R}^{(\pi)}(\rho_B)\big)\ &\le\ -\,2\log\cos\!\Big(\sqrt{\ \q\,\Lambda(\rho)\,}\Big).
\end{align}
(For small angles, $\q\,\Lambda(\rho)\ll 1$, the right-hand side gives $-\log F \le \q\,\Lambda(\rho)+O(\Lambda(\rho)^2)$.)

Define the past-sequential reflected sum
\begin{equation}
\label{eq:SRseq-342}
S^{(\q)}_{R,\mathrm{seq}}(A_1{:}\cdots{:}A_{\q}\!\mid B)
:= \min_{\pi}\ \sum_{i=1}^{\q}
S_R \big(A_{\pi(i)}{:}A_{\pi(<i)} \,\big|\, B \big).
\end{equation}
We use the convention (``conditional reflected entropy'')
\begin{equation}\label{eq:SR-cond-conv}
S_R(X{:}Y\mid Z)\ :=\ S_R(XZ:YZ)\ -\ S(Z),
\end{equation}
{\it i.e.}, the reflected entropy across the bipartition $(XZ):(YZ)$ with the natural normalization that removes the spectator $Z$.

Since $S_R(U{:}V)\ge I(U{:}V)$ for any bipartition, applying it to $(U,V)=(XZ,YZ)$ gives
\[
S_R(X{:}Y\mid Z)=S_R(XZ:YZ)-S(Z)\ \ge\ I(XZ:YZ)-S(Z)=I(X{:}Y\mid Z),
\]
and hence, applying this term-wise to \eqref{eq:def-Imulti-342} yields the non-holographic inequality
\begin{equation}
\label{eq:right-bridge-342}
I_{\mathrm{multi}}(A_1{:}\cdots{:}A_{\q}\!\mid B)\ \le\ S^{(\q)}_{R,\mathrm{seq}}(A_1{:}\cdots{:}A_{\q}\!\mid B).
\end{equation}
Although we cannot claim any universal relation between $S^{(\q)}_{R,\mathrm{seq}}$ and the reflected multi-entropy $S_{R\,(\q-1)}$, 
the inequality \eqref{eq:right-bridge-342} is entirely non-holographic.

\subsubsection{Comments on Holographic Bounds}\label{subsec:Markov-recovery-q=4}
We now combine the inequalities~(\ref{inequality1}), (\ref{inequality12}), and~(\ref{inequality13}) derived in subsection~\ref{subsub:q=4-bound} with the quantum Markov recovery framework reviewed in the previous subsection to establish a connection between the Multipartite Markov gap and recovery via the rotated Petz map.

Before analyzing more general, non-holographic settings in the next subsection, we briefly revisit the bounds obtained under the holographic assumption.  
In this case, the Multipartite Markov gap \( MG^{M(\q-1=3)}(A: B: C) \) admits the following fidelity-based lower bound\footnote{Below, one can also consider other partitions, \textit{e.g.,} $MG^{M(\q-1=3)}(A: B: C) 
\ge MG(AC:B)$ and $MG^{M(\q-1=3)}(A: B: C) 
\ge MG(A: BC) $, and the related  fidelity-based lower bounds.}:
\begin{align}
MG^{M(\q-1=3)}(A: B: C) 
\ge MG(C: AB) 
\ge -\sup_{t\in \mathbb{R}} \log F\left(\rho_{ABCC^{*}}, \mathcal{R}_{C \rightarrow CC^{*}}^{P,t}\left(\rho_{ABC}\right)\right).
\end{align}

This bound implies that when the Multipartite Markov gap is small, the rotated Petz map \( \mathcal{R}_{C \rightarrow CC^{*}}^{P,t} \) approximately reconstructs \( \rho_{ABCC^*} \) from \( \rho_{ABC} \). Thus, for holographic states, the Multipartite Markov gap serves as a diagnostic for the feasibility of ``holographic'' Markov recovery.

\subsubsection{Non-Holographic Setting: Qubit Example 1 (GHZ State)}\label{subsub:Exeample-GHZ}

In contrast to the holographic setting discussed in subsection~\ref{subsec:Markov-recovery-q=4}, when the underlying state is non-holographic, the inequalities~(\ref{inequality1}), (\ref{inequality12}), and~(\ref{inequality13}) may no longer hold. Consequently, the Multipartite Markov gap \( MG^{M(\q-1=3)}(A: B: C) \) may fail to indicate the feasibility of Markov recovery.

In general, the Multipartite Markov gap is difficult to evaluate analytically for non-holographic states. The easiest one which one can compute is R\'enyi versions such as the \((m=2,n=2)\) Multipartite Markov gap. However, since these are distinct from the conventional Markov gap, it becomes challenging to assess whether a meaningful relation exists between them in non-holographic contexts. Moreover, it remains unclear whether any form of Markov recovery can be directly associated with the Multipartite Markov gap in such cases.

To meaningfully relate the Multipartite Markov gap to quantum recovery beyond the holographic regime, one would need a more direct connection—such as an expression involving the conditional mutual information, akin to~\eqref{eq:q=3-fidelity-bound-1} and~\eqref{eq:q=3-fidelity-bound-2}—rather than relying solely on the geometric inequalities mentioned above.

In this subsection and the next  subsection \ref{subsub:Examples-recovery-qubit}, we explore whether such a relation persists even in non-holographic settings. To do so, we examine specific qubit examples and evaluate both the Multipartite Markov gap and relevant recovery maps.

Here, we analyze the multipartite GHZ state for general \( \q \), and explicitly construct a recovery map in the \( \q = 4 \) qubit system. Aside from GHZ and simple bipartite states, evaluating the Multipartite Markov gap is computationally difficult due to challenges in computing the associated partition functions. Hence, the GHZ state remains a rare, tractable (albeit trivial) example in which the Multipartite Markov gap vanishes.

In the next subsection, we turn to more non-trivial examples and analyze them using numerical methods.

\paragraph{\underline{Multipartite Markov gap of \texorpdfstring{$\q$}{q}-partite GHZ state}}

$\vspace{-7mm}$\\

We first show the Multipartite Markov gap of $\q$-partite GHZ state. The $\q$-partite GHZ state is given by
\begin{equation}\label{eq:def-q-partie-GHZ}
	\ket{\mathrm{GHZ}_{\q}}=\frac{1}{\sqrt{2}}\big(\ket{\underbrace{00\cdots 0}_{\q}}+\ket{\underbrace{11\cdots 1}_{\q}}\big).
\end{equation}

By using the techniques reviewed in subsection \ref{subsec:reflected}, we can straightforwardly compute $(m,n)$ R\'enyi reflected multi entropy of $\q$-partite GHZ state, and the result is given by
\begin{equation}\label{eq:renyi-multi-entropy-for-reflected-qGHZ}
	S_{R\, (q-1)}^{(m,n)}\left(A_1: A_2: \cdots: A_{\q-1}\right)=S_{n\,(m)}^{(\q-1 )}\left(A_{1}A_{1}^{*}: A_{2}A_{2}^{*}: \cdots: A_{\q-1}A_{\q-1}^{*}\right)=\frac{1-n^{q-2}}{1-n}\cdot  \frac{1}{n^{\q-3}} \log2.
\end{equation}
In the limit $m,n\to 1$, the reflected multi entropy of $\q$-partite GHZ state is given by
\begin{equation}
	\begin{aligned}
		S_{R\, (q-1)}\left(A_1: A_2: \cdots: A_{\q-1}\right)&=\lim_{m\to 1}\lim_{n\to 1} S_{R\, (q-1)}^{(m,n)}\left(A_1: A_2: \cdots: A_{\q-1}\right)\\
	&=(\q-2)\log 2.
	\end{aligned}
\end{equation}
Since R\'enyi Entanglement entropy for subsystem $A_{i}\, (i=1,2,\cdots, \q)$ is simply given by 
\begin{equation}
	S_{n}(A_{i}) =\log 2 \qquad i=1,2,\cdots, A_{\q},
\end{equation}
and the $(m,n)$ R\'enyi Multipartite Markov gap of the $\q$-partite GHZ state becomes,
\begin{equation}
	\begin{aligned}
		&MG^{M(\q-1)}_{m,n}(A_{1}:A_{2}:\cdots:A_{\q-1})\\
		&= S_{R\, (\q-1)}^{(m,n)}(A_{1}:A_{2}:\cdots:A_{\q-1}) - \left[ S_{n}(A_{1}) + S_{n}(A_{2})+ \cdots + S_{n}(A_{\q-1}) - S_{n}(A_{1}A_{2}\cdots A_{\q-1}) \right]\\
		&=\left[\frac{1-n^{q-2}}{1-n}\cdot  \frac{1}{n^{\q-3}} -(\q-2) \right]\log 2,
	\end{aligned}
\end{equation}
where in the last step, we used $S_{n}(A_{1}A_{2}\cdots A_{\q-1})=S_{n}(A_{\q})$.

In the limit $m,n\to 1$, the Multipartite Markov gap of the $\q$-partite GHZ state vanishes:
\begin{equation}\label{MGGHZ}
	\begin{aligned}
		&MG^{M(\q-1)}(A_{1}:A_{2}:\cdots:A_{\q-1})=0.
	\end{aligned}
\end{equation}
 This result is reminiscent of the conventional Markov gap, which vanishes for the conventional GHZ state \cite{Hayden:2021gno}.
We note that $(m,n)$ R\'enyi Multipartite Markov gap of the $\q$-partite GHZ state does not vanish unless $n=1$, unlike the $\q=3$ case where it vanishes for any $n$.

As we mentioned, apart from this GHZ state and bipartite states, it is difficult to compute the Multipartite Markov gap; thus, the $\q$-partite GHZ state is currently the only known computable (almost trivial) example that exhibits a vanishing Multipartite Markov gap.

\paragraph{\underline{Recovery map in \texorpdfstring{$\q=4$}{q=4}-partite GHZ state}}

$\vspace{-7mm}$\\

As we have seen, the Multipartite Markov gap for the $\q$-partite GHZ state vanishes.\footnote{One may also consider the more general state
\begin{equation}
	\ket{\mathrm{GHZ}_{\q=4}}_{ABCD} = \cos\theta \ket{0000}_{ABCD} + e^{i\phi} \sin\theta \ket{1111}_{ABCD}, \qquad \theta \in [0, 2\pi), \ \phi \in [0, \pi)
\end{equation}
as a deformation of the original GHZ state.
}
In this case, we expect recovery operations to succeed, just as in the conventional case $\q=3$ and, indeed, they do.

Let us now consider explicit recovery operations for the $\q$-partite GHZ state. Although this state is structurally simple and the recovery process is essentially trivial, it nonetheless provides valuable intuition for understanding the structure of quantum recovery maps.

For concreteness, we focus on the $\q=4$ case. In this setting, both the Multipartite and conventional Markov gaps vanish:
\begin{equation}
	\begin{aligned}
		0&=MG^{M(\q-1=3)}(A: B: C)= MG(A: BC)=MG(B: AC)= MG(C: AB).
	\end{aligned}
\end{equation}
In this case, the inequalities \eqref{inequality1}, \eqref{inequality12} and  \eqref{inequality13} are trivially satisfied. Thus, from the discussion of subsection \ref{subsub:review-Markov-recovery}, we can consider recovery maps:\\
 for $\forall t\in \mathbb{R},$
\begin{equation}
	 MG(A: BC)=0 \quad \rightarrow \quad \rho_{AA^{*}BC}=\mathcal{R}^{P,t}_{A\to AA^{*}} \left[ \rho_{ABC} \right], \quad \rho_{ABB^{*}CC^{*}}=\mathcal{R}^{P,t}_{BC\to BB^{*}CC^{*}} \left[ \rho_{ABC} \right],
\end{equation}
\begin{equation}
	 MG(B: AC)=0 \quad \rightarrow \quad \rho_{ABB^{*}C}=\mathcal{R}^{P,t}_{B\to BB^{*}} \left[ \rho_{ABC} \right], \quad \rho_{AA^{*}BCC^{*}}=\mathcal{R}^{P,t}_{AC\to AA^{*}CC^{*}} \left[ \rho_{ABC} \right],
\end{equation}
\begin{equation}
	 MG(C: AB)=0 \quad \rightarrow \quad \rho_{ABCC^{*}}=\mathcal{R}^{P,t}_{C\to CC^{*}} \left[ \rho_{ABC} \right], \quad \rho_{AA^{*}BB^{*}C}=\mathcal{R}^{P,t}_{AB\to AA^{*}BB^{*}} \left[ \rho_{ABC} \right].
\end{equation}

Thus, following subsection \ref{subsub:review-Markov-recovery}, we can explicitly construct the rotated Petz map. For example, for the case $\rho_{AA^{*}BC}$, the recovery map \eqref{eq:recovery-map}, acting only on $A$, is given by
\begin{equation}\label{eq:Markov-recovery-map-q=4-AA*}
	\begin{aligned}
		\mathcal{R}^{P,t}_{A\to AA^{*}} [\mathcal{O}_{ABC}] &= (\rho_{AA^{*}})^{1/2+it/2}  (\rho_{A})^{-1/2-it/2}(\mathcal{O}_{ABC})(\rho_{A})^{-1/2+it/2} (\rho_{AA^{*}})^{1/2-it/2}\\
		&= \left[ \ket{00}_{AA^{*}} \bra{00}+ \ket{11}_{AA^{*}} \bra{11}  \right] (\mathcal{O}_{ABC}) \left[ \ket{00}_{AA^{*}} \bra{00}+ \ket{11}_{AA^{*}} \bra{11}  \right]\\
		&=  \ket{0}_{A} \bra{0}  (\mathcal{O}_{ABC})  \ket{0}_{A} \bra{0}   \otimes \ket{0}_{A^{*}}\bra{0} + \ket{1}_{A} \bra{1}  (\mathcal{O}_{ABC})  \ket{1}_{A} \bra{1}   \otimes \ket{1}_{A^{*}}\bra{1},
	\end{aligned}
\end{equation} 
and this recovery map indeed recovers the original state $\rho_{AA^{*}BC}$ from $\rho_{ABC}$:
\begin{equation}
	\begin{aligned}
		\mathcal{R}^{P,t}_{A\to AA^{*}} [\rho_{ABC}]
		&=  \ket{0}_{A} \bra{0}  (\rho_{ABC})  \ket{0}_{A} \bra{0}   \otimes \ket{0}_{A^{*}}\bra{0} + \ket{1}_{A} \bra{1}  (\rho_{ABC})  \ket{1}_{A} \bra{1}   \otimes \ket{1}_{A^{*}}\bra{1}\\
		&=\frac{1}{2} \left( \ket{0000}_{ABCA^{*}} \bra{0000}+ \ket{1111}_{ABCA^{*}} \bra{1111} \right)\\
		&=\rho_{AA^{*}BC}.
	\end{aligned}
\end{equation} 
For the remaining states $\rho_{ABB^{*}C}$ and $\rho_{ABCC^{*}}$, we can obtain the corresponding recovery maps by replacing the projection operators $\ket{i}_{A} \bra{i}$ and $\ket{i}_{A^{*}} \bra{i}$ with those acting on the appropriate subsystems $B,B^{*}$ and $C,C^{*}$, respectively.
Similarly, for $\rho_{ABB^{*}CC^{*}}$, $\rho_{AA^{*}BCC^{*}}$, and $\rho_{AA^{*}BB^{*}C}$, we can obtain the recovery maps, which exactly recover the states.
Therefore, in the case of $\q$-partite GHZ states, the fidelity between the original state and the state recovered by the rotated Petz map is identically 1. 

This outcome is hardly surprising, given the inherently simple and symmetric structure of the $\q$-partite GHZ states. The (almost trivial) information is uniformly encoded into the subsystems, and the recovery maps (operations) simply extend it to additional subsystems without disturbing the uniformity of the encoding. After all, the GHZ state merely repeats the classical bit values $0$ and $1$ across all subsystems, making its structure exceptionally straightforward.

\subsubsection{Non-Holographic Setting: Qubit Example 2 (Other examples)}
\label{subsub:Examples-recovery-qubit}

 As noted at the beginning of subsection~\ref{subsub:Exeample-GHZ}, it is difficult to analytically evaluate the Multipartite Markov gap. However, once the replica indices \( m, n \) are fixed, we can compute the \((m=2,n=2)\) R\'enyi Multipartite Markov gap, which partially, though not fully, captures the behavior of the true gap.

To estimate the Multipartite Markov gap and assess its consistency with the Markov recovery problem, we proceed as follows. First, we compute the \((m=2,n=2)\) R\'enyi reflected multi-entropy and the corresponding R\'enyi Multipartite Markov gap for a set of four-qubit states. Next, we evaluate the fidelity between original and recovered states to quantify the effectiveness of recovery under erasure errors in the current setting. Finally, we discuss how the R\'enyi Multipartite Markov gap correlates with recovery fidelity.

We focus on various four-qubit states on system $ABCD$ investigated in \cite{enriquez_2024_yr5wp-mxc59} as in \cite{Iizuka:2025caq}. The states are given by
\begin{align}
\ket{\text{GHZ}_4} &= \frac{1}{\sqrt{2}} (\ket{0000} + \ket{1111}), \label{eq:state-GHZ} \\
\ket{\mathrm{W}_4} &= \frac{1}{2} (\ket{0001} + \ket{0010} + \ket{0100} + \ket{1000}), \label{eq:state-W} \\
\ket{D[4, (2,2)]} &= \frac{1}{\sqrt{6}} (\ket{0011} + \ket{1100} + \ket{0101} + \ket{1010} + \ket{0110} + \ket{1001}), \label{eq:state-D} \\
\ket{\text{HS}} &= \frac{1}{\sqrt{6}} (\ket{0011} + \ket{1100} + \omega(\ket{0101} + \ket{1010}) + \omega^2(\ket{0110} + \ket{1001})),  \\
& \qquad \qquad   \omega = \exp\left(\frac{2\pi i}{3}\right),  \notag \\
\ket{C_1} &= \frac{1}{2} (\ket{0000} + \ket{0011} + \ket{1100} - \ket{1111}), \label{eq:state-C1} \\
\ket{\Psi_1^{\text{max}}} &= 0.630\ket{0000} + 0.281\ket{1100} + 0.202\ket{1010} + 0.24\ket{0110} + 0.232e^{0.494\pi i}\ket{1110} \notag \\
&\quad + 0.059\ket{1001} + 0.282\ket{0101} + 0.346e^{-0.362\pi i}\ket{1101} + 0.218e^{0.626\pi i}\ket{1011} \notag \\
&\quad + 0.304\ket{0011} + 0.054e^{-0.725\pi i}\ket{0111} + 0.164e^{0.372\pi i}\ket{1111}, \label{eq:state-Psi-max} \\
\ket{BSSB_4} &= \frac{1}{2\sqrt{2}} (\ket{0110} + \ket{1011} + i(\ket{0010} + \ket{1111}) + (1+i)(\ket{0101} + \ket{1000})), \label{eq:state-BSSB} \\
\ket{HD} &= \frac{1}{\sqrt{6}} (\ket{1000} + \ket{0100} + \ket{0010} + \ket{0001} + \sqrt{2}\ket{1111}), \label{eq:state-HD} \\
\ket{YC} &= \frac{1}{2\sqrt{2}} (\ket{0000} - \ket{0011} - \ket{0101} + \ket{0110} + \ket{1001} + \ket{1010} + \ket{1100} + \ket{1111}), \label{eq:state-YC} \\
\ket{L} &= \frac{1}{2\sqrt{3}} \left((1+\omega)(\ket{0000} + \ket{1111}) + (1-\omega)(\ket{0011} + \ket{1100}) \right. \notag \\
&\quad \left. + \omega^2(\ket{0101} + \ket{1010} + \ket{1001} + \ket{0110})\right). \label{eq:state-L}
\end{align}

These four-qubit states are identified as maximally entangled with respect certain entanglement measures \cite{enriquez_2024_yr5wp-mxc59}.

For these states, we numerically evaluate $(m=2,n=2)$ R\'enyi reflected multi-entropy and  $(m=2,n=2)$ R\'enyi Multipartite Markov gap.
Table \ref{tab:Summary-qubits-Rearrange} summarizes the numerical results. Since we consider the $(m=2,n=2)$ R\'enyi Multipartite Markov gap, it takes the non-zero value even for the GHZ state, and in particular it becomes negative only for the GHZ state. 

\begin{table}[t]
\centering
\begin{tabular}{|c|c|c|}
\hline \textbf{States} & $S_{R\, (\q-1=3)}^{(m=2,n=2)}(A:B:C)$ & $MG^{M(\q-1=3)}_{m=2,n=2}(A:B:C)$ \\
\hline $\left|\mathrm{GHZ}_4\right\rangle$ & $\frac{3 \log 2}{2} \approx 1.03972$ & $-\frac{\log 2}{2} \approx -0.346574$ \\
\hline $\left|\mathrm{C}_1\right\rangle$ & \multirow{2}{*}{$\frac{5 \log 2}{2} \approx 1.73287$} & \multirow{2}{*}{$\frac{\log 2}{2} \approx 0.346574$} \\
$|Y C\rangle$ & & \\
\hline $|D[4,(2,2)]\rangle$ & $\frac{3 \log 2}{2} + \frac{5 \log 3}{2} - \frac{\log 5}{2} - \frac{\log 11}{2} \approx 1.78258$ & $\frac{5 \log 3}{2} - \frac{\log 2}{2} - \frac{\log 5}{2} - \frac{\log 11}{2} \approx 0.396291$ \\
\hline $\left|\Psi_1^{\max }\right\rangle$ & $1.79672$ & $0.483767$ \\
\hline $\left|BSSB_4\right\rangle$ & $\frac{1}{2}(9 \log 2 - 2 \log 3) \approx 2.02055$ & $\frac{1}{2}(5 \log 2 - 2 \log 3) \approx 0.634256$ \\
\hline $|\mathrm{HS}\rangle$ & \multirow{3}{*}{$2 \log 3 \approx 2.19722$} & \multirow{3}{*}{$2 \log 3 - 2 \log 2 \approx 0.81093$} \\
$|H D\rangle$ & & \\
$|L\rangle$ & & \\
\hline $\left|\mathrm{W}_4\right\rangle$ & $2 \log 5 - \frac{\log 17}{2} \approx 1.80227$ & $-6 \log 2 + 4 \log 5 - \frac{\log 17}{2} \approx 0.862262$ \\
\hline
\end{tabular}
\caption{Values of the $(m=2,n=2)$ R\'enyi reflected multi-entropy $S_{R\, (\q-1=3)}^{(m=2,n=2)}(A:B:C)$ and the $(m=2,n=2)$ R\'enyi Multipartite Markov gap $MG^{M(\q-1=3)}_{m=2,n=2}(A:B:C)$ of various four-qubit states. To obtain values of the $(m=2,n=2)$ R\'enyi Multipartite Markov gap, we need to compute the $(m=2,n=2)$ R\'enyi reflected multi-entropy. However, the values of the $(m=2,n=2)$ R\'enyi reflected multi-entropy themselves are not essential to the following discussions on the Markov recovery.}
\label{tab:Summary-qubits-Rearrange}
\end{table}

\paragraph{\underline{Relation to Approximate Markov Recovery Problem}}

$\vspace{-7mm}$\\

Next, we investigate the Multipartite Markov gap in the $(m=2,n=2)$ R\'enyi case and its relation to the approximate Markov recovery problem. To carry out this analysis, we consider the rotated Petz maps as {\it trial} recovery maps in the current context of the Markov recovery problem. While one could explore more optimal recovery maps that may provide a tighter relation to the $(m=2,n=2)$ R\'enyi Multipartite Markov gap, we restrict our attention to the rotated Petz maps for simplicity.

We numerically compute the following quantities, which characterize the Markov recovery problem in our setting:
\begin{itemize}
	\item Markov gaps, which are equal to relative entropy differences,
		\begin{equation}\label{eq:MG-RelativeEd-1}
			\begin{aligned}
				MG(A:BC)&=S(\rho_{ABCB^{*}C^{*}}||\rho_{A}\otimes \rho_{BCB^{*}C^{*}})-S(\rho_{ABC}||\rho_{A}\otimes \rho_{BC})\\
			&=S(\rho_{ABCA^{*}}||\rho_{AA^{*}}\otimes \rho_{BC})-S(\rho_{ABC}||\rho_{A}\otimes \rho_{BC}),
			\end{aligned}
		\end{equation}
		\begin{equation}\label{eq:MG-RelativeEd-2}
			\begin{aligned}
				MG(AC:B)&=S(\rho_{ABCB^{*}}||\rho_{AC}\otimes \rho_{BB^{*}})-S(\rho_{ABC}||\rho_{AC}\otimes \rho_{B})\\
			&=S(\rho_{ABCA^{*}C^{*}}||\rho_{ACA^{*}C^{*}}\otimes \rho_{B})-S(\rho_{ABC}||\rho_{AC}\otimes \rho_{B}),
			\end{aligned}
		\end{equation}
		\begin{equation}\label{eq:MG-RelativeEd-3}
			\begin{aligned}
				MG(AB:C)&=S(\rho_{ABCC^{*}}||\rho_{AB}\otimes \rho_{CC^{*}})-S(\rho_{ABC}||\rho_{AB}\otimes \rho_{C})\\
			&=S(\rho_{ABCA^{*}B^{*}}||\rho_{ABA^{*}B^{*}}\otimes \rho_{C})-S(\rho_{ABC}||\rho_{AB}\otimes \rho_{C}),
			\end{aligned}
		\end{equation}
	\item the fidelities 
	\begin{equation}\label{eq:fidelity-of-recovery-1}
		\sup_{t\in \mathbb{R}} F\left(\rho_{ABCA^{*}}, \mathcal{R}_{A \rightarrow AA^{*}}^{P,t}\left(\rho_{ABC}\right)\right),  \quad \sup_{t\in \mathbb{R}} F\left(\rho_{ABCB^{*}}, \mathcal{R}_{BC \rightarrow BCB^{*}C^{*}}^{P,t}\left(\rho_{ABC}\right)\right),
	\end{equation}
	\begin{equation}\label{eq:fidelity-of-recovery-2}
		\sup_{t\in \mathbb{R}} F\left(\rho_{ABCA^{*}}, \mathcal{R}_{B \rightarrow BB^{*}}^{P,t}\left(\rho_{ABC}\right)\right),  \quad \sup_{t\in \mathbb{R}} F\left(\rho_{ABCA^{*}C^{*}}, \mathcal{R}_{AC \rightarrow ACA^{*}C^{*}}^{P,t}\left(\rho_{ABC}\right)\right),
	\end{equation}
	\begin{equation}\label{eq:fidelity-of-recovery-3}
		\sup_{t\in \mathbb{R}} F\left(\rho_{ABCC^{*}}, \mathcal{R}_{C \rightarrow CC^{*}}^{P,t}\left(\rho_{ABC}\right)\right),  \quad \sup_{t\in \mathbb{R}} F\left(\rho_{ABCA^{*}B^{*}}, \mathcal{R}_{AB \rightarrow ABA^{*}B^{*}}^{P,t}\left(\rho_{ABC}\right)\right).
	\end{equation}
\end{itemize}

\begin{table}[t]
\centering
\begin{tabular}{|c|c|c|c|c|c|c|c|}
\hline
\textbf{States} & $MG^{M(\q-1=3)}_{m=2,n=2}$ & \multicolumn{6}{c|}{Fidelity for local recovery $\alpha \to \alpha \alpha^{*}$ ($\alpha=A,BC,B,AC,C,AB$)} \\
\hline
 & & $A$ & $BC$ & $B$ & $AB$ & $C$ & $AB$ \\
\hline
$\left|\mathrm{GHZ}_4\right\rangle$ & \multirow{1}{*}{$-0.346574$} & $1$ & $1$ & $1$ & $1$ & $1$ & $1$ \\
\hline
$\left|\mathrm{C}_1\right\rangle$ & \multirow{2}{*}{$0.346574$} & $1$ & $1$ & $1$ & $1$ & $1$ & $1$ \\
\cline{1-1} \cline{3-8}
$|Y C\rangle$ & & $1$ & $1$ & $1$ & $1$ & $1$ & $1$ \\
\hline
$|D[4,(2,2)]\rangle$ & \multirow{1}{*}{$0.396291$} & $0.806584$ & $0.814815$ & $0.806584$ & $0.814815$ & $0.806584$ & $0.814815$ \\
\hline
$\left|\Psi_1^{\max }\right\rangle$ & \multirow{1}{*}{$0.483767$} & $0.423589$ & $0.245369$ & $0.392004$ & $0.228274$ & $0.405128$ & $0.314966$ \\
\hline
$\left|BSSB_4\right\rangle$ & \multirow{1}{*}{$0.634256$} & $0.25$ & $0.0625$ & $0.213386$ & $0.159349$ & $0.213387$ & $0.159349$ \\
\hline
$|\mathrm{HS}\rangle$ & \multirow{3}{*}{$0.81093$} & $0.342069$ & 0.318998 & $0.342069$ & 0.318998 & $0.342069$ & 0.318998 \\
\cline{1-1} \cline{3-8}
$|H D\rangle$ & & $0.770295$ & $0.770295$ & $0.770295$ & $0.770295$ & $0.770295$ & $0.770295$ \\
\cline{1-1} \cline{3-8}
$|L\rangle$ & & $0.366628$ & $0.366628$ & $0.366628$ & $0.366628$ & $0.366628$ & $0.366628$ \\
\hline
$\left|\mathrm{W}_4\right\rangle$ & \multirow{1}{*}{$0.862262$} & $0.791543$ & $0.798224$ & $0.791543$ & $0.798224$ & $0.791543$ & $0.798224$ \\
\hline
\end{tabular}
\caption{Values of
 the $(m=2,n=2)$ R\'enyi Multipartite Markov gap $MG^{M(\q-1=3)}_{m=2,n=2}(A:B:C)$ and
 the fidelities between the original states and the recovered states by the rotated Petz map,  (\ref{eq:fidelity-of-recovery-1}), (\ref{eq:fidelity-of-recovery-2}) and (\ref{eq:fidelity-of-recovery-3}), for the various four-qubit states. We estimate the maximal value within the range $t\in [-300,300]$.}
\label{tab:Summary-fidelities}
\end{table}

\begin{table}[h]
\centering
\begin{tabular}{|c|c|c|c|c|c|}
\hline
\textbf{States} & $MG^{M(\q-1=3)}_{m=2,n=2}$ & \raisebox{-1ex}{\shortstack{Exact\\recovery}} & $MG(A:BC)$ & $MG(AC:B)$ & $MG(AB:C)$ \\
\hline
$\left|\mathrm{GHZ}_4\right\rangle$ & \multirow{1}{*}{$-0.346574$} & $\bigcirc$ & $0$ & $0$ & $0$ \\
\hline
$\left|\mathrm{C}_1\right\rangle$ & \multirow{2}{*}{$0.346574$} & $\bigcirc$ & $ 0$ & $ 0$ & $ 0$ \\
\cline{1-1} \cline{3-6}
$|Y C\rangle$ & & $\bigcirc$ & $ 0$ & $ 0$ & $ 0$ \\
\hline
$|D[4,(2,2)]\rangle$ & \multirow{1}{*}{$0.396291$} & $\times$ & $0.308065$ & $0.308065$ & $0.308065$ \\
\hline
$\left|\Psi_1^{\max }\right\rangle$ & \multirow{1}{*}{$0.483767$} & $\times$ & $0.168936$ & $0.240868$ & $0.182715$ \\
\hline
$\left|BSSB_4\right\rangle$ & \multirow{1}{*}{$0.634256$} & $\times$ & $0$ & $0.14584$ & $0.14584$ \\
\hline
$|\mathrm{HS}\rangle$ & \multirow{3}{*}{$0.81093$} & $\times$ & $0.126469$ & $0.126469$ & $0.126469$ \\
\cline{1-1} \cline{3-6}
$|H D\rangle$ & & $\times$ & $0.27031$ & $0.27031$ & $0.27031$ \\
\cline{1-1} \cline{3-6}
$|L\rangle$ & & $\times$ & $0.27031$ & $0.27031$ & $0.27031$ \\
\hline
$\left|\mathrm{W}_4\right\rangle$ & \multirow{1}{*}{$0.862262$} & $\times$ & $0.338266$ & $0.338266$ & $0.338266$ \\
\hline
\end{tabular}
\caption{Values of the $(m=2,n=2)$ R\'enyi Multipartite Markov gap $MG^{M(\q-1=3)}_{m=2,n=2}(A:B:C)$ and the conventional Markov gap $MG$ of various four-qubit states for three possible divisions. These Markov gaps are related to the relative entropy differences (\ref{eq:MG-RelativeEd-1}), (\ref{eq:MG-RelativeEd-2}) and (\ref{eq:MG-RelativeEd-3}).}
\label{tab:Summary-three-MG}
\end{table}

In Tables \ref{tab:Summary-fidelities} and \ref{tab:Summary-three-MG}, we summarize the numerical results.

From Table~\ref{tab:Summary-three-MG}, we observe that for the states $\left|\mathrm{GHZ}_4\right\rangle$ \eqref{eq:state-GHZ}, $\left|\mathrm{C}_1\right\rangle$ \eqref{eq:state-C1}, and $|YC\rangle$ \eqref{eq:state-YC}, which exhibit negative or relatively small values of the $(m=2,n=2)$ R\'enyi Multipartite Markov gap $MG^{M(\q-1=3)}_{m=2,n=2}(A:B:C)$, the recovery fidelities reach 1, indicating perfect recovery. As shown in Table~\ref{tab:Summary-fidelities}, the conventional Markov gaps also vanish for these states, reinforcing the correlation between small $(m=2,n=2)$ holographic gaps and successful recovery. These states likewise exhibit small Multipartite Markov gaps in Table~\ref{tab:Summary-qubits-Rearrange}.

On the other hand, for other states having relatively large values of the $(m=2,n=2)$ R\'enyi Multipartite Markov gap in the tables, the fidelities deviate from $1$ and the conventional Markov gaps basically take non-zero values except in the specific case of $\left|BSSB_4\right\rangle$. This means that, for these states having relatively larger values of the $(m=2,n=2)$ R\'enyi Multipartite Markov gap than those of $\left|\mathrm{GHZ}_4\right\rangle$, $\left|\mathrm{C}_1\right\rangle$ and $|YC\rangle$, the recovery maps approximately recover the input state with non-unity fidelities as shown in Table \ref{tab:Summary-fidelities}.

A more detailed analysis reveals several intriguing patterns in the data. We observe a clear threshold behavior around $MG^{M(\q-1=3)}_{m=2,n=2} \approx 0.346574$, which appears to set the boundary between exact and approximate recovery regimes. States with gap values at or below this threshold achieve perfect recovery, while those exceeding it exhibit degraded performance.  This value may represent a fundamental limit for exact recovery from the perspective of the $(m=2,n=2)$ R\'enyi Multipartite Markov gap.

Moreover, the relationship between the magnitude of the $(m=2,n=2)$ R\'enyi Multipartite Markov gap and recovery fidelity is not strictly monotonic. While there is a general trend of decreasing fidelity with increasing $(m=2,n=2)$ R\'enyi Multipartite Markov gap size, notable exceptions exist. For example, $\left|\mathrm{W}_4\right\rangle$ has the largest gap value ($0.862262$) yet retains relatively high fidelity ($\approx 0.79$), whereas $\left|BSSB_4\right\rangle$—with a smaller gap ($0.634256$)—exhibits significantly lower fidelities, as low as $0.0625$ for a certain operation, namely $\alpha = BC$ case. This suggests that the Multipartite Markov gap captures important structural information about quantum states, but additional factors related to the specific entanglement structure may also affect recovery effectiveness.

We also observe asymmetries in recovery performance across different subsystem partitions. For symmetric states like $|HD\rangle$ and $|L\rangle$, all recovery operations yield identical fidelities, reflecting the underlying symmetry. By contrast, asymmetric states such as $\left|\Psi_1^{\max}\right\rangle$ and $\left|BSSB_4\right\rangle$ show marked variation in recovery fidelity depending on which subsystem or combination is being recovered.

The case of $\left|BSSB_4\right\rangle$ warrants particular attention due to its anomalous behavior. Despite having a moderate $(m=2,n=2)$ R\'enyi Multipartite Markov gap ($0.634256$), it exhibits a vanishing conventional Markov gap $MG(A:BC) = 0$ while showing poor recovery fidelities. This suggests that the multipartite and conventional Markov gaps capture complementary aspects of multipartite correlations, and that neither alone fully characterizes the feasibility of state recovery. In particular, the $(m=2,n=2)$ R\'enyi Multipartite Markov gap appears to be more sensitive to certain forms of multipartite entanglement that remain invisible to conventional tripartite measures.

Therefore, from these observations, one can see that for states having smaller $(m=2,n=2)$ R\'enyi Multipartite Markov gaps below around $0.346574$, we can exactly recover the input states by using the rotated Petz map in the Markov recovery problem. More broadly, our analysis here demonstrates that the $(m=2,n=2)$ R\'enyi Multipartite Markov gap serves as a meaningful quantitative indicator for assessing the quality of the recovery operations in non-holographic multipartite systems.

In summary, we find that states with $(m=2,n=2)$ R\'enyi Multipartite Markov gaps below approximately $0.346574$ can be exactly recovered via the rotated Petz map in the Markov recovery framework. More generally, our analysis demonstrates that this gap serves as a useful quantitative indicator of recovery quality in non-holographic multipartite systems. While the $(m=2,n=2)$ R\'enyi Multipartite Markov gap may not capture all relevant aspects of quantum correlations, it offers valuable insight into the boundary between exact and approximate recoverability and how the quality of recovery degrades beyond this regime.

\subsection{Summary and comparison between the conventional Markov Gap and the Multipartite Markov Gap}

\paragraph{\underline{Summary}}

As we have seen, the Multipartite Markov gap has different properties from the conventional tripartite Markov gap. Nevertheless, they also share notable similarities between them. In what follows, we summarize the differences and similarities between our Multipartite Markov gap and the conventional Markov gap below.
\begin{itemize}
    	\setlength{\itemsep}{0em}
    \item Building blocks
    	\begin{itemize}
    	\setlength{\itemsep}{0.5em}
    	\item{Conventional Markov gap: $MG(A:B)$    ({\it i.e.,} $\q-1=2$ case)}: \\
    	Reflected entropy and Entanglement entropy (or mutual information)    	\item{Multipartite Markov gap: $MG^{M(\q-1)}(A_{1}:A_{2}:\cdots:A_{\q-1})$}:\\
    	Reflected multi-entropy and Entanglement entropy (or mutual information)
    	\end{itemize}

    \item Non-negativity

    	\begin{itemize}
    	\setlength{\itemsep}{0.5em}
    	\item Conventional Markov gap: $MG(A:B) \geq 0.$   R\'enyi Markov gap $MG_{m,n}(A:B)$ can be negative.

    	\item Multipartite Markov gap: 
    	$MG^{M(\q-1)}(A_{1}:A_{2}:\cdots:A_{\q-1})\geq 0$ for holographic states, but unknown for non-holographic states

    	\end{itemize}

    \item Value for GHZ state

    	\begin{itemize}
    	\setlength{\itemsep}{0.5em}
    	\item Conventional Markov gap: $MG(A:B)=MG_{m,n}(A:B)=0 $
    	
    	   	\item Multipartite Markov gap: $\begin{dcases}
	MG^{M(\q-1)}(A_{1}:A_{2}:\cdots:A_{\q-1})=0,\\
	MG^{M(\q-1)}_{m,n}(A_{1}:A_{2}:\cdots:A_{\q-1})=\left[\frac{1 - n^{\q-2}}{1 - n} \cdot \frac{1}{n^{\q - 3}} - (\q - 2)\right] \log 2,
\end{dcases}
$    	\\
$MG^{M(\q-1=3)}_{m,n}(A:B:C)=\left(\frac{1}{n} - 1\right)\log 2$

    	\end{itemize}

    \item Relation to the Markov recovery problem

    	\begin{itemize}
    	\setlength{\itemsep}{0.5em}
    	\item{Conventional Markov gap: $MG(A:B)$    ($\q-1=2$ case)} Directly related
    	 \item Multipartite Markov gap: {$MG^{M(\q-1)}(A_{1}:A_{2}:\cdots:A_{\q-1})$} Related in the sense of characterizing exactly recoverable cases   	
    	\end{itemize}

       \item \textbf{Genuineness}
    \begin{itemize}
        \setlength{\itemsep}{0.3em}
        \item Conventional Markov gap:  Yes. For bipartite or product states, $MG(A:B) = MG_{m,n}(A:B) = 0$, hence it detects genuine tripartite entanglement only
        \item Multipartite Markov gap:  No. For generic $(\q-1)$-partite states, $MG^{M(\q-1)}(A_{1}:A_{2}:\cdots:A_{\q-1}) > 0$
    \end{itemize}
    	
\end{itemize}

\paragraph{\underline{Comparison}}
One notable property of the conventional Markov gap $MG(A:B)$ is that it vanishes for a mixed state $\rho_{AB} = \mathrm{Tr}_C[\ket{\psi}_{ABC} \bra{\psi}]$ when the underlying pure state is partially separable, i.e., $\ket{\psi}_{ABC} = \ket{\psi_1}_A \otimes \ket{\psi_2}_{BC}$.

However, as illustrated in Figure~\ref{fig:Holographic-proof-positivity-for-biseparable-case-q=4}, the Multipartite Markov gap $MG^{M(\q-1=3)}(A:B:C)$, introduced in this work, can remain nonzero even for a mixed state $\rho_{ABC} = \mathrm{Tr}_D[\ket{\psi}_{ABCD} \bra{\psi}]$ 
when the underlying pure state $\ket{\psi}_{ABCD}$ is partially 
separable pure state $\ket{\psi}_{ABCD} = \ket{\psi_1}_A \otimes \ket{\psi_3}_{BCD}$. In such cases, we find
\begin{align}
    MG^{M(3)}(A:B:C) = MG(B:C),
\end{align}
where $MG(B:C)$ denotes the conventional Markov gap for the reduced state $\rho_{BC} = \mathrm{Tr}_A \rho_{ABC}$.

This indicates that the multipartite Markov gap lacks the feature of detecting genuine multipartite entanglement, as it can be nonzero even for $\tilde{\q} < \q$ partite entangled states. 

Owing to this limitation, one may seek an alternative generalization of the Markov gap that effectively isolates the genuine multipartite component. In the next section, we propose such a generalization, constructed from the reflected multi-entropy, which vanishes for partially separable pure states.

\section{Genuine Reflected Multi-Entropy: Capturing genuine multipartite entanglement}
\label{genuineMG}
In this section, we propose an alternative generalization of the Markov gap for $\q$-partite pure states, such that the resulting quantity vanishes for any $\q'$-partite entangled pure states with $\q' < \q$.
This construction is motivated by the notion of genuine multi-entropy~\cite{Iizuka:2025ioc, Iizuka:2025caq}, and we refer to the resulting quantity as the \textit{genuine reflected multi-entropy}.

Note that, throughout this section, we do not assume holography; rather, our construction is solely guided by the requirement that the quantity captures genuine multipartite features.

\subsection{Review of the genuine multi-entropy for $\q=3,4$}
Before proceeding, we briefly review the genuine multi-entropy, as introduced in \cite{Iizuka:2025ioc,Iizuka:2025caq}. 
The genuine $\q$-partite multi-entropy is defined as a specific linear combination of $\tilde{\q}$-partite multi-entropies for $\tilde{\q} \leq \q$, such that it vanishes for all $\q'$-partite entangled states with $\q' < \q$. Although we primarily focus on the cases $\q=3$ and $\q=4$ in this paper for simplicity, the definition and results can be naturally generalized to arbitrary $\q$. For a detailed construction in the general $\q$ case, we refer the reader to our earlier works \cite{Iizuka:2025ioc,Iizuka:2025caq}.

For the $\q=3$ case, the R\'enyi genuine multi-entropy $\text{GM}_n^{(3)}(A: B: C)$ is given by
\begin{equation}\label{eq:genuine-multi-entropy-q=3}
\text{GM}_n^{(3)}(A: B: C)\coloneqq S_n^{(3)}(A: B: C)-\frac{1}{2}\left(S_n^{(2)}(A: B C)+S_n^{(2)}(B: C A)+S_n^{(2)}(C: A B)\right).
\end{equation}
This combination was originally pointed out in \cite{Penington:2022dhr, Harper:2024ker} 
and analyzed by \cite{Harper:2024ker, Liu:2024ulq, Iizuka:2025caq, Iizuka:2025ioc}.

For the $\q=4$ case, the R\'enyi genuine multi-entropy $\text{GM}_n^{(4)}(A: B: C: D)$ is given by \cite{Iizuka:2025caq, Iizuka:2025ioc}
\begin{equation}\label{eq:genuine-multi-entropy-q=4}
\begin{aligned}
&\text{GM}_n^{(4)}(A: B: C: D)\\
&\coloneqq S_n^{(4)}(A: B: C: D)-\frac{1}{3}\left(S_n^{(3)}(A B: C: D)+S_n^{(3)}(A C: B: D)\right. \\
&\left.\quad \, +S_n^{(3)}(A D: B: C)+S_n^{(3)}(B C: A: D)+S_n^{(3)}(B D: A: C)+S_n^{(3)}(C D: A: B)\right) \\
&\quad +a\left(S_n^{(2)}(A B: C D)+S_n^{(2)}(A C: B D)+S_n^{(2)}(A D: B C)\right) \\
&\quad +\left( \frac{1}{3}-a \right)\left(S_n^{(2)}(A B C: D)+S_n^{(2)}(A B D: C)+S_n^{(2)}(A C D: B)+S_n^{(2)}(B C D: A)\right).
\end{aligned}
\end{equation}
where $a (\in \mathbb{R})$ is a free parameter.
These R\'enyi genuine multi-entropies vanish for all lower $\q'$-partite entangled states with $\q' < \q$ by construction. One can directly check this property by considering pure states such as 
\begin{equation}
	\ket{\psi}_{ABC}=\ket{\psi_{1}}_{A}\otimes \ket{\psi_{2}}_{BC},
\end{equation}
\begin{equation}
	\ket{\phi}_{ABCD}=\ket{\phi_{1}}_{A}\otimes \ket{\phi_{3}}_{BCD},
\end{equation}
for $\q=3$ and $\q=4$, respectively.

\subsubsection{Relation between \texorpdfstring{$\q=3$}{q=3} R\'enyi-2  genuine multi-entropy and tripartite Markov gap}

Although both the Markov gap \eqref{eq:def-Markov-gap} and the $\q=3$ genuine multi-entropy \eqref{eq:genuine-multi-entropy-q=3} are measures of tripartite entanglement, they are distinct and yield different values for the same state. However, in the special case of R\'enyi indices $(m=2,n=2)$, they are related. We briefly review this relation here. Motivated by this relation, we can extend the Markov gap to ``genuine reflected multi-entropy" for general $\q$ later.

We start with the $\q=3$ R\'enyi-2 genuine multi-entropy \eqref{eq:genuine-multi-entropy-q=3}. By using the relation between the following quantities \cite{Penington:2022dhr,Berthiere:2020ihq,Berthiere:2023gkx,Liu:2024ulq}
\begin{equation}\label{eq:renyi-two-relation-q=2}
	\begin{aligned}
		S_{R}^{(m=2,n=2)}(A_{1}:A_{2})&=2\left[ S_{2}^{(3)}(A_{1}:A_{2}:A_{3})-S_{2}\left(A_1 A_2\right) \right]\\
		&=2\left[ S_{2}^{(3)}(A_{1}:A_{2}:A_{3})-S_{2}^{(2)}\left(A_1 A_2:A_{3}\right) \right],
	\end{aligned}
\end{equation}
that connects the $(m=2,n=2)$ R\'enyi reflected entropy with R\'enyi-2 multi-entropy, and substituting the relation into \eqref{eq:genuine-multi-entropy-q=3}, we obtain the following relation
\begin{equation}
\label{eq:relation-(q=3-n=2)-GM-and-n=2=MG}
	\begin{aligned}
		\text{GM}_2^{(3)}(A: B: C) &= S_2^{(3)}(A: B: C)-\frac{1}{2}\left(S_2^{(2)}(A: B C)+S_2^{(2)}(B: C A)+S_2^{(2)}(C: A B)\right)\\
        &=\frac{1}{2}S_{R}^{(m=2,n=2)}(A:B)-\frac{1}{2}I_2^{(2)}(A:B) \\& =\frac{1}{2} MG_{m=2, n=2}(A: B) \,,
	\end{aligned}
\end{equation}
where $I_2^{(2)}(A:B)=S_2^{(2)}(A: B C)+S_2^{(2)}(B: C A)-S_2^{(2)}(C: A B)$ is the R\'enyi-2 mutual information\footnote{By using the relation \eqref{eq:def-renyi-EE}, this R\'enyi-2 mutual information can be written as the conventional form $I_2^{(2)}(A:B)=S_{2}(A)+S_{2}(B)-S_{2}(AB)$.}.
Note that this relation does not hold for values of $n$ other than $n=2$ (and $m=2$).

\subsection{Prescriptions for constructing ``genuine reflected multi-entropy''}
The purpose of this section is to propose another generalization of the Markov gap for $\q$-partite pure states. We generalize the Markov gap based on the genuine R\'enyi multi-entropy. Since the genuine R\'enyi multi-entropy $\text{GM}_{n}^{(\q)}(A_1: \dots: A_{\q})$  vanishes for lower $\q'$-partite entangled pure states with $\q'<\q$ for any $n$, we impose the same condition on the genuine R\'enyi reflected multi-entropy $GS_{R\, (\q-1)}^{(m,n)}(A_1:\dots:A_{\q-1})$ for a given $\q$-partite pure state.
\begin{description}
  \item[\underline{Prescription A:}] The genuine R\'enyi reflected multi-entropy $ GS_{R\, (\q-1)}^{(m,n)}(A_1:\dots:A_{\q-1}) $ vanishes for all pure states that are $\q’$-partite entangled with $\q’ < \q$, for any choice of $m$ and $n$.
\end{description}

Readers may wonder what the difference is between the genuine R\'enyi reflected multi-entropy and the genuine R\'enyi multi-entropy. The key distinction lies in whether or not the reflected multi-entropy is used. Since the Markov gap contains the reflected entropy, we also impose prescription B to construct the genuine R\'enyi $(m,n)$ reflected multi-entropy $GS_{R\, (\q-1)}^{(m,n)}(A_1:\dots:A_{\q-1})$.
\begin{description}
  \item[\underline{Prescription B:}]  The genuine R\'enyi reflected multi-entropy $GS_{R\, (\q-1)}^{(m,n)}(A_1:\dots:A_{\q-1})$ is given by a linear combination of the R\'enyi reflected multi-entropies $S_{R\, (\q'-1)}^{(m,n)}$ for $\q'\le\q$ and the R\'enyi multi-entropies $S_n^{(\q' )}$ for $\q'<\q$. We treat $\q-1$ subsystems $A_1,\dots,A_{\q-1}$ symmetrically, where $A_\q$ is the subsystem traced out to obtain a $\q-1$ partite mixed state on $A_1,\dots,A_{\q-1}$ from a $\q$-partite pure state.
\end{description}
In other words, the genuine reflected multi-entropy contains the reflected multi-entropy, which is different from the genuine multi-entropy. Note that, in the construction of the reflected multi-entropy, the subsystem $ A_{\q} $ is traced out and thus plays a special role compared to the other $ \q-1 $ subsystems $ A_1, \dots, A_{\q-1} $. This necessitates treating $ A_{\q} $ differently in the definition.

Additionally, in the case of $\q=3,m=2,n=2$, the genuine R\'enyi $n=2$ multi-entropy $\text{GM}_2^{(3)}(A: B: C)$ is related to the R\'enyi $m=2,n=2$ version of the Markov gap $MG_{m=2, n=2}(A: B)$ as reviewed in eq.~(\ref{eq:relation-(q=3-n=2)-GM-and-n=2=MG}). Motivated by this, we also impose a similar relation between $GS_{R\, (\q-1)}^{(m=2,n=2)}(A_1:\dots:A_{\q-1})$ and $\text{GM}_{n=2}^{(\q)}(A_1: \dots: A_{\q})$ for $\q \ge 4$.
\begin{description} 
  \item[\underline{Prescription C:}] The genuine R\'enyi reflected multi-entropy $GS_{R\, (\q-1)}^{(m=2,n=2)}(A_1:\dots:A_{\q-1})$ for $m=2,n=2$ is related to the genuine R\'enyi multi-entropy $\text{GM}_{n=2}^{(\q)}(A_1: \dots: A_{\q})$ for $n=2$ as follows:
  \begin{align}
  GS_{R\, (\q-1)}^{(m=2,n=2)}(A_1:\dots:A_{\q-1})=2 \text{GM}_{n=2}^{(\q)}(A_1: \dots: A_{\q}).
  \end{align}
  for $\q \ge 3$. 
\end{description}

Under the prescriptions A, B, and C, the genuine R\'enyi reflected multi-entropy $GS_{R\, (\q-1)}^{(m,n)}(A_1:\dots:A_{\q-1})$ is determined. 
We then define the genuine reflected multi-entropy by taking the limits $m \to 1$ and $n \to 1$ as follows:
\begin{equation}\label{eq:def-q-multi-partite-Markov-gap}
	GS_{R\, (\q-1)}(A_1:\dots:A_{\q-1}) \coloneqq \lim_{m \to 1} \lim_{n \to 1} GS_{R\, (\q-1)}^{(m,n)}(A_1:\dots:A_{\q-1}).
\end{equation}
Later, we explicitly present the expression of $GS_{R\, (\q-1=3)}^{(m,n)}(A:B:C)$ for $\q=4$, and verify that it satisfies all three above prescriptions.

\subsection{Relation between  R\'enyi-2 multi-entropy \texorpdfstring{$S_{2}^{(\q)}$}{S_2^q} and \texorpdfstring{$(m=2,n=2)$}{(m=2,n=2)} reflected multi-entropy \texorpdfstring{$S_{R\, (\q-1)}^{(m=2,n=2)}$}{S_{R (q-1)}^{(m=2,n=2)}} }

Before presenting the explicit expression of the genuine reflected multi-entropy, we generalize the relation \eqref{eq:renyi-two-relation-q=2} to arbitrary $\q$, which is essential for establishing prescription C.
In general, the R\'enyi multi-entropy and the reflected multi-entropy are distinct measures of multipartite correlations, but in the special case of $(m=2, n=2)$, a clear relation between them emerges.
In what follows, we start from the $(m=2, n=2)$ R\'enyi reflected multi-entropy as defined in \eqref{eq:def-Zn(m)^{q-1}}, \eqref{eq:def-renyi-multi-entropy-for-reflected}, and \eqref{eq:def-renyi-reflected-multi-entropy}, 
\begin{equation}
\label{mn22reflectedmulti}
	\begin{aligned}
		S_{R\, (\q-1)}^{(m=2,n=2)}\left(A_1: A_2: \cdots: A_{\q-1}\right)
&=  S_{n=2\,(m=2)}^{(\q-1 )}\left(A_{1}A_{1}^{*}: A_{2}A_{2}^{*}: \cdots: A_{\q-1}A_{\q-1}^{*}\right)\\
&=-  \frac{1}{2^{\q-3}} \log \left[ \frac{Z_{n=2 \, (m=2)}^{(\q -1)}}{\left(Z_{1 \, (m=2)}^{(\q -1)}\right)^{2^{\q-2}}}.
	\right] \end{aligned}
\end{equation}
Let us now focus on the numerator \( Z_{n=2 \, (m=2)}^{(\q -1)} \) inside the logarithm of the expression above, and examine the corresponding diagrams that represent the contraction patterns of the reduced density matrices.
For the cases \(\q = 3\) and \(\q = 4\), the diagrams for \( Z_{n=2 \, (m=2)}^{(\q -1=2)} = Z_{n=2 \, (m=2)} \) and \( Z_{n=2 \, (m=2)}^{(\q -1=3)} \) are shown in Figures \ref{fig:Renyi-reflected-multi-q=3-n=2} and \ref{fig:Renyi-reflected-multi-q=4-n=2}, respectively.

\begin{figure}[t]
    \centering
    \begin{subfigure}[b]{0.6\linewidth}
        \includegraphics[width=\linewidth]{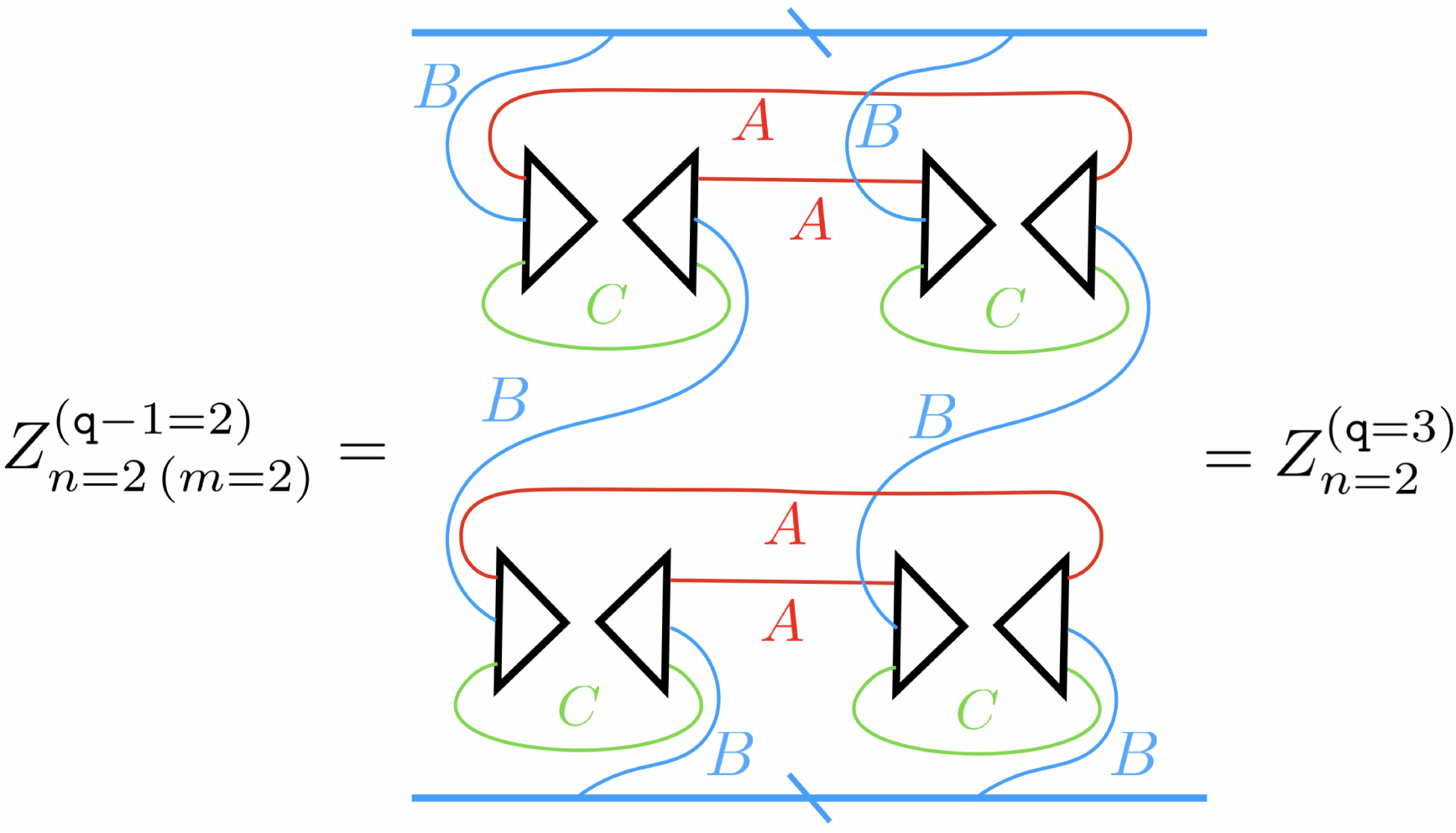}
        \caption{$\q=3$ case}
        \label{fig:equivalence-upper}
    \end{subfigure}

    \vskip\baselineskip

    \begin{subfigure}[b]{0.85\linewidth}
        \includegraphics[width=\linewidth]{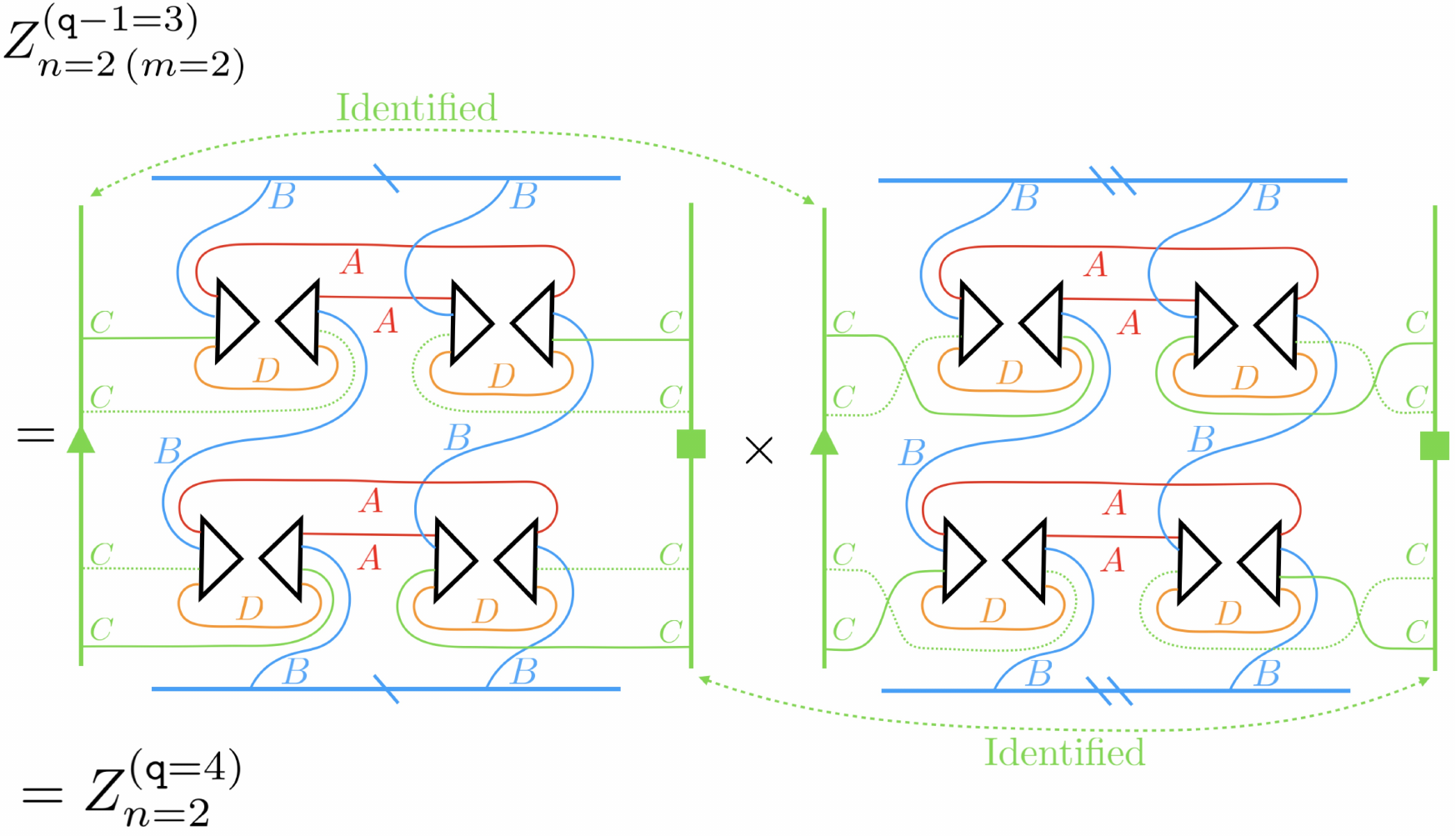}
        \caption{$\q=4$ case}
        \label{fig:equivalence-lower}
    \end{subfigure}
    
    \caption{Diagrammatic proof of the equivalences between $Z_{n=2 \, (m=2)}^{(\q-1)}$ and $Z_{n=2}^{(\q)}$ ($\q=3$ and 4 case). We can rearrange the diagrams in Figure \ref{fig:Renyi-reflected-multi-q=3-n=2} ($\q=3$) and Figure \ref{fig:Renyi-reflected-multi-q=4-n=2} ($\q=4$) cases like these figures.}
    \label{fig:equivalence-Renyi-reflected-multi-and-multi-n=2}
\end{figure}

By appropriately rearranging these diagrams, one can see that the contraction structures of \( Z_{n=2 \, (m=2)}^{(\q -1)} \) appearing in the R\'enyi-2 reflected multi-entropy are identical to those of \( Z_{n=2}^{(\q)} \) that appear in the R\'enyi multi-entropy expression \eqref{eq:def-Zn(m)^{q}}, at least for \(\q=3,4\), as illustrated in Figure \ref{fig:equivalence-Renyi-reflected-multi-and-multi-n=2}.
This observation suggests a structural equivalence between the two quantities in this special case $m=2, n=2$. By extending this diagrammatic correspondence to general \(\q\), we arrive at the following relation:
\begin{equation}\label{eq:relation-partion-function-reflected-multi-and-multi}
	\left.Z_{n=2 \, (m=2)}^{(\q -1)}\right|_{(A_{1}A_{1}^{*}: A_{2}A_{2}^{*}: \cdots: A_{\q-1}A_{\q-1}^{*})}=\left.Z_{n=2 }^{(\q)}\right|_{(A_{1}: A_{2}: \cdots: A_{\q-1}:A_{\q})}.
\end{equation} 

On the other hand, the normalization factor appearing in the denominator of the logarithm in \eqref{mn22reflectedmulti} is given by the expression \eqref{Z1mq-1norm}, and thus,
\begin{align}
\label{normalizationfactord}
Z_{1 , (m=2)}^{(\q -1)} = \tr_{A_{1},A_{2},\cdots,A_{\q-1}}\left[(\rho_{A_{1},A_{2},\cdots,A_{\q-1}})^{2}\right].
\end{align}

Therefore, by combining \eqref{mn22reflectedmulti}, \eqref{eq:relation-partion-function-reflected-multi-and-multi}, and \eqref{normalizationfactord}, the $(m=2, n=2)$ R\'enyi reflected multi-entropy can be recast into the following form,
and we obtain the desired relation:
\begin{equation}
	\begin{aligned}
		&S_{R\, (\q-1)}^{(m=2,n=2)}\left(A_1: A_2: \cdots: A_{\q-1}\right) \\
&=-  \frac{1}{2^{\q-3}} \log \frac{ Z_{n=2 }^{(\q)} }{\left(\tr_{A_{1},A_{2},\cdots,A_{\q-1}}\left[(\rho_{A_{1},A_{2},\cdots,A_{\q-1}})^{2}\right]\right)^{2^{\q-2}}} \\
&=-  \frac{1}{2^{\q-3}} \log \frac{ Z_{n=2 }^{(\q)} }{ (Z_{1}^{(\q)})^{2^{\q-1}} } + \frac{1}{2^{\q-3}} \log \frac{\left(\tr_{A_{1},A_{2},\cdots,A_{\q-1}}\left[(\rho_{A_{1},A_{2},\cdots,A_{\q-1}})^{2}\right]\right)^{2^{\q-2}} }{ (Z_{1}^{(\q)})^{2^{\q-1}} }  \\
&=2\left(-  \frac{1}{2^{\q-2}} \log \frac{ Z_{n=2 }^{(\q)} }{ (Z_{1}^{(\q)})^{2^{\q-1}} } \right) 
-2\left(- \log \frac{\tr_{A_{1},A_{2},\cdots,A_{\q-1}}\left[(\rho_{A_{1},A_{2},\cdots,A_{\q-1}})^{2}\right]}{ \left(\tr_{A_{1},A_{2},\cdots,A_{\q-1}}\left[\rho_{A_{1},A_{2},\cdots,A_{\q-1}}\right]\right)^{2} }\right)  \\
&=2\left[S_{n=2}^{(\q)}\left(A_1: A_2: \cdots: A_{\q}\right)-S_{n=2}^{(\q=2)}\left(A_1 A_2 \cdots A_{\q-1}:A_{\q}\right)\right],
\label{eq:renyi-two-relation}
	\end{aligned}
\end{equation}
where we used the following
\begin{align}
Z_{1}^{(\q)} = \tr_{A_{1},A_{2},\cdots,A_{\q-1},A_{\q}}\left[\rho_{A_{1},A_{2},\cdots,A_{\q-1},A_{\q}}\right] = \tr_{A_{1},A_{2},\cdots,A_{\q-1}}\left[\rho_{A_{1},A_{2},\cdots,A_{\q-1}}\right].
\end{align}
In the final equality, we used the definition of the multi-entropy given in \eqref{eq:def-renyi-multi-entropy}.
Note that, for the $\q = 3$ case, the relation \eqref{eq:renyi-two-relation} reduces to the known relation \eqref{eq:renyi-two-relation-q=2}.

\subsection{The genuine reflected multi-entropy for $\q=4$}
We now present the explicit expression of the {\it genuine} R\'enyi reflected multi-entropy $ GS_{R\, (\q-1=3)}^{(m,n)}(A:B:C) $ for the case \(\q=4\).
Subsequently, we verify that \( GS_{R\, (\q-1=3)}^{(m,n)}(A:B:C) \) satisfies the three prescriptions introduced above.

For a given four-partite pure state $\ket{\psi}_{ABCD}$, we construct the genuine R\'enyi $(m,n)$ reflected multi-entropy $GS_{R\, (\q-1=3)}^{(m,n)}(A:B:C)$ as follows.  
First, following prescription B, we consider a general linear combination of R\'enyi reflected multi-entropies $S_{R\, (\q'-1)}^{(m,n)}$ for $\q' \leq 4$, and R\'enyi multi-entropies $S_n^{(\q')}$ for $\q' < 4$, treating the $\q - 1 = 3$ subsystems $A$, $B$, and $C$ symmetrically within the combination.  
Each coefficient in the linear combination is then determined so that prescriptions A and C are satisfied.

Although the computation is somewhat tedious, the procedure itself is straightforward.  
Carrying out this computation, we obtain the following expression for $GS_{R\, (\q-1=3)}^{(m,n)}(A:B:C)$:
\begin{equation}\label{eq:def-Renyi-2-q=4-multi-partite-Markov-gap}
\begin{aligned}
		&\;\;\;\;\;GS_{R\, (\q-1=3)}^{(m,n)}(A:B:C)\\
		&= S_{R\, (\q-1=3)}^{(m,n)}(A:B:C)
		-\frac{1}{2} \left[ S_{R\, (\q-1=2)}^{(m,n)}(AB:C) + S_{R\, (\q-1=2)}^{(m,n)}(AC:B) + S_{R\, (\q-1=2)}^{(m,n)}(A:BC) \right] \\
		& \quad + \frac{1}{3} \left( S_{n}^{(3)}(AB: C: D) + S_{n}^{(3)}(AC: B: D) + S_{n}^{(3)}(BC: A: D) \right)\\
		& \quad -\frac{2}{3} \left( S_{n}^{(3)}(AD: B: C) + S_{n}^{(3)}(BD: A: C) + S_{n}^{(3)}(CD: A: B) \right) \\
		& \quad + 2a \left( S_{n}^{(2)}(AB: CD) + S_{n}^{(2)}(AC: BD) + S_{n}^{(2)}(AD: BC) \right) \\
		& \quad + 2\left( \frac{1}{3}-a \right) \left( S_{n}^{(2)}(ABC: D) + S_{n}^{(2)}(ABD: C) + S_{n}^{(2)}(ACD: B) + S_{n}^{(2)}(BCD: A) \right)\\
		& \quad - S_{n}^{(2)}(ABC: D),
	\end{aligned}
\end{equation}
where $a \in \mathbb{R}$ is a free parameter, and the R\'enyi reflected multi-entropies appearing in this expression are computed using the canonical purification of $(\rho_{ABC})^m = \left( \tr_D\left[ \ket{\psi}_{ABCD} \bra{\psi}\right] \right)^m$, obtained by tracing out the subsystem $D$.  
By construction, $GS_{R\, (\q-1=3)}^{(m,n)}(A:B:C)$, as defined in \eqref{eq:def-Renyi-2-q=4-multi-partite-Markov-gap}, satisfies prescription B.

Furthermore, we can rewrite $GS_{R\, (\q-1=3)}^{(m,n)}(A:B:C)$ in terms of the $\q=3$ R\'enyi genuine multi-entropy given in \eqref{eq:genuine-multi-entropy-q=3}, and the R\'enyi tripartite mutual information, as follows:
\begin{equation}\label{eq:def-another-form-Renyi-2-q=4-multi-partite-Markov-gap}
	\begin{aligned}
		&\;\;\;\;\;GS_{R\, (\q-1=3)}^{(m,n)}(A:B:C)\\
		&= S_{R\, (\q-1=3)}^{(m,n)}(A:B:C)
		-\frac{1}{2} \left[ S_{R\, (\q-1=3)}^{(m,n)}(AB:C) + S_{R\, (q-1=3)}^{(m,n)}(AC:B) + S_{R\, (\q-1=3)}^{(m,n)}(A:BC) \right] \\
		& \quad + \frac{1}{3} \left( \text{GM}_{n}^{(3)}(AB: C: D) + \text{GM}_{n}^{(3)}(AC: B: D) + \text{GM}_{n}^{(3)}(BC: A: D) \right)\\
		& \quad -\frac{2}{3} \left( \text{GM}_{n}^{(3)}(AD: B: C) + \text{GM}_{n}^{(3)}(BD: A: C) + \text{GM}_{n}^{(3)}(CD: A: B) \right) \\
		& \quad + \left( \frac{1}{6}-2a \right)I_{n}^{(3)}(A:B:C),
	\end{aligned}
\end{equation}
where $I_{n}^{(3)}(A:B:C)$ is the R\'enyi tripartite mutual information defined by
\begin{equation}
	\begin{aligned}
		I_{n}^{(3)}(A:B:C)&\coloneqq I_{n}^{(2)}(A:B)+I_{n}^{(2)}(A:C)-I_{n}^{(2)}(A:BC)\\
		&=S_{n}(A)+S_{n}(B)+S_{n}(C)-\left[ S_{n}(AB)+S_{n}(AC)+S_{n}(BC) \right]+S_{n}(ABC).
	\end{aligned}
\end{equation}

By substituting eq.~(\ref{eq:renyi-two-relation}) for $\q=3,4$ into eq.~(\ref{eq:def-Renyi-2-q=4-multi-partite-Markov-gap}) for $(m=2,n=2)$, one can explicitly confirm
\begin{align}
GS_{R\, (\q-1=3)}^{(m=2,n=2)}(A:B:C)=2 \, \text{GM}_{n=2}^{(4)}(A: B: C: D),
\end{align}
which shows that $GS_{R\, (\q-1=3)}^{(m=2,n=2)}(A:B:C)$, given in \eqref{eq:def-Renyi-2-q=4-multi-partite-Markov-gap}, satisfies prescription~C.  
Since the genuine multi-entropy $\text{GM}_{n=2}^{(4)}(A: B: C: D)$ vanishes for pure states that are $\q'$-partite entangled with $\q' < 4$, the same holds for $GS_{R\, (\q-1=3)}^{(m=2,n=2)}(A:B:C)$ by construction.

However, it is not immediately obvious whether $GS_{R\, (\q-1=3)}^{(m,n)}(A:B:C)$ vanishes for lower entangled pure states for arbitrary values of $m$ and $n$.  
We will now explicitly confirm that this is indeed the case.

In our construction of $GS_{R\, (\q-1=3)}^{(m,n)}(A:B:C)$, the $\q - 1 = 3$ subsystems $A$, $B$, and $C$ are treated symmetrically, while the subsystem $D$ is treated asymmetrically.  
To examine the behavior under lower entangled states, we consider the following three cases:

\begin{enumerate}
    \item $\ket{\psi}_{ABCD} = \ket{\psi_3}_{ABC} \otimes \ket{\psi_1}_D$,
    \item $\ket{\psi}_{ABCD} = \ket{\psi_3}_{ABD} \otimes \ket{\psi_1}_C$,
    \item general $\q'$-partite entangled pure states with $\q' < 4$.
\end{enumerate}

\subsubsection{Case 1: $GS_{R\, (\q-1=3)}^{(m,n)}(A:B:C)=0$ for $\ket{\psi}_{ABCD}=|\psi_3\rangle_{ABC}\otimes|\psi_1\rangle_{D}$}
Consider a tripartite entangled pure state $\ket{\psi}_{ABCD} = \ket{\psi_3}_{ABC} \otimes \ket{\psi_1}_D$.  
For this state, the R\'enyi (reflected) multi-entropies simplify as follows, 
\begin{align}
&S_{R\, (\q-1=3)}^{(m,n)}(A:B:C)=2S_{n}^{(3)}(A: B: C), \;\;\; S_{R\, (\q-1=2)}^{(m,n)}(AB:C)=2S_{n}^{(2)}(AB: C),\\
&S_{n}^{(3)}(AB: C: D)=S_{n}^{(2)}(AB: C),\;\;\;S_{n}^{(3)}(AD: B: C)=S_{n}^{(3)}(A: B: C),\\
&S_{n}^{(2)}(AB:CD)=S_{n}^{(2)}(AB:C),\;\;\;S_{n}^{(2)}(ABC:D)=0,\;\;\;S_{n}^{(2)}(ABD:C)=S_{n}^{(2)}(AB:C).
\end{align}
This is because the subsystem $D$ is completely unentangled with the rest, tracing it out has no effect on the reduced density matrix of $ABC$.  
Moreover, any reflected or ordinary R\'enyi multi-entropy that involves $D$ factorizes appropriately.

Substituting these relations into $GS_{R\, (\q-1=3)}^{(m,n)}(A:B:C)$ (\ref{eq:def-Renyi-2-q=4-multi-partite-Markov-gap}), we obtain
\begin{align}
GS_{R\, (\q-1=3)}^{(m,n)}(A:B:C)=0
\end{align}
for $\ket{\psi}_{ABCD}=|\psi_3\rangle_{ABC}\otimes|\psi_1\rangle_{D}$. Note that $\ket{\psi}_{ABCD}=|\psi_3\rangle_{ABC}\otimes|\psi_1\rangle_{D}$ includes bi-partite entangled states such as $|\psi_3\rangle_{ABC}=|\psi_2\rangle_{AB}\otimes|\psi_1\rangle_{C}$.

\subsubsection{Case 2: $GS_{R\, (\q-1=3)}^{(m,n)}(A:B:C)=0$ for $\ket{\psi}_{ABCD}=|\psi_3\rangle_{ABD}\otimes|\psi_1\rangle_{C}$}
Next, consider another tripartite entangled pure state $\ket{\psi}_{ABCD} = \ket{\psi_3}_{ABD} \otimes \ket{\psi_1}_C$.  
For this state, the R\'enyi (reflected) multi-entropies are again simplified due to the product structure between subsystem $C$ and the rest, and we have  
\begin{align}
&S_{R\, (\q-1=3)}^{(m,n)}(A:B:C)=S_{R\, (\q-1=2)}^{(m,n)}(A:B), \;\;\; S_{R\, (\q-1=2)}^{(m,n)}(AB:C)=0,\\
&S_{R\, (\q-1=2)}^{(m,n)}(AC:B)=S_{R\, (\q-1=2)}^{(m,n)}(A:B),\;\;\;S_{n}^{(3)}(AB: C: D)=S_{n}^{(2)}(AB: D),\\
&S_{n}^{(3)}(AC: B: D)=S_{n}^{(3)}(A: B: D),\;\;\;
S_{n}^{(2)}(AB:CD)=S_{n}^{(2)}(AB:D),\\
&S_{n}^{(2)}(ABC:D)=S_{n}^{(2)}(AB:D),\;\;\;S_{n}^{(2)}(ABD:C)=0.
\end{align}
Substituting these relations into $GS_{R\, (\q-1=3)}^{(m,n)}(A:B:C)$ (\ref{eq:def-Renyi-2-q=4-multi-partite-Markov-gap}), we also obtain
\begin{align}
GS_{R\, (\q-1=3)}^{(m,n)}(A:B:C)=0
\end{align}
for $\ket{\psi}_{ABCD}=|\psi_3\rangle_{ABD}\otimes|\psi_1\rangle_{C}$, which includes bi-partite entangled states such as $|\psi_3\rangle_{ABD}=|\psi_2\rangle_{AD}\otimes|\psi_1\rangle_{B}$. Due to the symmetry between $A,B,C$, the genuine R\'enyi reflected multi-entropy $GS_{R\, (\q-1=3)}^{(m,n)}(A:B:C)$ also vanishes for $\ket{\psi}_{ABCD}=|\psi_3\rangle_{ACD}\otimes|\psi_1\rangle_{B}$ and $\ket{\psi}_{ABCD}=|\psi_3\rangle_{BCD}\otimes|\psi_1\rangle_{A}$.

\subsubsection{Case 3: $GS_{R\, (\q-1=3)}^{(m,n)}(A:B:C)=0$ for any $\q'$-partite entangled pure states with $\q' < 4$}
One important property of the R\'enyi multi-entropy is that it is additive under tensor products of pure states~\cite{Penington:2022dhr, Gadde:2023zni}.  
Since the R\'enyi reflected multi-entropy is defined as the R\'enyi multi-entropy of a canonically purified state, it inherits this additivity under tensor products of pure states.  
Our construction of $GS_{R\, (\q-1=3)}^{(m,n)}(A:B:C)$ is given by a linear combination of R\'enyi (reflected) multi-entropies.  
Therefore, $GS_{R\, (\q-1=3)}^{(m,n)}(A:B:C)$ is also additive under tensor products of pure states.

By combining this additivity with the vanishing properties discussed above, we conclude that the genuine R\'enyi reflected multi-entropy $GS_{R\, (\q-1=3)}^{(m,n)}(A:B:C)$ vanishes for any $\q'$-partite entangled pure state with $\q' < 4$.  
This confirms that $GS_{R\, (\q-1=3)}^{(m,n)}(A:B:C)$, as defined in \eqref{eq:def-Renyi-2-q=4-multi-partite-Markov-gap}, satisfies prescription~A.  
For example, we have
\begin{align}
GS_{R\, (\q-1=3)}^{(m,n)}(A:B:C) = 0
\end{align}
for the state $\ket{\psi}_{ABCD} = \ket{\psi_2}_{AB} \otimes \ket{\psi_2}_{CD}$, which is a tensor product of bipartite entangled pure states.

Using the same procedure, by considering a general linear combination of the R\'enyi reflected multi-entropies $S_{R\, (\q'-1)}^{(m,n)}$ for $\q' \leq \q$ and the R\'enyi multi-entropies $S_n^{(\q')}$ for $\q' < \q$, with symmetric treatment of the $\q-1$ subsystems $A_1, \dots, A_{\q-1}$, and by determining the coefficients in the combination so that prescriptions A and C are satisfied, one can systematically construct the genuine R\'enyi reflected multi-entropy $GS_{R\, (\q-1)}^{(m,n)}(A_1 : \dots : A_{\q-1})$ for arbitrary $\q$.

\subsection{Computable example: $GS_{R\, (\q-1=3)}^{(m,n)}(A:B:C)$ of the GHZ state}

As a computable example, let us compute $GS_{R\, (\q-1=3)}^{(m,n)}(A:B:C)$ of $\ket{\mathrm{GHZ}_{4}}$ (\ref{eq:def-q-partie-GHZ}). The $(n,\mathtt{q})$-R\'enyi multi-entropy of GHZ state is given by
\begin{align}\label{eq:renyi-multi-entropy-qGHZ}
S_n^{(\mathtt{q})}=\frac{1-n^{\mathtt{q}-1}}{1-n}\frac{1}{n^{\mathtt{q}-2}}\log 2.
\end{align}
By using (\ref{eq:renyi-multi-entropy-for-reflected-qGHZ}) and (\ref{eq:renyi-multi-entropy-qGHZ}), one can obtain $GS_{R\, (\q-1=3)}^{(m,n)}(A:B:C)$ of $\ket{\mathrm{GHZ}_{4}}$ as
\begin{align}
GS_{R\, (\q-1=3)}^{(m,n)}(A:B:C)=\left( \frac{1}{6}-2a \right)\log2.
\end{align}
Interestingly, $GS_{R\, (\q-1=3)}^{(m,n)}(A:B:C)$ of $\ket{\mathrm{GHZ}_{4}}$ does not depend on either $m$ or $n$. In particular, this nonzero term comes from $\left( \frac{1}{6}-2a \right)I_{n}^{(3)}(A:B:C)$ in eq.~(\ref{eq:def-another-form-Renyi-2-q=4-multi-partite-Markov-gap}), and
\begin{align}\label{GSa112}
\left.GS_{R\, (\q-1=3)}^{(m,n)}(A:B:C)\right|_{a=\frac{1}{12}}=0,
\end{align}
for $\ket{\mathrm{GHZ}_{4}}$. It would be interesting to consider deeply the significance of the fact that $MG^{M(\q-1)}(A_{1}:A_{2}:\cdots:A_{\q-1})$ (\ref{MGGHZ}) and $\left.GS_{R\, (\q-1=3)}^{(m,n)}(A:B:C)\right|_{a=\frac{1}{12}}$ (\ref{GSa112}) vanish for the GHZ state.

\section{Discussion}
\label{sec:discussion}
In this paper, we propose two multipartite generalizations of the Markov gap originally defined for tripartite pure states: the \emph{Multipartite Markov gap} and the \emph{genuine reflected multi-entropy}.

For the first generalization, under the holographic assumption that the dual of the $\q$-partite multi-entropy is given by the minimal multiway cut  that divides the bulk into $\q$ subregions~\cite{Gadde:2022cqi,Gadde:2023zzj}, we introduce a new, simple lower bound on the reflected multi-entropy in terms of entanglement entropies, derived using holographic entropy computations. This bound differs from the one proposed in~\cite{Yuan:2024yfg}. Based on this inequality, we define the Multipartite Markov gap, which is guaranteed to be non-negative for holographic states. Moreover, we show that the Multipartite Markov gap in $\q$-partite subsystems is bounded from below by the conventional Markov gap between two coarse-grained subsystems obtained by partitioning the $(\q - 1)$ subsystems involved in its definition. This bound establishes a connection between the Multipartite Markov gap and the Markov recovery problem for quantum states defined on the $(\q - 1)$ subsystems. In holographic settings, this corresponds to the bulk reconstruction problem.

To explore whether a similar connection holds in non-holographic systems, we study the $(m=2,n=2)$ R\'enyi Multipartite Markov gap and its relation to the Markov recovery problem in various four-qubit states. We find that when the $(m=2,n=2)$ R\'enyi Multipartite Markov gap is below a certain threshold (approximately $0.346574$), exact recovery is possible using the rotated Petz map. Conversely, when the gap exceeds this threshold, approximate recovery becomes possible, although the fidelity between the recovered and original states may vary in a non-monotonic manner with respect to the gap. Nevertheless, since a small $(m=2,n=2)$ R\'enyi Multipartite Markov gap below the threshold ensures exact recovery, this quantity serves as a useful indicator for identifying quantum states that admit exact recovery.

Subsequently, based on quantum information-theoretic considerations without invoking holography, we introduce another multipartite generalization of the Markov gap: the genuine reflected multi-entropy. This quantity in $\q$-partite subsystems is defined by requiring the following three properties: (1) it vanishes for all possible $\q'$-partite entangled pure states with $\q' < \q$; (2) it is expressed as a linear combination of reflected multi-entropy and original multi-entropy; and (3) the R\'enyi-$(m=2, n=2)$ genuine reflected multi-entropy $GS_{R\, (\q-1)}^{(m,n)}$ equals twice the R\'enyi-2 genuine multi-entropy~\cite{Iizuka:2025ioc}, {\it i.e.},
$GS_{R\, (\q-1)}^{(m=2,n=2)} = 2\, \mathrm{GM}^{(\q)}_{n=2}$. 

By construction, the $\q$-partite genuine reflected multi-entropy thus measures only genuine $\q$-partite entanglement. While this quantity is closely related to genuine multi-entropy and indeed proportional to it in the R\'enyi-2 case, for general R\'enyi indices the two differ, as the genuine reflected multi-entropy involves both reflected multi-entropy and original multi-entropy.

Finally, we conclude this paper by presenting several open questions.
\begin{itemize}
    \item The definition of the Multipartite Markov gap relies heavily on the inequality~\eqref{eq:general-q-lower-bound-reflected}, which is derived using holography as discussed in Section~\ref{sec:MMG}. For the tripartite case $\q=3$, the corresponding inequality~\eqref{eq:inequaitli-q=3} can be derived from strong subadditivity, without invoking holography.

    On the other hand, it remains unclear whether the inequality~\eqref{eq:general-q-lower-bound-reflected} can be derived from a strong-subadditivity-like property in the absence of holography. Due to the complexity of evaluating the partition function required for computing the general R\'enyi reflected multi-entropy, it is currently difficult to test this directly through analytic examples. However, for R\'enyi indices with $(m=2, n=2)$, numerical evaluation is possible, as demonstrated in Subsection~\ref{subsub:Examples-recovery-qubit}.

    As seen in Table~\ref{tab:Summary-qubits-Rearrange}, the $(m=2,n=2)$ R\'enyi Multipartite Markov gaps are positive for various four-qubit states except for the GHZ state. This suggests that the inequality~\eqref{eq:ineq-q=4-reflected-multientropy} holds in these cases.

    We expect that the $(m=2,n=2)$ R\'enyi version of inequality~\eqref{eq:ineq-q=4-reflected-multientropy} holds generally, except perhaps for the GHZ family or specific linear combinations of GHZ and other states, implying that the Multipartite Markov gap generically takes nonzero values.

    It would be interesting to investigate whether this inequality holds in more general scenarios (e.g., arbitrary $(m,n)$ R\'enyi cases), and equivalently, whether the Multipartite Markov gap is always non-negative.

    \item As discussed in Subsection~\ref{subsub:lower-bound-q}, we have shown under holographic assumptions that the Multipartite Markov gap is bounded from below by the lower $\q$ Multipartite Markov gap. However, it is highly nontrivial whether such a lower bound exists for general non-holographic states or for general R\'enyi indices.

    In Subsection~\ref{subsec:Markov-recovery}, we focused on the $(m=2, n=2)$ R\'enyi Multipartite Markov gaps and the conventional Markov gaps for various four-qubit states, as summarized in Table~\ref{tab:Summary-three-MG}. To test the lower bound conjecture, we examine the $(m=2,n=2)$ R\'enyi conventional Markov gaps instead of the $m,n \to 1$ case. In Table~\ref{tab:Summary-three-Renyi-MG}, we report the values of the $(m=2,n=2)$ R\'enyi Multipartite Markov gaps and the conventional $(m=2,n=2)$ R\'enyi Markov gaps for three bipartitions.

    From this data, we observe that—except for the GHZ state—the $(m=2,n=2)$ R\'enyi Multipartite Markov gaps are consistently larger than the conventional ones. This supports the conjecture that a lower bound exists for general states, excluding the GHZ family or special mixtures involving GHZ.

    Further investigation is required to determine whether such a lower bound universally exists for general non-holographic quantum states and more general R\'enyi settings.

    \item In this paper, we proved the {\it holography-independent} inequalities relating the (rotated) Petz recovery map, the conditional multi-information and the past-sequential reflected sum as shown in section \ref{subsec:nonholo-seqpetz}. However, the proof was obtained by combining tripartitions of the system and thus it is not a direct proof in the genuinely multipartite setting, {\it i.e.}, the inequality is not related to the reflected multi-entropy. To make connection between our multipartite Markov gap and the recovery operation, we need to find the more optimal inequality between the past-sequential reflected sum and the reflected multi-entropy since handling a many-body system partition by partition is inefficient.  We currently lack an efficient one-shot method that treats the whole system at once.  We leave this problem as a future problem.

\end{itemize}
Exploring these open questions promises to deepen our understanding of multipartite entanglement and quantum recoverability, and may ultimately uncover new structural principles governing the interplay between entropy, geometry, and holography. 
	 
    \begin{table}[t]
\centering
\begin{tabular}{|c|c|c|c|c|}
\hline
\textbf{States} & $MG^{M(\q-1=3)}_{m,n}$ & $MG_{m,n}(A:BC)$ & $MG_{m,n}(AC:B)$ & $MG_{m,n}(AB:C)$ \\
\hline
$\left|\mathrm{GHZ}_4\right\rangle$ & $-0.346574$ & $0$ & $0$ & $0$ \\
\hline
$\left|\mathrm{C}_1\right\rangle$ & \multirow{2}{*}{$0.346574$} & $0$ & $0$ & $0$ \\
\cline{1-1} \cline{3-5}
$|Y C\rangle$ & & $0$ & $0$ & $0$ \\
\hline
$|D[4,(2,2)]\rangle$ & $0.396291$ & $0.351398$ & $0.351398$ & $0.351398$ \\
\hline
$\left|\Psi_1^{\max }\right\rangle$ & $0.483767$ & $0.179666$ & $0.241916$ & $0.223649$ \\
\hline
$\left|BSSB_4\right\rangle$ & $0.634256$ & $0$ & $0.182322$ & $0.182322$ \\
\hline
$|\mathrm{HS}\rangle$ & \multirow{3}{*}{$0.81093$} & $0.251314$ & $0.251314$ & $0.251314$ \\
\cline{1-1} \cline{3-5}
$|H D\rangle$ & & $0.251314$ & $0.251314$ & $0.251314$ \\
\cline{1-1} \cline{3-5}
$|L\rangle$ & & $0.251314$ & $0.251314$ & $0.251314$ \\
\hline
$\left|\mathrm{W}_4\right\rangle$ & $0.862262$ & $0.117783$ & $0.117783$ & $0.117783$ \\
\hline
\end{tabular}
\caption{Values of the $(m=2,n=2)$ R\'enyi Multipartite Markov gap $MG^{M(\q-1=3)}_{m=2,n=2}(A:B:C)$ and the conventional $(m=2,n=2)$ R\'enyi Markov gaps $MG_{m=2,n=2}$ of various four-qubit states for three possible divisions.}
\label{tab:Summary-three-Renyi-MG}
\end{table}

\section*{Acknowledgments}
The work of N.I. and A.M. was supported in part by MEXT KAKENHI Grant-in-Aid for Transformative Research Areas A “Extreme Universe” No. 21H05184. The work of N.I. was also supported in part by NSTC of Taiwan Grant Number 114-2112-M-007-025-MY3.

\appendix

\section{List of identities for entanglement entropy}
In this appendix, we list identities for entanglement entropy, which we use in the main body of this paper. 

\newpage
\begin{itemize}
	\item 
    \leavevmode\vspace{-1.9em}
    \begin{flalign}\label{eq:q=4-identity-1}
\begin{aligned}
		 &S(A) + S(B)+ S(C) - S(ABC)\\
        &= \frac{1}{3}\big[I^{(2)} (A: B C)+I^{(2)}(B: A C)+I^{(2)}(C: A B) \\
        &\hspace{6cm}+ I^{(2)}(A: B)+I^{(2)}(B: C)+I^{(2)}(A: C) \big]
	\end{aligned}
&&
\end{flalign}

\item 
  \leavevmode\vspace{-1.9em}
\begin{flalign}\label{eq:q=4-identity-2}
\begin{aligned}
		 S(A) + S(B)+ &S(C) - S(ABC)\\
        &= \left[ I^{(2)}(A:B)+I^{(2)}(A:C)+I^{(2)}(B:C)-I^{(3)}(A:B:C)\right].
	\end{aligned}
&&
\end{flalign}
Here, $I^{(3)}(A:B:C)$ is tripartite information defined by
\begin{equation}
	\begin{aligned}
		I^{(3)}(A: B: C)&=S(A) + S(B) + S(C) - S(AB) - S(AC) - S(BC) + S(ABC)\\
		&=I^{(2)}(A: B)+I^{(2)}(A: C)-I^{(2)}(A: B C).
	\end{aligned}
\end{equation}

\item
    \leavevmode\vspace{-1.9em}
 \begin{equation}\label{eq:q=5-identity-1}
 	\begin{aligned}
 		&  S(A)+S(B)+S(C)+S(D)-S(ABCD)\\
		&= \frac{1}{4} \bigg[I^{(2)}(A:BCD)+I^{(2)}(B:CDA)+I^{(2)}(C:DAB)+I^{(2)}(D:ABC)\\
		&  \qquad\qquad \qquad \quad + I^{(2)}(B:CD)+ I^{(2)}(C:DA)+ I^{(2)}(D:AB)+ I^{(2)}(A:BC)\\
		&  \qquad \qquad\qquad \qquad\qquad \qquad \quad + I^{(2)}(C:D)+ I^{(2)}(D:A)+ I^{(2)}(A:B)+ I^{(2)}(B:C)  \bigg]
 	\end{aligned}
 \end{equation}

\item 
    \leavevmode\vspace{-1.9em}
 \begin{equation}\label{eq:q=6-identity-1}
	\begin{aligned}
		& S(A)+S(B)+S(C)+S(D)+S(E)-S(ABCDE)\\
		&= \frac{1}{5} \bigg[I^{(2)}(A:BCDE)+I^{(2)}(B:CDEA)+I^{(2)}(C:DEAB)+I^{(2)}(D:EABC)+I^{(2)}(E:ABCD)\\
		&\qquad \quad + I^{(2)}(B:CDE)+ I^{(2)}(C:DEA)+ I^{(2)}(D:EAB)+ I^{(2)}(E:ABC)+ I^{(2)}(A:BCD)\\
		&\qquad \qquad \quad + I^{(2)}(C:DE)+ I^{(2)}(D:EA) + I^{(2)}(E:AB)+ I^{(2)}(A:BC)+ I^{(2)}(B:CD)\\
		 &\qquad \qquad \qquad \quad + I^{(2)}(D:E)+I^{(2)}(E:A)+I^{(2)}(A:B)+I^{(2)}(B:C)+I^{(2)}(C:D)\bigg]
	\end{aligned}
\end{equation}

\newpage
\item 
    \leavevmode\vspace{-1.9em}
\begin{equation}\label{eq:q-identity-1}
	\begin{aligned}
		&S(A_{1}) + S(A_{2})+ \cdots + S(A_{\q-1}) - S(A_{1}A_{2}\cdots A_{\q-1})\\
		&=\frac{1}{\q-1} \bigg[ \big\{ I^{(2)}(A_{1}:A_{2}A_{3}\cdots A_{\q-1})+I^{(2)}(A_{2}:A_{3}A_{4}\cdots A_{\q-1}A_{1})+\cdots\\
		&\hspace{20em}\cdots +I^{(2)}(A_{\q-1}:A_{1}A_{2}\cdots A_{\q-3}A_{\q-2})\big\}\\
		&\qquad \qquad  +\big\{ I^{(2)}(A_{2}:A_{3}A_{4}\cdots A_{\q-1})+I^{(2)}(A_{3}:A_{4}A_{5}\cdots A_{\q-1}A_{1})+\cdots\\
		&\hspace{20em}\cdots+I^{(2)}(A_{1}:A_{2}A_{3}\cdots A_{\q-3}A_{\q-2})\big\}\\
		&\qquad \qquad   +\big\{ I^{(2)}(A_{3}:A_{4}A_{5}\cdots A_{\q-1})+I^{(2)}(A_{4}:A_{5}A_{6}\cdots A_{\q-1}A_{1})+\cdots\\
		& \hspace{20em} \cdots+I^{(2)}(A_{2}:A_{3}A_{4}\cdots A_{\q-3}A_{\q-2})\big\}\\
		&\qquad \qquad   + \cdots\\
		&\qquad \qquad   +\big\{ I^{(2)}(A_{k}:A_{k+1}A_{k+2}\cdots A_{\q-1})+I^{(2)}(A_{k+1}:A_{k+2}A_{k+3}\cdots A_{\q-1}A_{1})+\cdots\\
		&\hspace{18em}\cdots+I^{(2)}(A_{k-1}:A_{k}A_{k+2}\cdots A_{\q-3}A_{\q-2})\big\}\\
		&\qquad \qquad   + \cdots\\
		&\qquad \qquad   +\big\{ I^{(2)}(A_{\q-3}:A_{\q-2} A_{\q-1})+I^{(2)}(A_{\q-2}:A_{\q-1}A_{1})+\cdots+I^{(2)}(A_{\q-4}:A_{\q-3}A_{\q-2})\big\}\\
		&\qquad \qquad   +\big\{ I^{(2)}(A_{\q-2}:A_{\q-1})+I^{(2)}(A_{\q-1}:A_{1})+\cdots+I^{(2)}(A_{\q-3}:A_{\q-2})\big\}\bigg].
	\end{aligned}	
\end{equation}
Here, the above expression is constructed according to the following procedure:
\begin{enumerate}
\item We divide the subsystems $A_1, A_2, A_3, \dots, A_{\q-1}$ as $A_1 : A_2 A_3 \cdots A_{\q-1}$ and write down the corresponding mutual information. We then cyclically permute the subsystems and sum over all the resulting expressions of the same form.
\item We remove $A_1$ and consider the division $A_2 : A_3 \cdots A_{\q-1}$, compute its mutual information, and again sum over all cyclic permutations.
\item This procedure is repeated: we remove $A_2$, then $A_3$, and so on, until we reach the final division $A_{\q-2} : A_{\q-1}$. At each step, we compute the mutual information and sum over the corresponding cyclic permutations.
\end{enumerate}
The final result obtained in this way, after dividing by $\q - 1$, reproduces the original expression.

\end{itemize}

\bibliographystyle{JHEP}
\bibliography{reference}

\end{document}